\documentclass[preprint,showpacs,preprintnumbers,amsmath,amssymb,prb]{revtex4-1}
\usepackage{graphicx}
\usepackage{dcolumn}
\usepackage{bm}
\usepackage{epsfig}
\usepackage{subfigure}
\begin{document}

\preprint{PREPRINT}

\title{Relation between occupation in the first coordination shells and Widom
  line in Core-Softened Potentials}

\author{Evy Salcedo} \affiliation{Departamento de F\'{\i}sica,
  Universidade Federal de Santa Catarina, 88010-970, Florian\'opolis,
  SC, Brazil}

\author{Ney M. Barraz Jr.}
\affiliation{Universidade Federal da Fronteira Sul, \\ 85303-160, Laranjeiras do Sul, Paran\'a, PR, Brazil }

\author{Marcia C. Barbosa} \affiliation{Instituto de F\'{i}sica,
  Universidade Federal do Rio Grande do Sul, 91501-970, Porto Alegre,
  Rio Grande do Sul, Brazil}

\date{\today}
\begin{abstract}
  Three core-softened families of potentials are checked for the
  presence of density and diffusion anomalies. These potentials 
exhibit a repulsive core with a softening region and
at larger distances an attractive well.  We found that the region
in the 
  pressure-temperature phase diagram  in which the anomalies are
  present increases if the slope between the core-softened
scale and the attractive  part
of the potential decreases. The 
anomalous region also increases if the range of 
the core-softened or of the attractive part
of the potential decreases. We also show that the presence
of the density anomaly
is consistent with the  non monotonic  changes 
of the   radial distribution function at 
each one of the two scales when temperature and density are 
varied.  Then, using this anomalous behavior of
the structure  we
show that the pressures and the temperatures in which the 
radial distribution functions of the two length scales
are equal are  identified with the Widom line. 
\end{abstract}
\pacs{64.70.Pf, 82.70.Dd, 83.10.Rs, 61.20.Ja}
\maketitle

\section{\label{sec:introduction}Introduction}

Core-softened $(CS)$ potentials have been attracting attention due to
their connections with the anomalous behavior of liquid systems
including water. These potentials, U(r), exhibit a repulsive core with
a softening region limited by $r_1<r<r_2$ where
$d(rf)/dr>0$ with $f=-dU/dr$~\cite{De91}.  Despite their 
simplicity, these models 
originate from the
desire of constructing a simple two-body isotropic potential capable
of describing the complicated features of systems interacting via
anisotropic potentials ~\cite{He70, Sc00, Bu02, Ca03,
Ca05, Bu03, Sk04, Fr02, Ba04, Ol05, He05a, He05b, Ja98, Wi02, Ma04,
Ku04, Xu05, Ol06a, Ol06b, Ol07, Ol08a, Ol08b, Ol09, Gr09, Lo07,
Fo08}. This procedure generates models that are
analytically tractable and computationally less expensive than the
atomistic models. Moreover, they are lead to conclusions that are more
universal and are related to families of atomistic
systems~\cite{Ke75,An76,Ts91,An00}.

One of the features that has been successfully described by many of
these models is the density anomaly. For water the specific volume at
ambient pressure starts to increase when cooled below $T\approx4
^oC$. The anomalous behavior of water was first suggested 300 years
ago~\cite{Wa64} and was confirmed by a number of
experiments~\cite{Ke75,An76}. Besides, between $0.1$ MPa and $190$ MPa
water also exhibits an anomalous increase of
compressibility~\cite{Sp76,Ka79} and, at atmospheric pressure, an
increase of isobaric heat capacity upon cooling~\cite{An82,To99}. For
the case of water the density anomaly is attributed to the presence of
hydrogen bonds between neighbor molecules. As the temperature
increases the bonds break and the density increases. However, other
systems such as Te, \cite{Th76} Ga, Bi,~\cite{Handbook}
S,~\cite{Sa67,Ke83}, Ge$_{15}$Te$_{85}$,~\cite{Ts91},
silica,~\cite{An00,Ru06b,Sh02,Po97} silicon~\cite{Sa03} and
BeF$_2$,~\cite{An00} show the same density anomaly without presenting
hydrogen bonds what suggests that the mechanism for the presence of
density anomaly might be more universal.

In compass with the presence of the density anomaly in water a few
years ago it was suggested that there are two liquid phases, a low
density liquid (LDL) and a high density liquid (HDL)~\cite{Po92}.  The
critical point ending this transition, found only in computer
simulations is located at the supercooled region beyond the line of
homogeneous nucleation and thus cannot be experimentally
measured. Even with this limitation, this hypothesis has been
supported by indirect experimental results~\cite{Mi98, Sp76,Xu05}. The
presence of two liquid phase and of second critical point is also
observed in certain  $CS$ potentials~\cite{Ca05, Bu03, Sk04, Fr02, Ba04, Ol05, He05a, He05b, Ja98, Wi02, Ma04,
Ku04, Xu05, Ol08a, Ol08b, Ol09, Gr09, Lo07,
Fo08}.

Which are the conditions for a CS potential to exhibit density anomaly
and two liquid phases? A definitive answer to this question is still
missing.  There are, however, a few clues. If a CS potential has
discontinuous forces it presents two liquid phases but no density
anomaly~\cite{Fr01} is observed. However, once the CS potential is
modified to have continuous forces, the anomalies appear~\cite{Ol08b}.

Recently it has been proposed that a CS potential exhibits density
anomaly if the two length scales identified with
the softened region would be accessible~\cite{Ol09,
Ol08a, Vi10} what can be understood as follows. The radial
distribution, $g(r)$ of a CS potential has peaks at $g(r_1)$ and
$g(r_2)$ where $r_1$ and $r_2>r_1$ are the two length scales of the CS
potential~\cite{Ol06b, De91}. If $\partial g(r) /\partial \rho
|_{r=r_1} \partial g(r) /\partial \rho |_{r=r_2}<1$ then the system
would have density anomaly.

In this paper we test if this link between the behavior of the
structure (radial distribution function) and the thermodynamic
anomalies holds for a number of two length scales potentials. We study
the pressure temperature phase diagram of a two Fermi
model~\cite{Ab11}. The advantage of this model is that by changing few
parameters is possible to vary the distance and the difference in
energy between the two length scales without introducing extra
scales. Moreover the length scales are well defined.

Hence, having identified a connection
between the density anomaly and the 
behavior of the structure, we
also test if the radial distribution 
function is also related to the
presence of two liquids predicted 
for these CS potentials.  We show 
that  the pressures and temperatures in which the
radial distribution function associated with one scale equals the
radial distribution function of the other scales is linked with
peaks in the constant pressure specific heat, namely the Widom line.

The remaining of this paper goes as follows. In Sec.~\ref{sec:model}
the model is introduced and the simulation details are presented.  In
Sec.~\ref{sec:results} the pressure-temperature phase diagram is
presented together with the behavior of the radial distribution
function with density and temperature. Conclusions are presented in
sec.~\ref{sec:conclusions}.

\section{\label{sec:model}    The Model}

Our system consists of $N$ identical particles interacting through a
continuous pair potential obtained by the addition of $3$ different
Fermi-Dirac distributions~\cite{Ab11},
\begin{equation}
  U=\sum_{i=1}^{3}
  \frac{\varepsilon_{i}}
        {\exp\left(\frac{r-r_{oi}}
        {\sigma_{i}}\right)+\alpha_i}.
  \label{eq:potential}
\end{equation}
The resulting expression describes a family of pair interaction
potentials discriminated by different choices of the parameters
$\{\varepsilon_{i}, r_{oi}, \sigma_{i}, \alpha_i\}$. Appropriated
choices of the parameters allow us to obtain potentials that go from
a smooth two length scales potential to a sharp, almost
discontinuous, square potential~\cite{Ma05,Ab11}.

\begin{figure}[ht]
\begin{centering}
\includegraphics[clip,width=5cm]{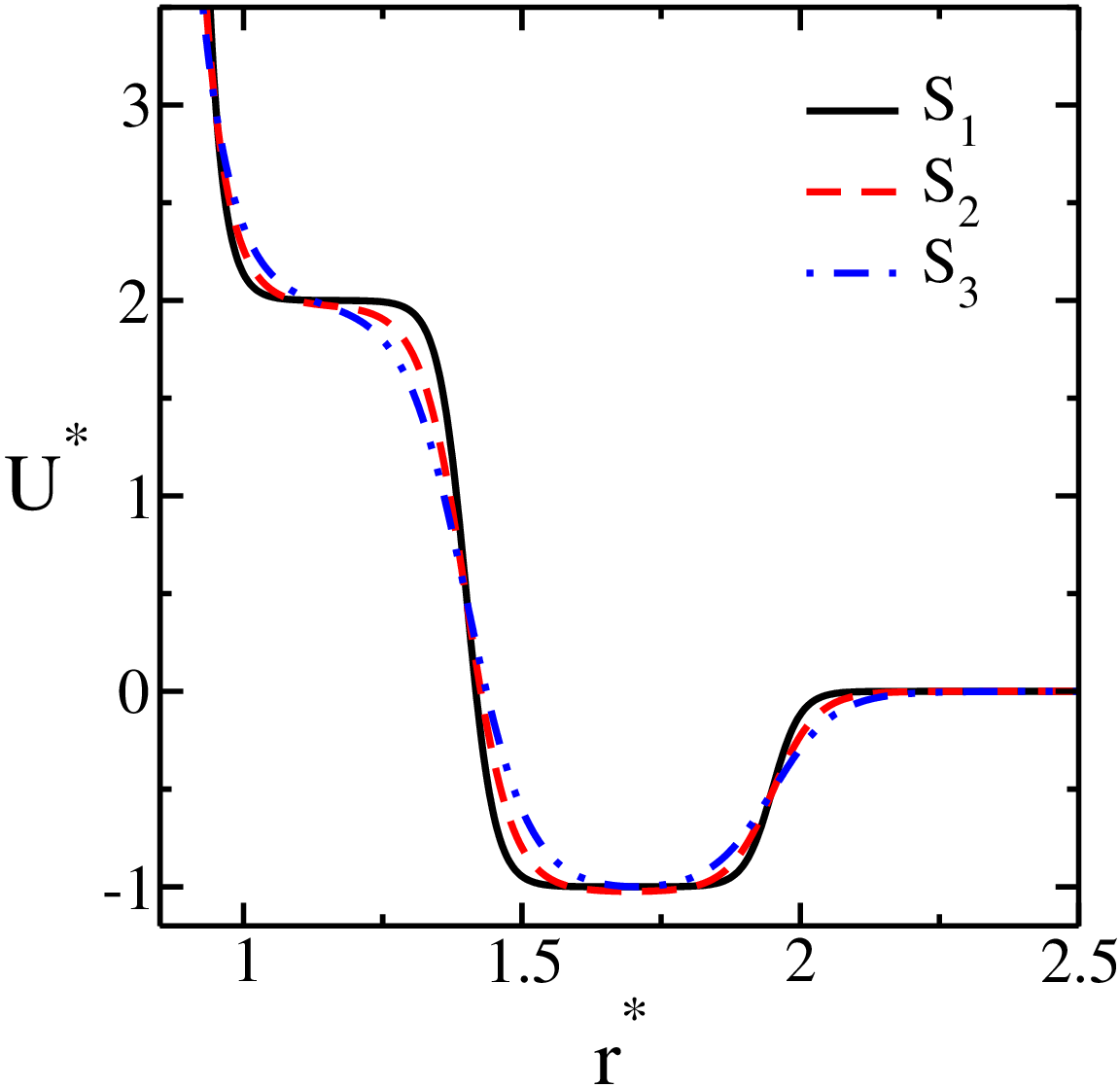}
\includegraphics[clip,width=5cm]{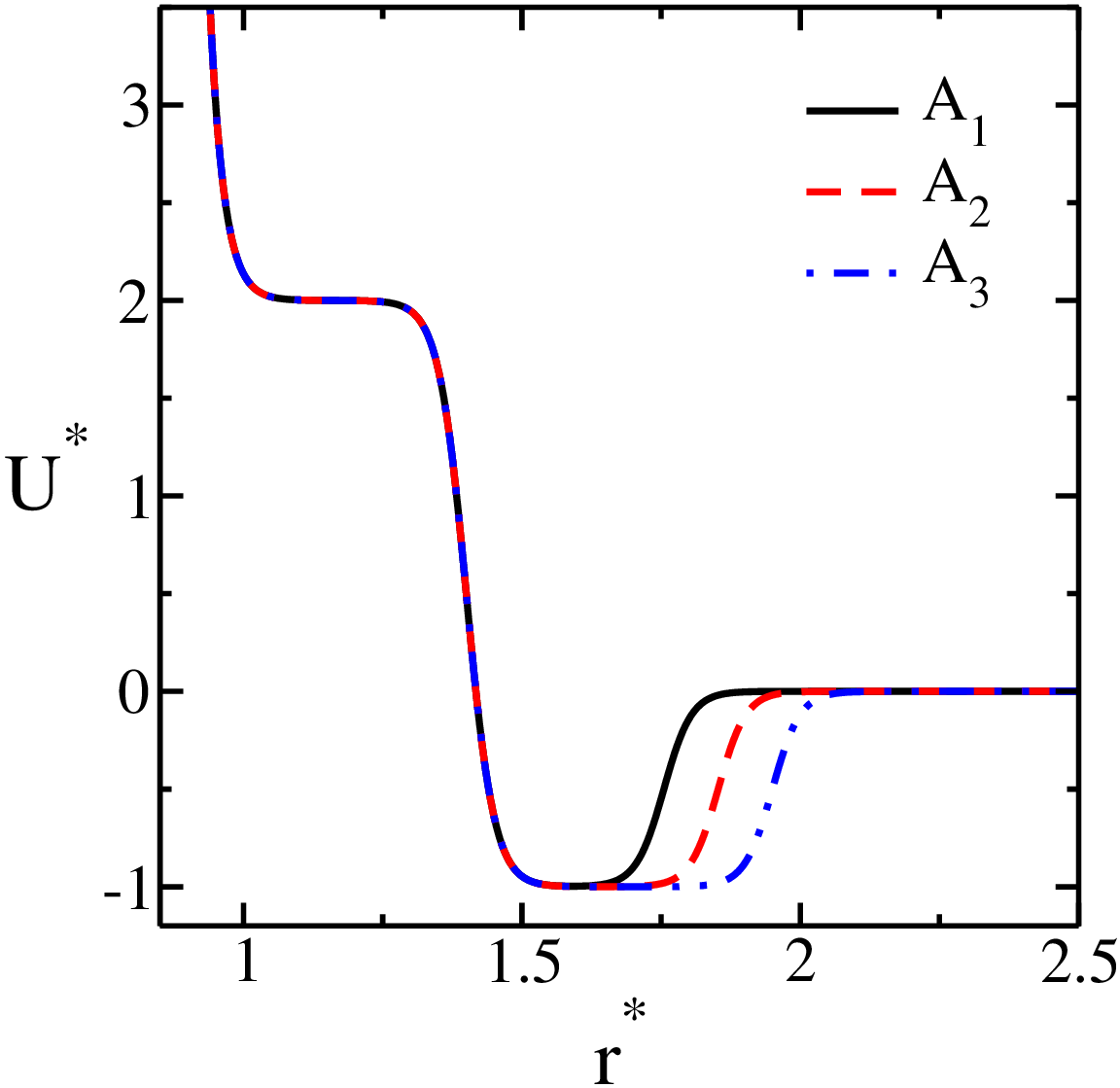}
\includegraphics[clip,width=5cm]{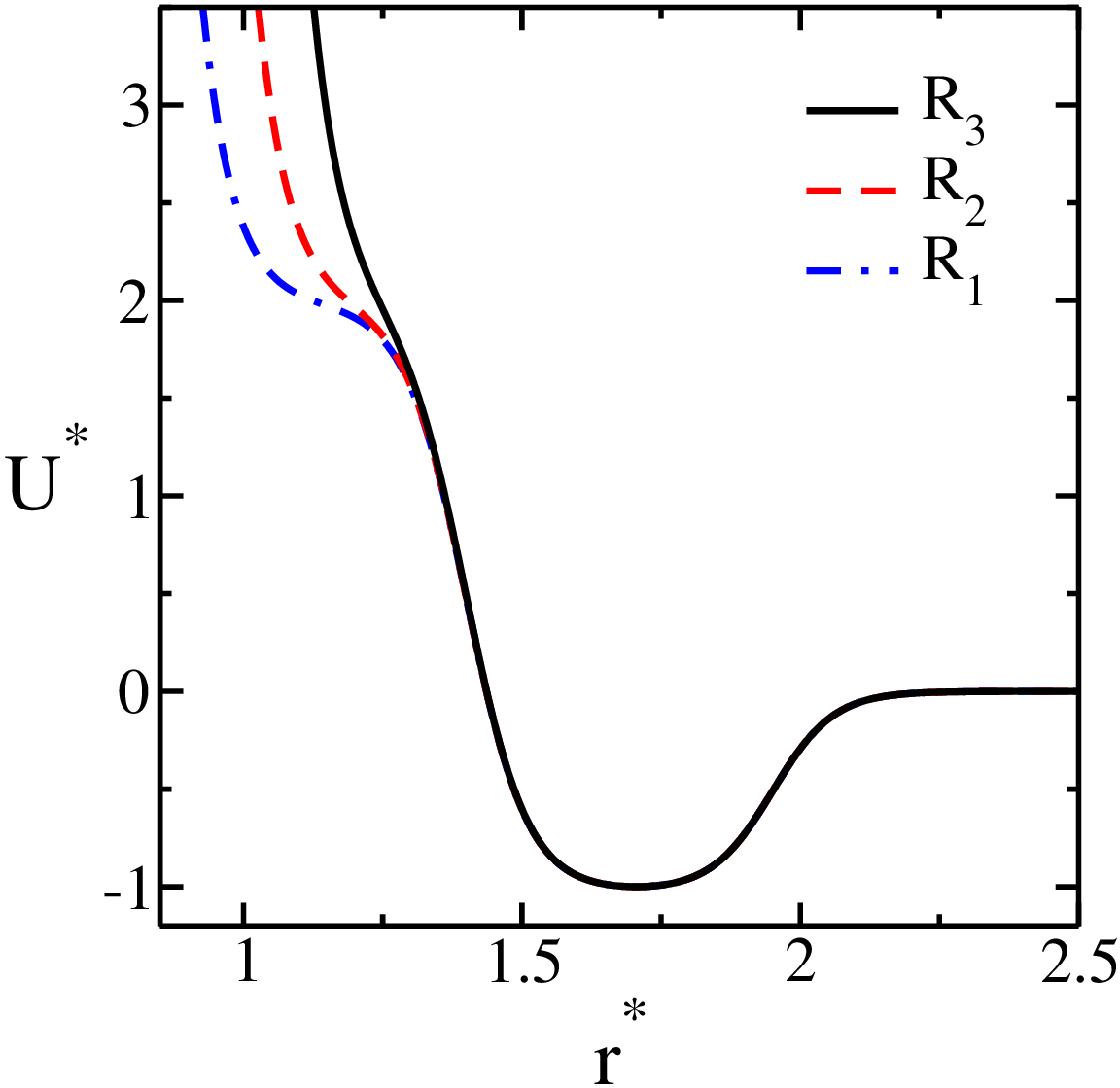}
\par\end{centering}
\caption{Interaction potential.}
\label{fig:potential}
\end{figure}

In the Table~\ref{table:ref} nine different sets of parameters are
shown, organized in three families named, $S$, $A$, and $R$. As shown
in Figure~\ref{fig:potential}, for each family a specific
characteristic of pair interaction potential is tuned. Then it is
possible to test the effect of changing the two length scales in the
pressure temperature phase diagram.

In the potentials $S$ the slope between the two length scales is
varied. Then it is possible to check if the slope between the two
length scales controls the location in the pressure-temperature phase
diagram of the density anomalous region as suggested by Yan et
al.~\cite{Ya08}.

In the case of the potentials $A$, the attractive length becomes
broader. Consequently using this potential we test if increasing the
range of the attraction leads to a decrease in the critical pressure
as proposed by Skibinsky et al.~\cite{Sk04}.

In the case of the potentials $R$, the repulsive length scale becomes
broader. Therefore this family of potentials is appropriated to
observe if the enlargement of the repulsive length scale leads to a
decrease in the liquid-liquid critical pressure and to an increase in
the liquid-liquid critical temperature as suggested by Skibinsky et
al.~\cite{Sk04}. In addition to verify the assumptions of Yan et
al.~\cite{Ya08} and of Skibinsky et al.~\cite{Sk04} related to
criticality, these three families of potentials are the perfect
scenario to check our hypothesis that the density anomaly region in
the pressure-temperature phase diagram is delimited by properties of
the radial distribution function at the two length scales.

\begin{table}[b]
\caption{Parameters for potentials $S$, $A$ and $R$ in reduced units of
$\varepsilon$ and $\sigma=r_{o1}/0.950$.}
  \begin{tabular}{c|ccc||ccc||ccc}
    \hline 
    Parameter            & $S_{1}$  & $S_{2}$  & $S_{3}$ & $A_{1}$  & $A_{2}$  & $A_{3}$ & $R_{1}$ & $R_{2}$  & $R_{3}$ \\
    \hline 
    $\varepsilon^*_{1}$  & $1.000$  & $1.000$  & $1.000$ & $1.000$  & $1.000$  & $1.000$ & $1.000$ & $1.000$  & $1.000$ \\
    $\varepsilon^*_{2}$  & $3.000$  & $3.000$  & $3.000$ & $3.000$  & $3.000$  & $3.000$ & $3.000$ & $3.000$  & $3.000$ \\
    $-\varepsilon^*_{3}$ & $1.000$  & $1.027$  & $1.023$ & $1.023$  & $1.023$  & $1.023$ & $1.023$ & $1.023$  & $1.023$ \\
    $r^*_{o1}$           & $0.950$  & $0.950$  & $0.950$ & $0.950$  & $0.950$  & $0.950$ & $0.950$ & $1.050$  & $1.150$ \\
    $r^*_{o2}$           & $1.400$  & $1.400$  & $1.400$ & $1.400$  & $1.400$  & $1.400$ & $1.400$ & $1.400$  & $1.400$ \\
    $r^*_{o3}$           & $1.950$  & $1.950$  & $1.950$ & $1.755$  & $1.8525$ & $1.950$ & $1.950$ & $1.950$  & $1.950$ \\
    $\sigma^*_{1,2,3}$   & $0.025$  & $0.040$  & $0.055$ & $0.025$  & $0.025$  & $0.025$ & $0.055$ & $0.055$  & $0.055$ \\
    $\alpha^*_{1}$       & $0.000$  & $0.000$  & $0.000$ & $0.000$  & $0.000$  & $0.000$ & $0.000$ & $0.000$  & $0.000$ \\      
    $\alpha^*_{2,3}$     & $1.000$  & $1.000$  & $1.000$ & $1.000$  & $1.000$  & $1.000$ & $1.000$ & $1.000$  & $1.000$ \\
    \hline 
  \end{tabular}
  \label{table:ref} 
\end{table}

The thermodynamic and dynamic behavior of the systems were obtained
using $NVT$ molecular dynamics using Nose-Hoover heat-bath with
coupling parameter $Q = 2$. The system is characterized by $500$
particles in a cubic box with periodic boundary conditions,
interacting with the intermolecular potential described above.

Standard periodic boundary conditions together with
predictor-corrector algorithm were used to integrate the equations of
motion with a time step $\Delta t^{*}=0.002$ and potential cut off
radius $r_{c}^{*}=2.5$. The initial configuration is set on solid or
liquid state and, in both cases, the equilibrium state was reached
after $t_{eq}^{*}=1000$. From this time on the physical quantities
were stored in intervals of $\Delta t_R^* = 1$ during $t_R^* =
1000$. The system is uncorrelated after $t_d^*=10$, from the
velocity auto-correlation function, and $50$ decorrelated samples
were used to get the average of the physical quantities. The
thermodynamic stability of the system was checked analyzing the
dependence of pressure on density, by the behavior of the energy and
also by visual analysis of the final structure, searching for
cavitation.

In what follows we take $\varepsilon_1$,
$\sigma=r_{o1}/0.950$ as fundamental units for energy and distance, respectively, and
all physical quantities are expressed in reduced units, namely
\begin{eqnarray}
T^{*}&\equiv & \frac{k_{B}T}{\varepsilon_1}\; \nonumber \\ \rho^{*}&\equiv
&\rho \sigma^3 \nonumber \\ P^*&\equiv& \frac{P\sigma^3}{\varepsilon_1} \nonumber \\ D^*&\equiv&
\frac{D(m/\varepsilon_1)^{1/2}}{\sigma}
\end{eqnarray}
where T, P and D are respectively temperature, pressure and diffusion
coefficient.  The diffusion coefficient is obtained from the
expression:
\begin{equation}
  D = \lim_{t \rightarrow \infty} \frac {\langle \left[ \vec{r}_j (t_0
      + t) - \vec{r}_j(t_0) \right]^2 \rangle_{t_0}} {6t}
  \label{eq:diffusion}
\end{equation}
where $\vec{r}_j(t)$ are the coordinates of particle $j$ at time $t$,
and $\langle \cdots \rangle_{t_0}$ denotes an average over all
particles and over all $t_0$.

The error associated with pressure and temperature are $\Delta
p^*\approx 0.005$ and $\Delta T^*\approx0.01$.

\section{\label{sec:results} Results}

\subsection{Pressure-Temperature Phase Diagram}

\begin{figure}[ht]
  \begin{centering}
    \begin{tabular}{ccc}
      \includegraphics[clip,width=5cm]{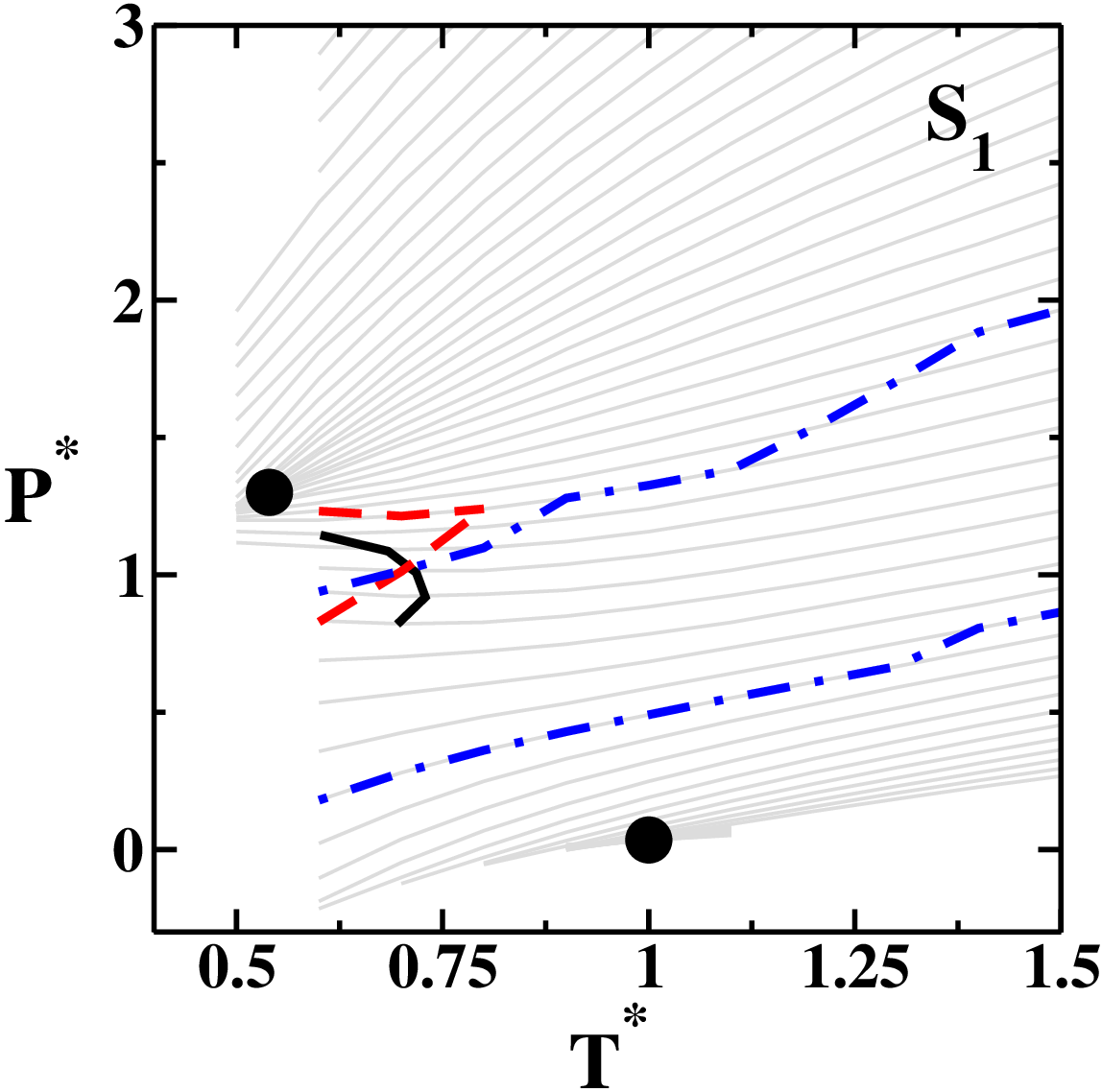}  & \includegraphics[clip,width=5cm]{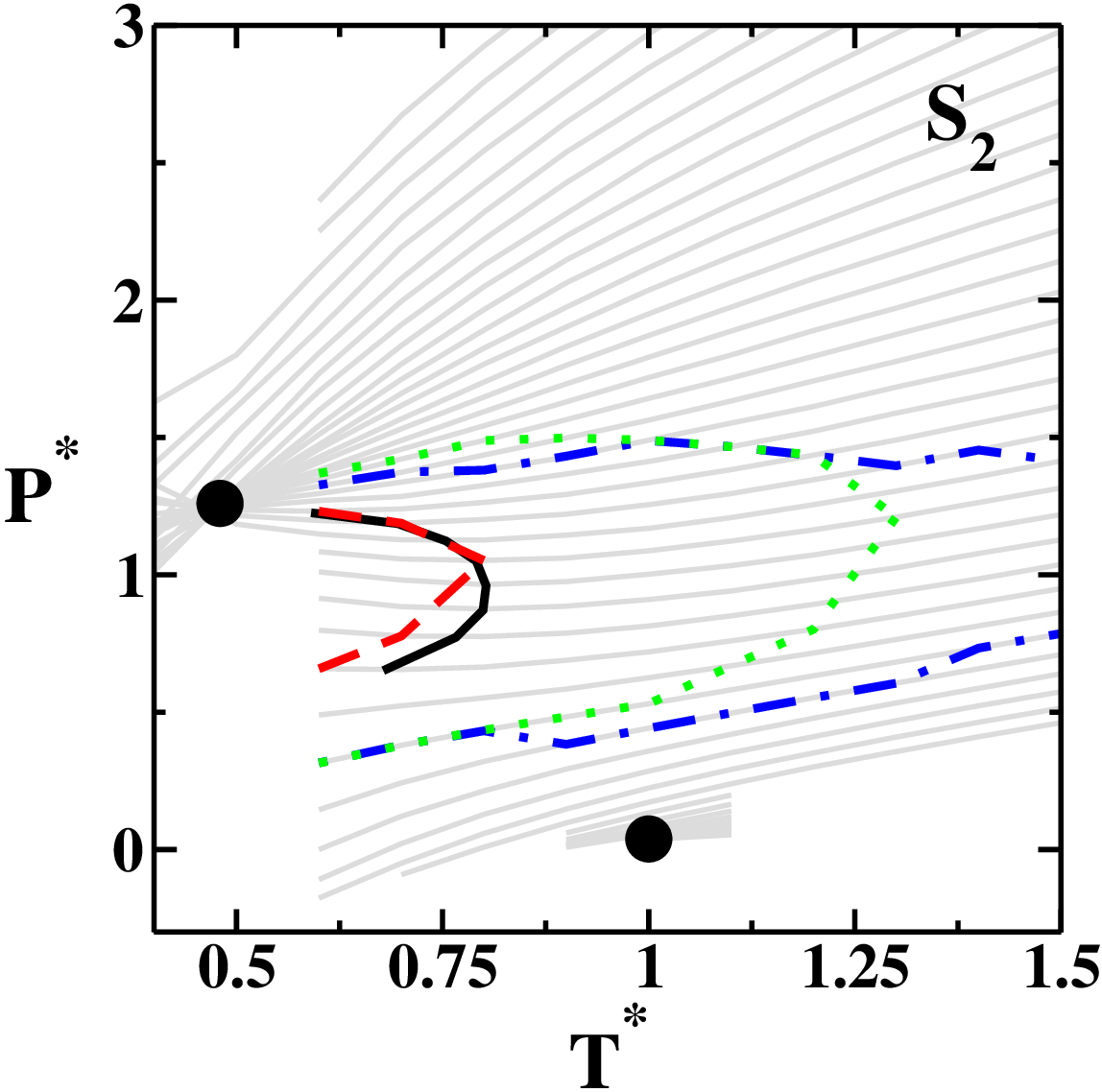}  & \includegraphics[clip,width=5cm]{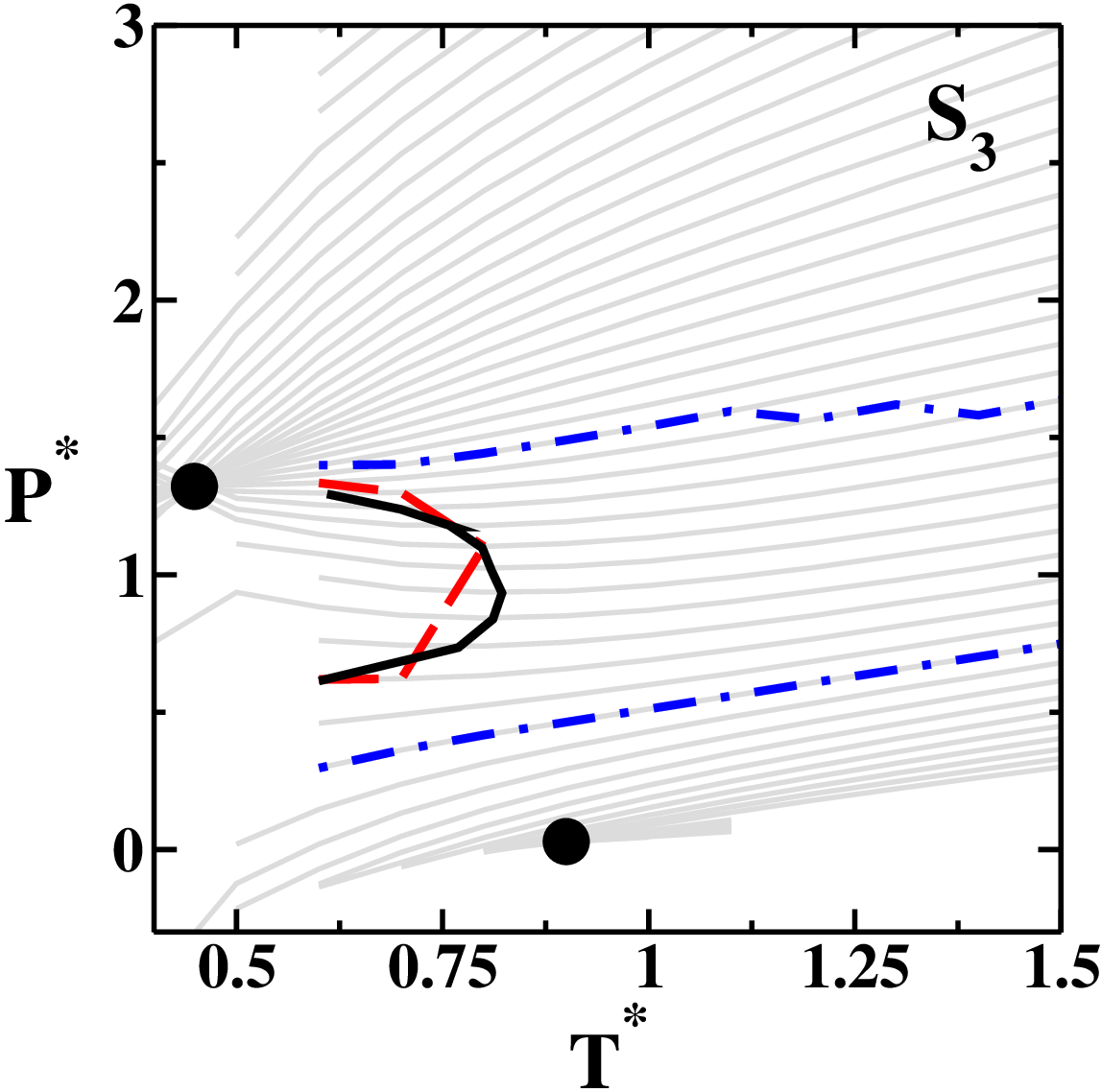} \\
      \includegraphics[clip,width=5cm]{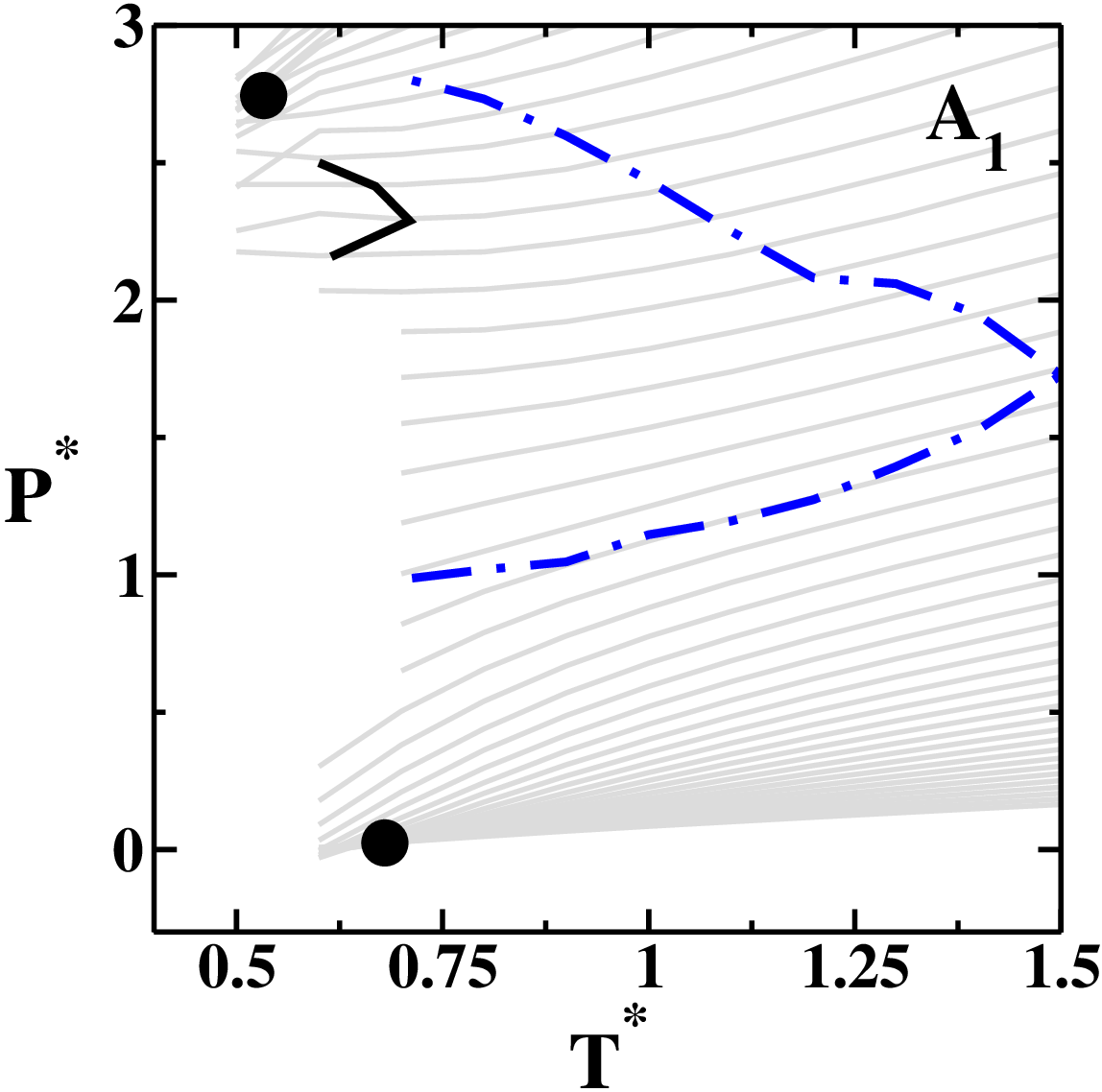}  & \includegraphics[clip,width=5cm]{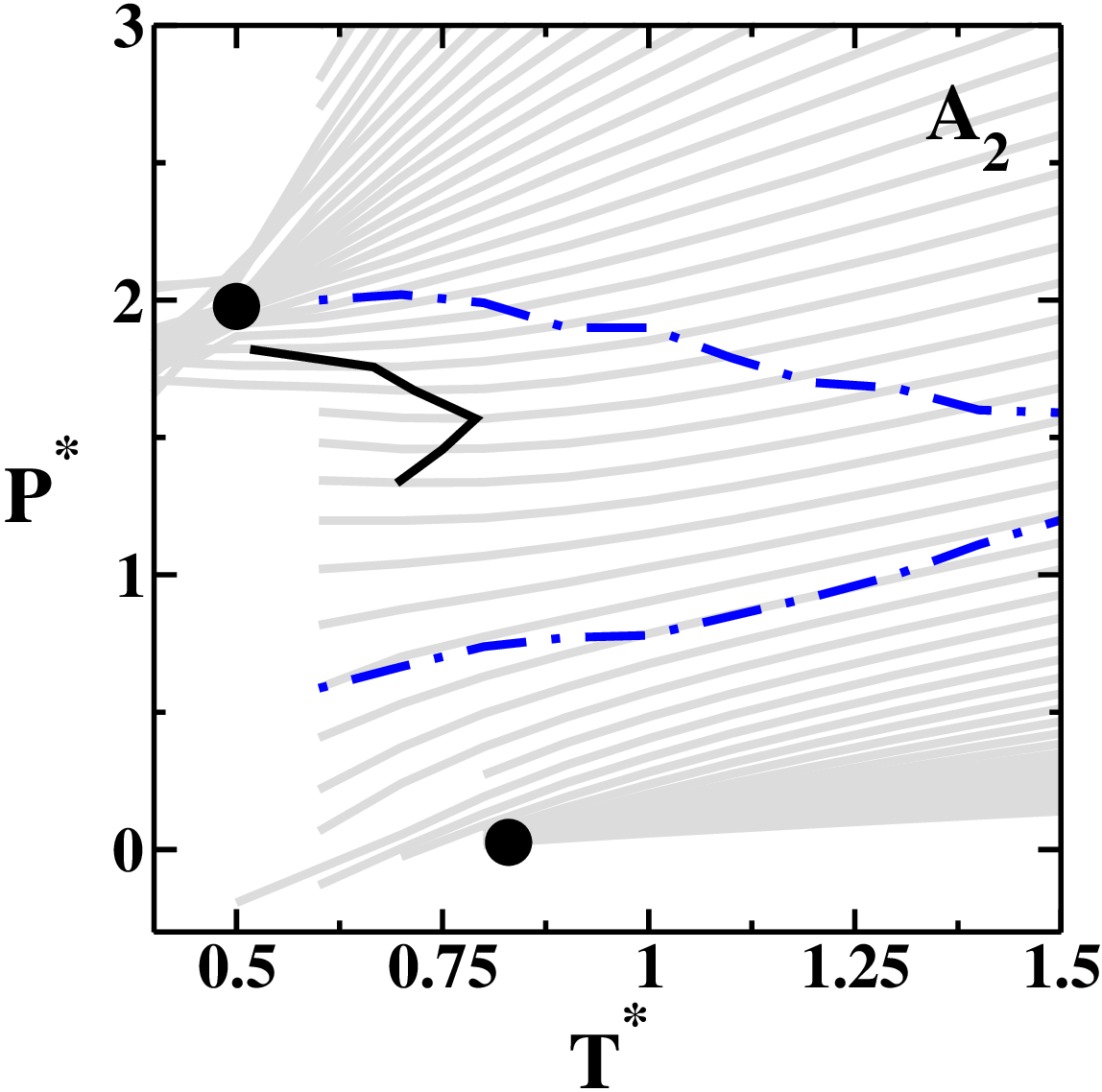}  & \includegraphics[clip,width=5cm]{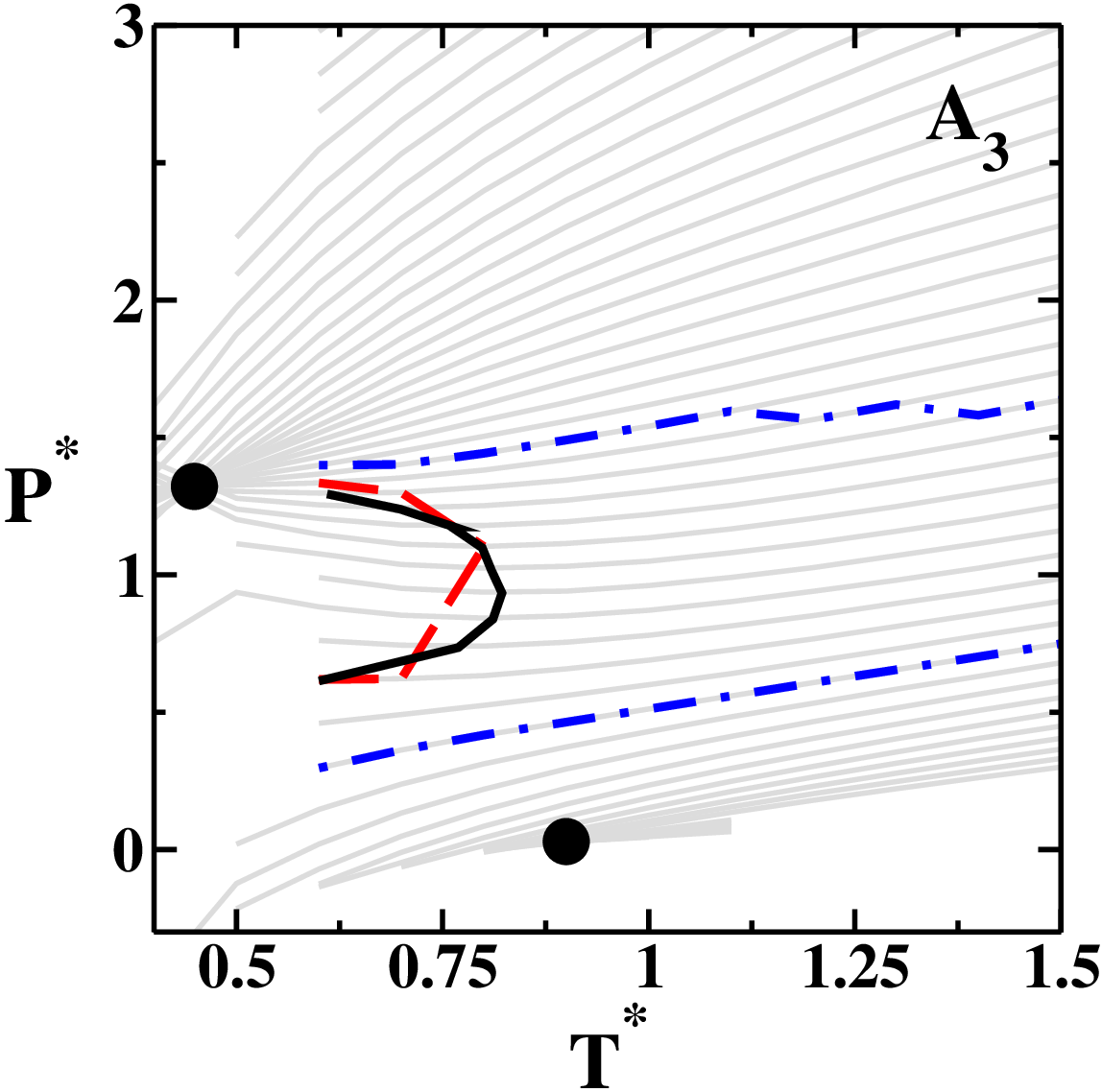} \\
      \includegraphics[clip,width=5cm]{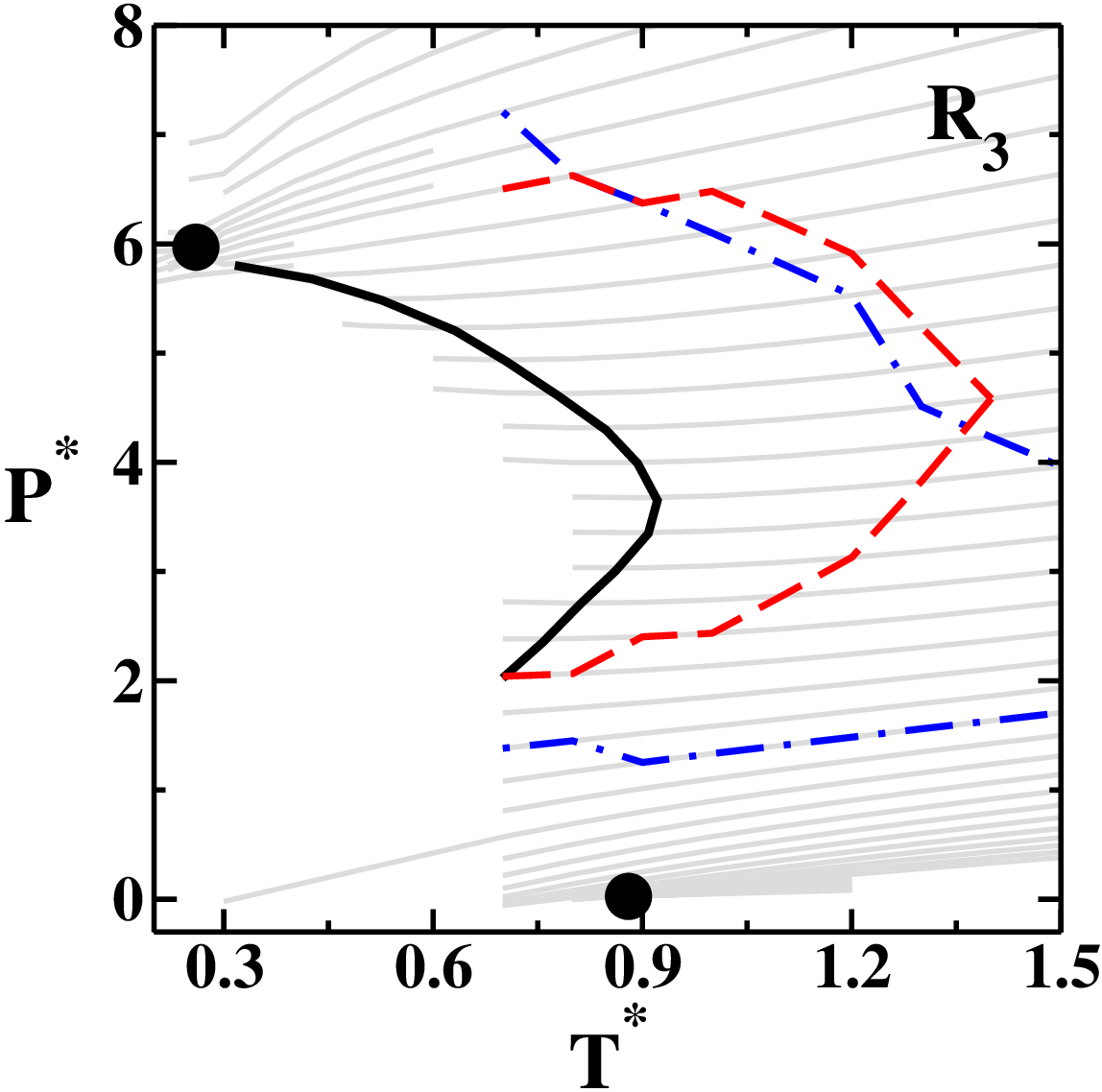}  & \includegraphics[clip,width=5cm]{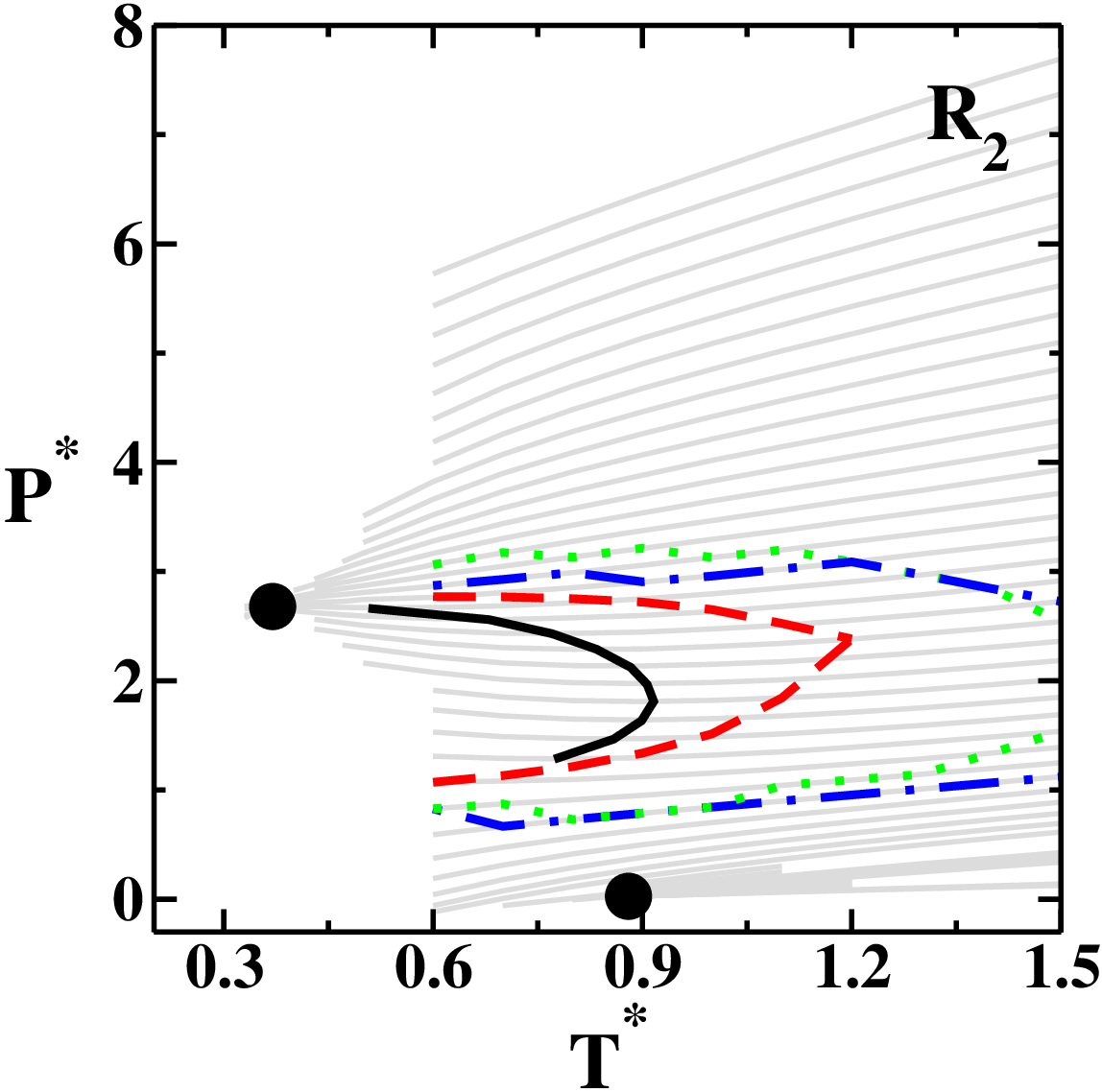}  & \includegraphics[clip,width=5cm]{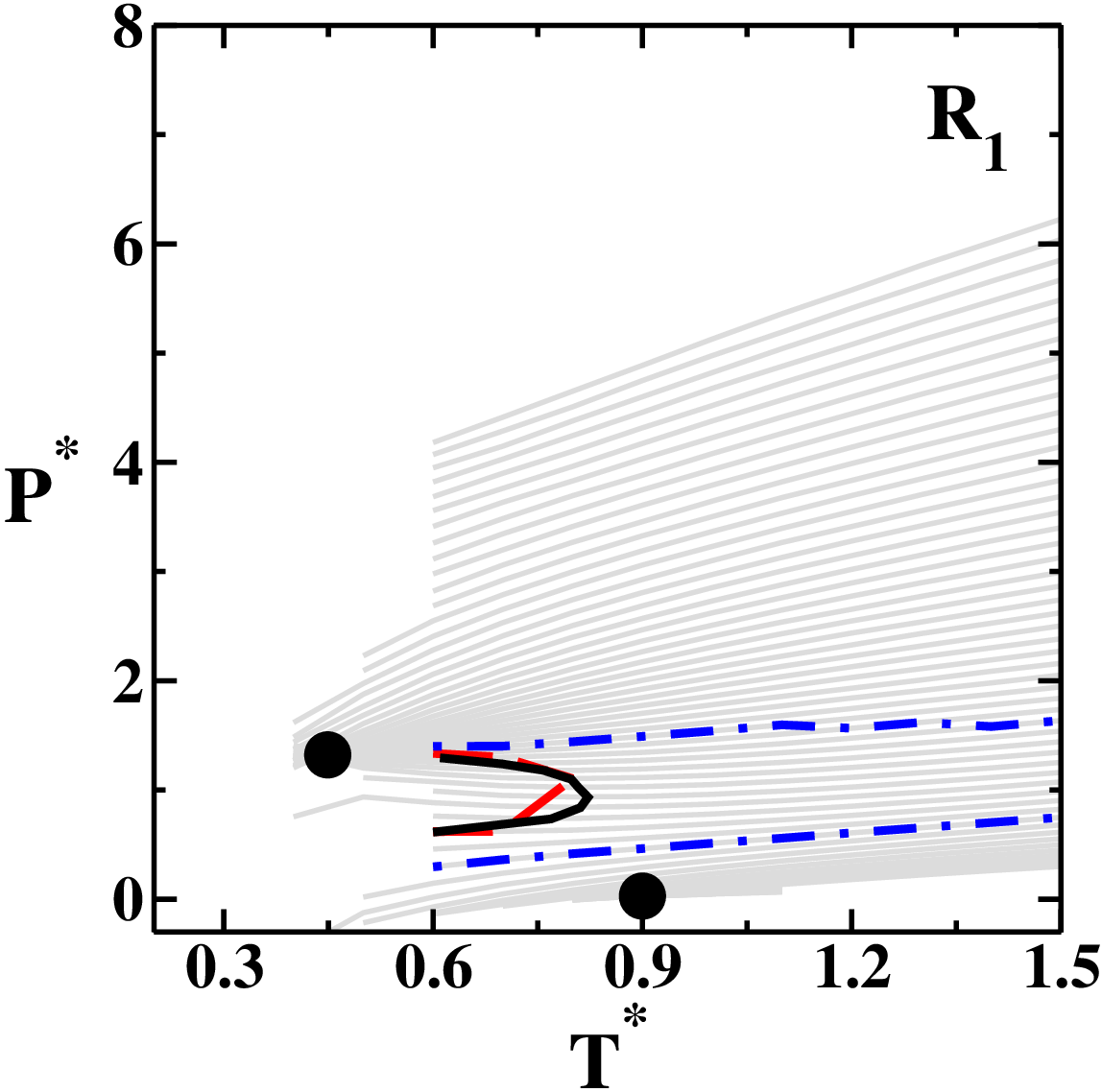} \\
    \end{tabular}
  \end{centering}
  \caption{Pressure-temperature phase diagrams for the $S$, $A$ and  $R$ families of potentials. The gray lines are isochores, the solid lines are the temperature of maximum density (TMD), the dashed lines are the extrema of diffusion and the dot-dashed line are the extrema of the translational order parameter. The filled circles are the liquid-gas (at high temperatures) and   the liquid-liquid (low temperatures)  critical points.}
  \label{fig:PT}
\end{figure}

Fig.~\ref{fig:PT} presents the pressure versus temperature phase
diagram obtained for the three families: $S$, $A$ and $R$. In all the
nine cases the system exhibits at high temperatures a fluid phase, at
intermediate temperatures and very low pressures a gas 
phase and at intermediate
pressures a low density liquid phase (LDL) while at very high
pressures a high density liquid phase (HDL). The coexistence
line between the gas and the low density liquid
phases (not shown) ends in a gas-LDL critical point illustrated
as a filled circle in  
Fig.~\ref{fig:PT}. The LDL-HDL coexistence line (not shown)
ends in a LDL-HDL critical point also shown as a filled circle.
  The two critical points are located in the pressure and
temperature phase diagram by the point in which the isochores meet. The
critical pressures and the critical temperatures values are confirmed
by analyzing the slope of the 
pressure versus density at
constant temperature phase diagram.  The maximum 
of these curves identify the critical point. For the other state points the slope of the pressure versus density phase diagram  is also used as
a check of stability.

\begin{figure}[ht]
  \begin{centering}
    \begin{tabular}{ccc}
      \includegraphics[clip,width=5cm]{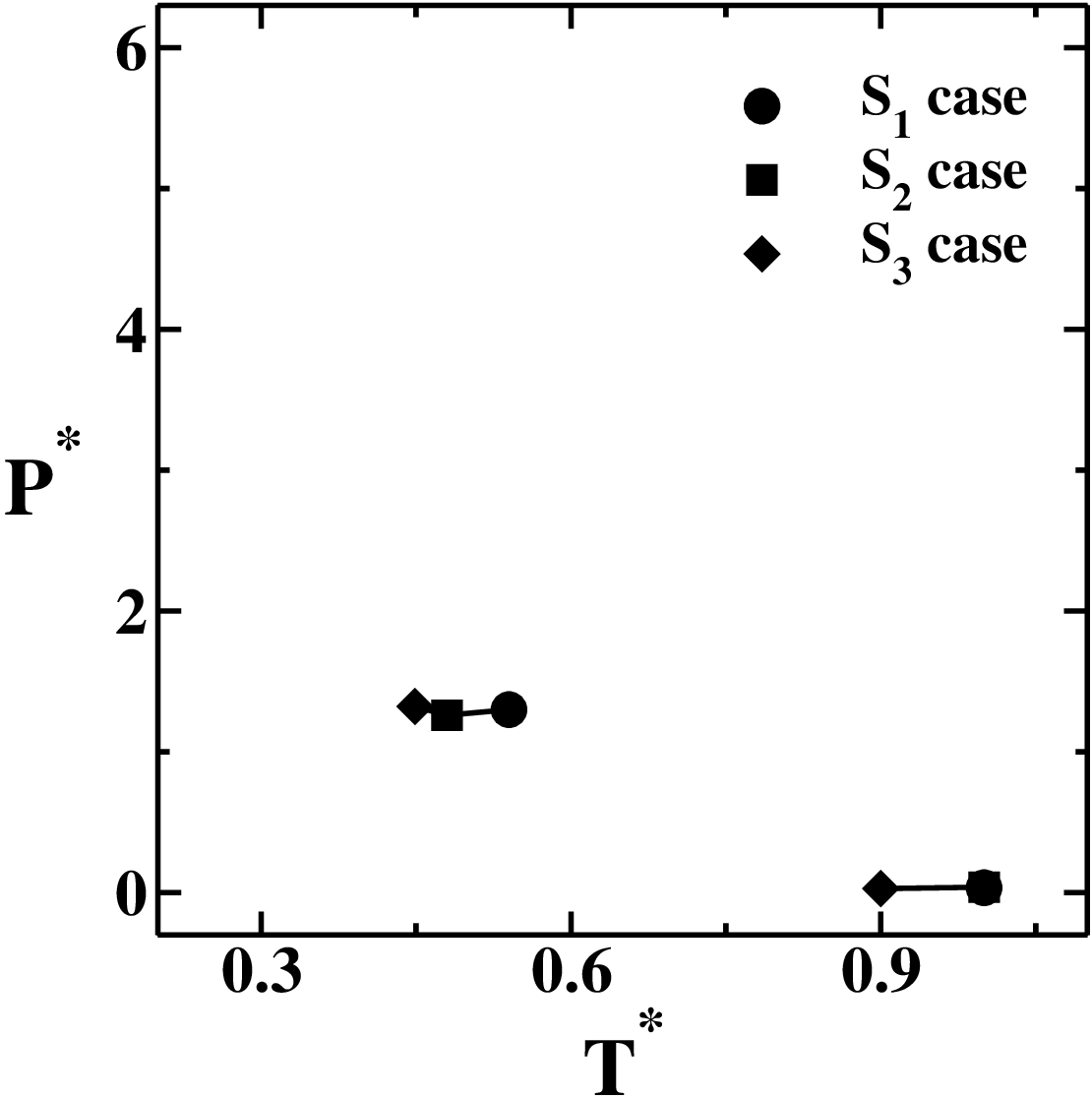} &
      \includegraphics[clip,width=5cm]{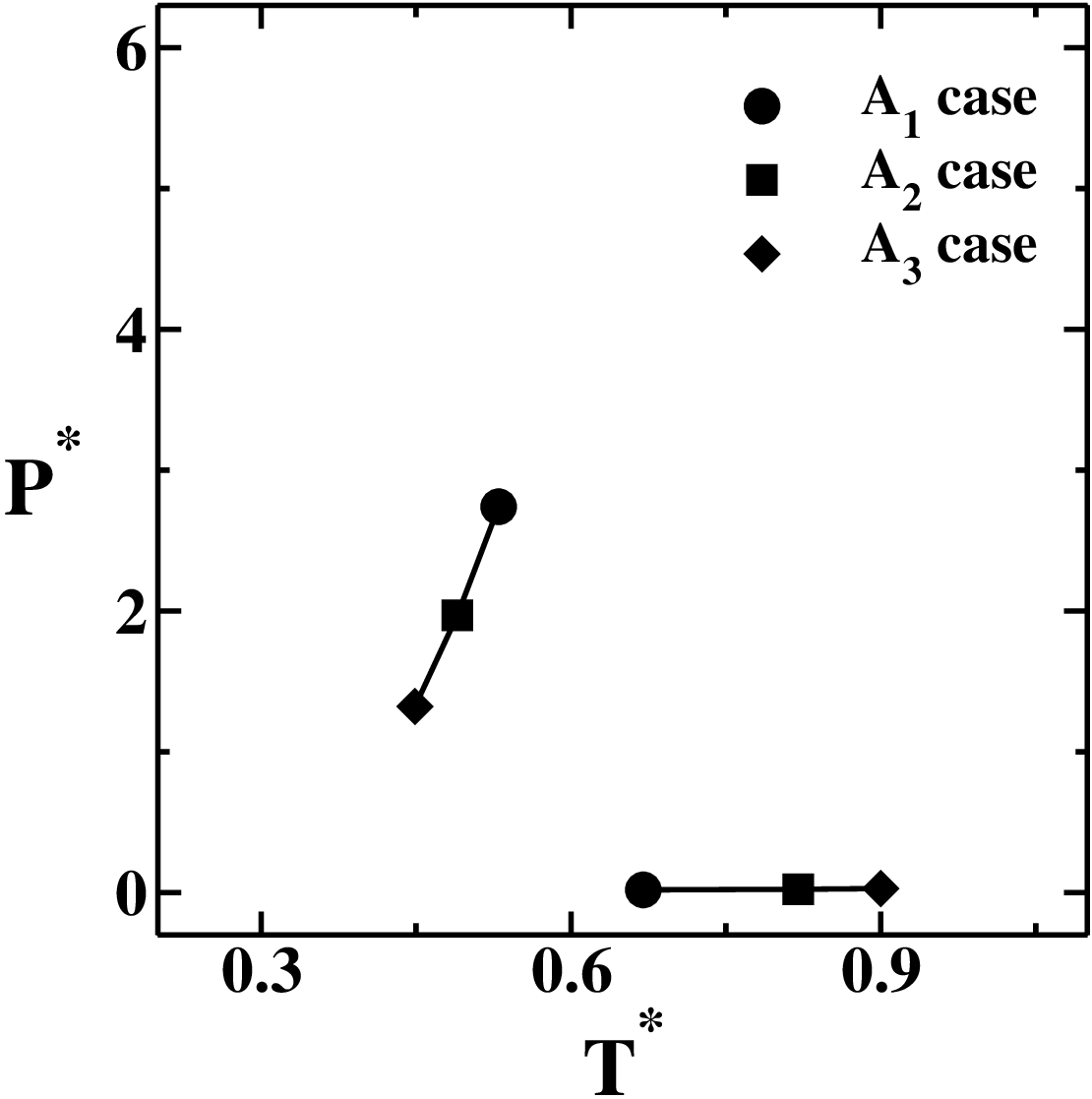} &
      \includegraphics[clip,width=5cm]{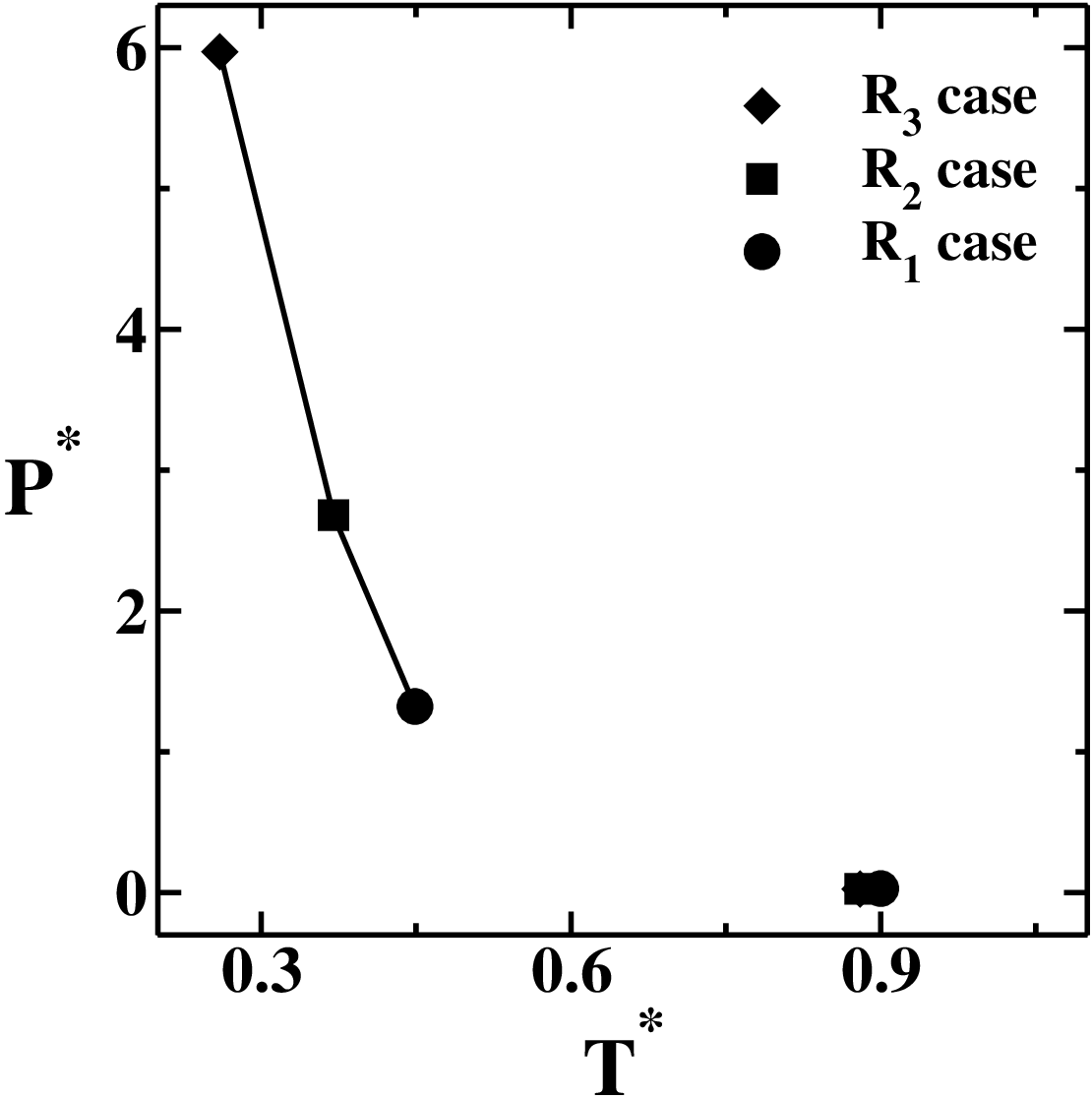}
    \end{tabular}
  \end{centering}
  \caption{Location of the critical points on pressure-temperature
    phase diagram for cases $S$, $A$ and $R$. }
  \label{fig:PT_CP} 
\end{figure}

The Table~\ref{table:CP_GL} and the Table~\ref{table:CP_LL} and the
Fig~\ref{fig:PT_CP} summarize the values of the first (liquid-gas) and
second (liquid-liquid) critical points and their changes in the $p-T$
phase diagram for the three families studied.
\begin{table}
  \caption{Liquid-gas critical point location for potentials $S$, $A$ and $R$.}
  \begin{tabular}{ccc|ccc|ccc}
    \hline
    Potential  & $T_{c_{1}}^*$  & $p_{c_{1}}^*$ &Potential  & $T_{c_{1}}^*$  & $p_{c_{1}}^*$ &Potential  & $T_{c_{1}}^*$  & $p_{c_{1}}^*$\\
    \hline
    $S_{1}$  & $0.05$  & $1.00$ &$A_{1}$  & $0.04$  & $0.68$ &$R_{1}$  & $0.02$  & $0.90$\\
    $S_{2}$  & $0.05$  & $0.99$ &$A_{2}$  & $0.03$  & $0.82$ &$R_{2}$  & $0.04$  & $0.88$\\
    $S_{3}$  & $0.04$  & $0.88$ &$A_{3}$  & $0.05$  & $0.90$ &$R_{3}$  & $0.04$  & $0.88$\\
    \hline
  \end{tabular}
  \label{table:CP_GL}
\end{table}
\begin{table}
  \caption{Liquid-liquid critical point location for potentials $S$, $A$ and $R$.}
  \begin{tabular}{ccc|ccc|ccc}
    \hline
    Potential  & $T_{c_{2}}^*$  & $p_{c_{2}}^*$ &Potential  & $T_{c_{2}}^*$  & $p_{c_{2}}^*$ &Potential  & $T_{c_{2}}^*$  & $p_{c_{2}}^*$\\
    \hline
    $S_{1}$  & $1.31$  & $0.54$ &$A_{1}$  & $2.75$  & \ $0.53$ &$R_{1}$  & $1.34$  & $0.44$\\
    $S_{2}$  & $1.26$  & $0.48$ &$A_{2}$  & $1.98$  & \ $0.50$ &$R_{2}$  & $2.67$  & $0.37$\\
    $S_{3}$  & $1.34$  & $0.44$ &$A_{3}$  & $1.33$  & \ $0.44$ &$R_{3}$  & $6.01$  & $0.26$\\
    \hline
  \end{tabular}
  \label{table:CP_LL}
\end{table}

Fig.~\ref{fig:PT} shows that in the family of potentials $S$ the
values of the pressure and temperature of the
liquid-gas and the liquid-liquid critical points  are not
sensitive to the change of slope as predicted by Yan et
al.~\cite{Ya08}. For the $A$ family, also illustrated in
Fig.~\ref{fig:PT} indicates that
  the temperature of the liquid-gas critical point 
increases  with the increase of the range of the attractive
scale, while the temperature and the pressure
of the  liquid-liquid critical point
decrease. This
result indicates that if the attractive scale increases the high 
density liquid requires less pressure to be formed  while the 
gas phases exist for higher temperatures as predicted
by Skibinstky et al.~\cite{Sk04,Ba11}. In the case $R$, shown as well
in the Fig.~\ref{fig:PT}, the
liquid-liquid critical pressure decreases with the increase of the
range of the repulsive scale. This result indicates that as the
repulsive scale becomes broader, it requires less pressure for the
high density liquid to be formed while the repulsive scale has
almost no effect in the low density liquid-gas coexistence line as
predicted also by Skibinstky et al.~\cite{Sk04,Ba11}. A summary
of the liquid-gas and liquid-liquid critical pressures
and temperatures are shown on Fig.~\ref{fig:PT_CP}.

\subsection{Density, Diffusion and Translational Anomalies}

\begin{table}
  \caption{Limit values for density ($\rho^*$), temperature ($T^*$)
    and pressure ($p^*$) of the thermodynamics anomalies on
    pressure-temperature diagram. Here the point $p_{l}$ represents
    the density, temperature and pressure of the point of the lowest
    pressure in the TMD line, $p_{m}$ represents the point of the
    highest temperature and $p_{h}$ represents the point of the
    highest pressure of the TMD line.}

\label{table:TMD} 
  \begin{tabular}{ccccc|ccccc|ccccc}
    \hline
    case     &          & $p_l$  & $p_{m}$ & $p_{h}$ & case  &          & $p_l$  & $p_m$  & $p_h$  & case &          & $p_l$  & $p_m$  & $p_h$\\
    \hline
             & $\rho^*$ & $0.34$ & $0.35$  & $0.38$  &       & $\rho^*$ & $0.32$ & $0.34$ & $0.39$ &      & $\rho^*$ & $0.31$ & $0.34$ & $0.39$\\
    $S_{1}$  & $T^*$    & $0.70$ & $0.73$  & $0.60$  &$S_2$  & $T^*$    & $0.68$ & $0.80$ & $0.60$ &$S_3$ & $T^*$    & $0.60$ & $0.82$ & $0.61$\\
             & $p^*$    & $0.70$ & $0.73$  & $1.14$  &       & $p^*$    & $0.65$ & $0.96$ & $1.27$ &      & $p^*$    & $0.61$ & $0.92$ & $1.29$\\
    \hline
             & $\rho^*$ & $0.42$ & $0.43$  & $0.45$  &       & $\rho^*$ & $0.37$ & $0.39$ & $0.42$ &      & $\rho^*$ & $0.31$ & $0.34$ & $0.39$\\
    $A_{1}$  & $T^*$    & $0.61$ & $0.71$  & $0.59$  &$A_2$  & $T^*$    & $0.51$ & $0.79$ & $0.69$ &$A_3$ & $T^*$    & $0.60$ & $0.82$ & $0.61$\\
             & $p^*$    & $2.15$ & $2.29$  & $2.51$  &       & $p^*$    & $1.33$ & $1.56$ & $1.82$ &      & $p^*$    & $0.61$ & $0.92$ & $1.29$\\
    \hline
             & $\rho^*$ & $0.31$ & $0.34$ & $0.39$ &       & $\rho^*$ & $0.33$ & $0.37$ & $0.42$ &       & $\rho^*$ & $0.34$ & $0.39$ & $0.47$\\
    $R_{1}$  & $T^*$    & $0.60$ & $0.82$ & $0.61$ & $R_2$ & $T^*$    & $0.77$ & $0.92$ & $0.50$ & $R_3$ & $T^*$    & $0.70$ & $0.92$ & $0.32$\\
             & $p^*$    & $0.61$ & $0.92$ & $1.29$ &       & $p^*$    & $1.28$ & $1.81$ & $2.67$ &       & $p^*$    & $2.04$ & $3.65$ & $5.76$\\
    \hline
  \end{tabular}
\end{table}

Fig.~\ref{fig:PT} shows the temperature of maximum density (TMD) 
for all the nine studied cases as a
solid thick lines. For all the potentials $S$, $A$ and $R$ the TMD lines
are observed. The limits of the TMD in the pressure-temperature phase
diagram are shown in the Table~\ref{table:TMD} where $p_{l}$
represents the values of $(\rho^*,T^*,p^*)$ for the point of the
lowest pressure in the TMD line, $p_{m}$ is the point with the highest
temperature and $p_{h}$ is the point with the highest pressure.

The three top graphs in Fig.~\ref{fig:PT} show that the effect of
decreasing the slope between the two length scales in the pair
interaction potential is to move the TMD to higher temperatures. This
result explains why the TMD is not observed in the discontinuous
square well (DSW) model~\cite{Sk04}.  As the slope increases the TMD
pressure and temperature approach the amorphous region and the system
becomes unstable. For slopes higher than $S_3$ case no anomalous behavior
is observed.

The middle graphs in Fig.~\ref{fig:PT} show that as the attractive
scale increases, the TMD moves to higher temperatures and lower
pressures as observed in potentials in which the attractive scale
becomes dominant~\cite{Si10}.

The bottom graphs in Fig.~\ref{fig:PT} show that as the repulsive
scale becomes broader, the density anomaly region in the pressure
temperature phase diagram goes to lower pressures and shrinks as
observed in potentials in which the repulsive scale becomes
dominant~\cite{Ba09}.

In addition in all the phase diagrams it is possible to observe that
the TMD maximum pressure never exceeds the critical
pressure~\cite{Sa11}. Our results indicate that the location in the
pressure temperature phase diagram of the density anomalous region
depends on the distance between the two length
scales~\cite{Ba09,Si10,Ba11}.

\begin{figure}[ht]
  \centering
  \begin{tabular}{ccc}
    \includegraphics[clip,width=5cm]{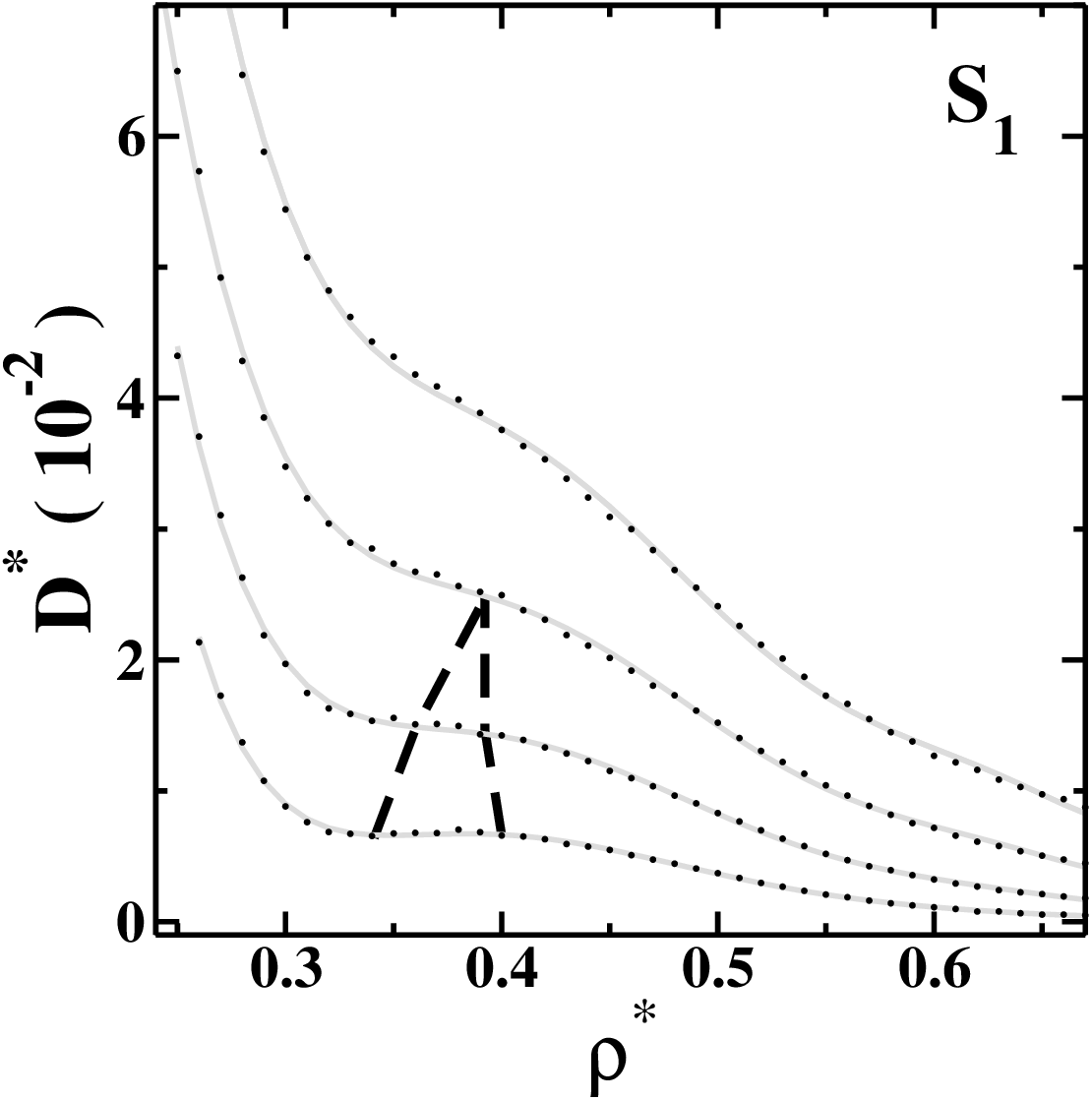}  & \includegraphics[clip,width=5cm]{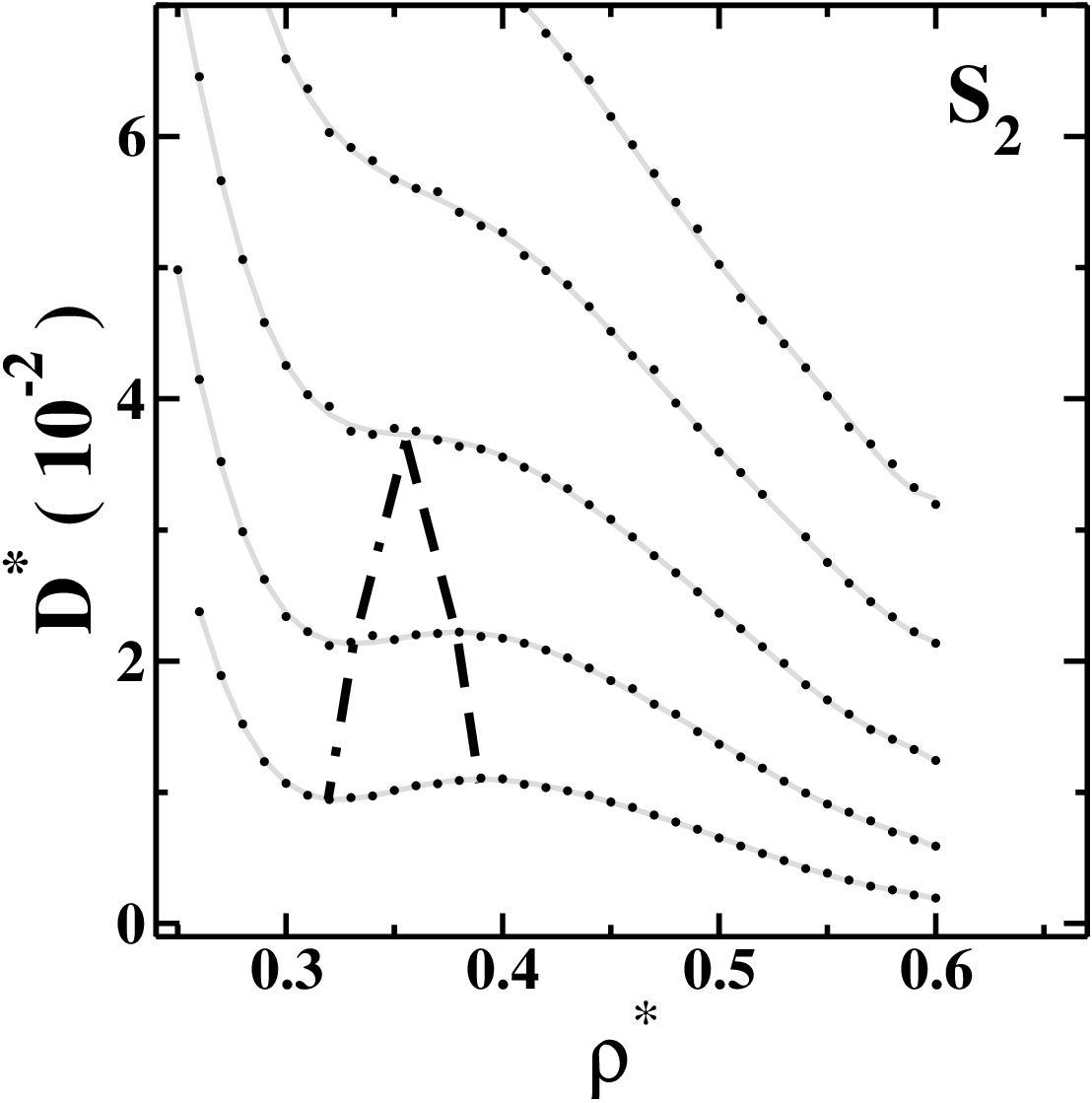}  & \includegraphics[clip,width=5cm]{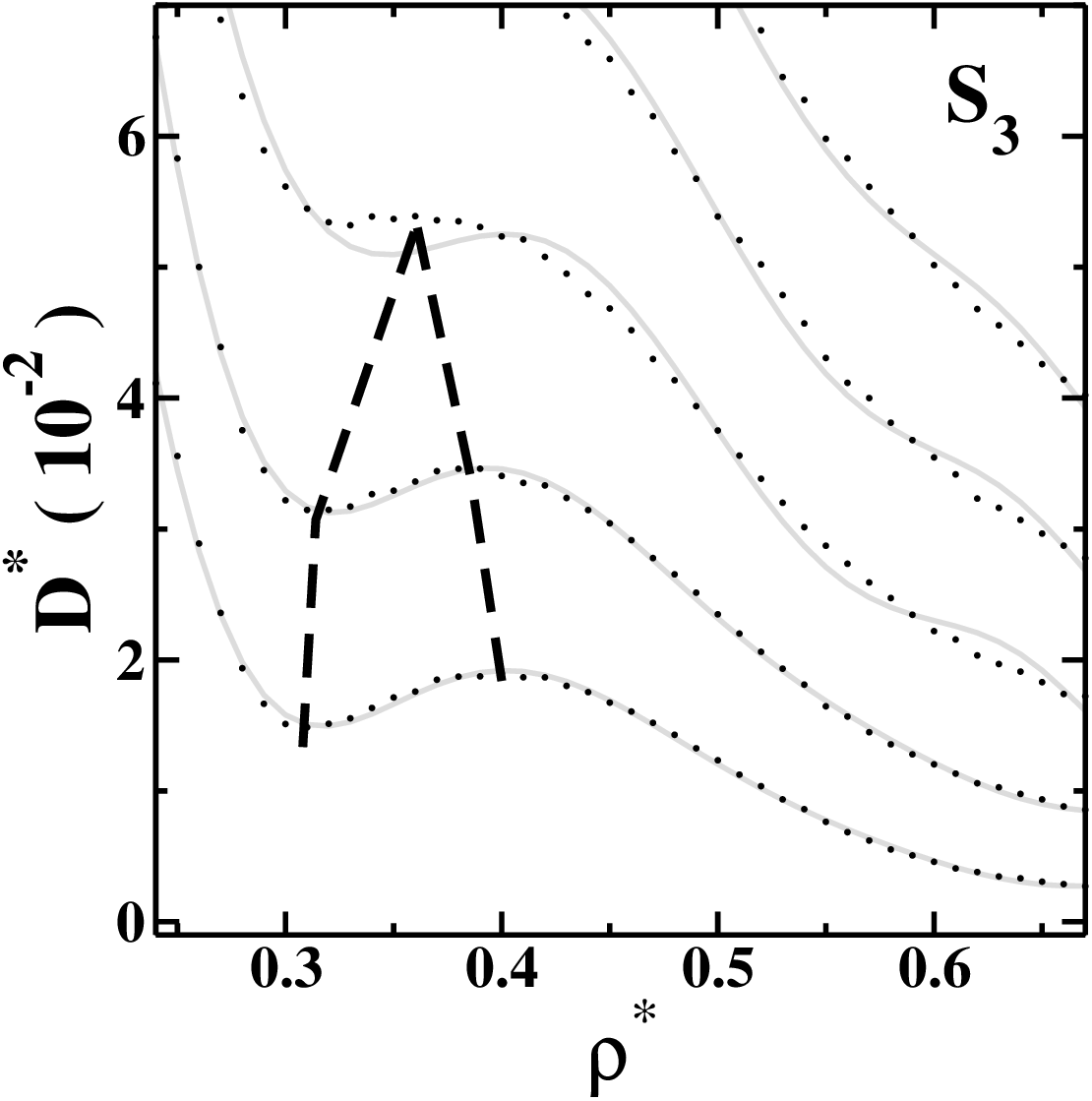} \\
    \includegraphics[clip,width=5cm]{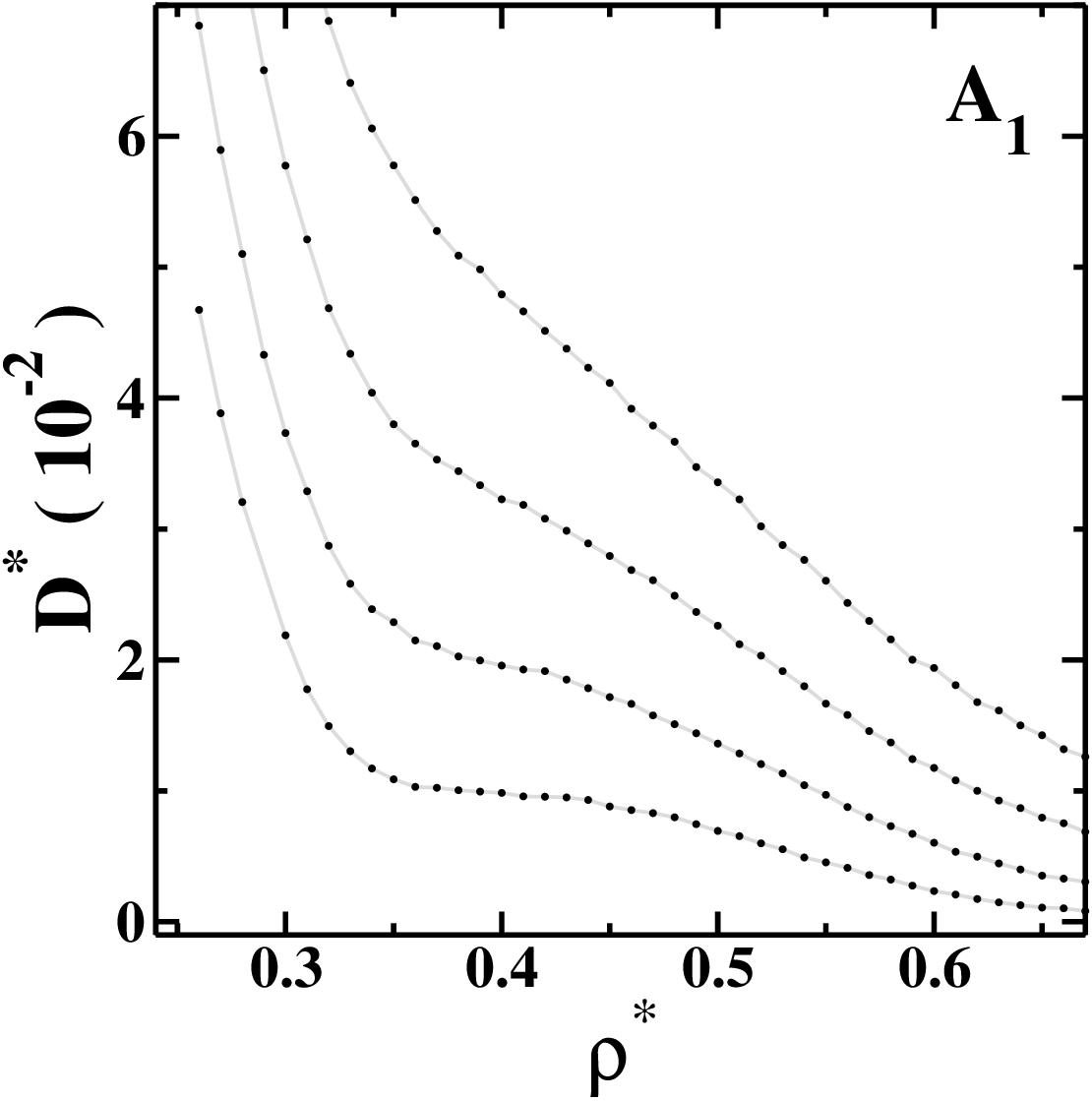}  & \includegraphics[clip,width=5cm]{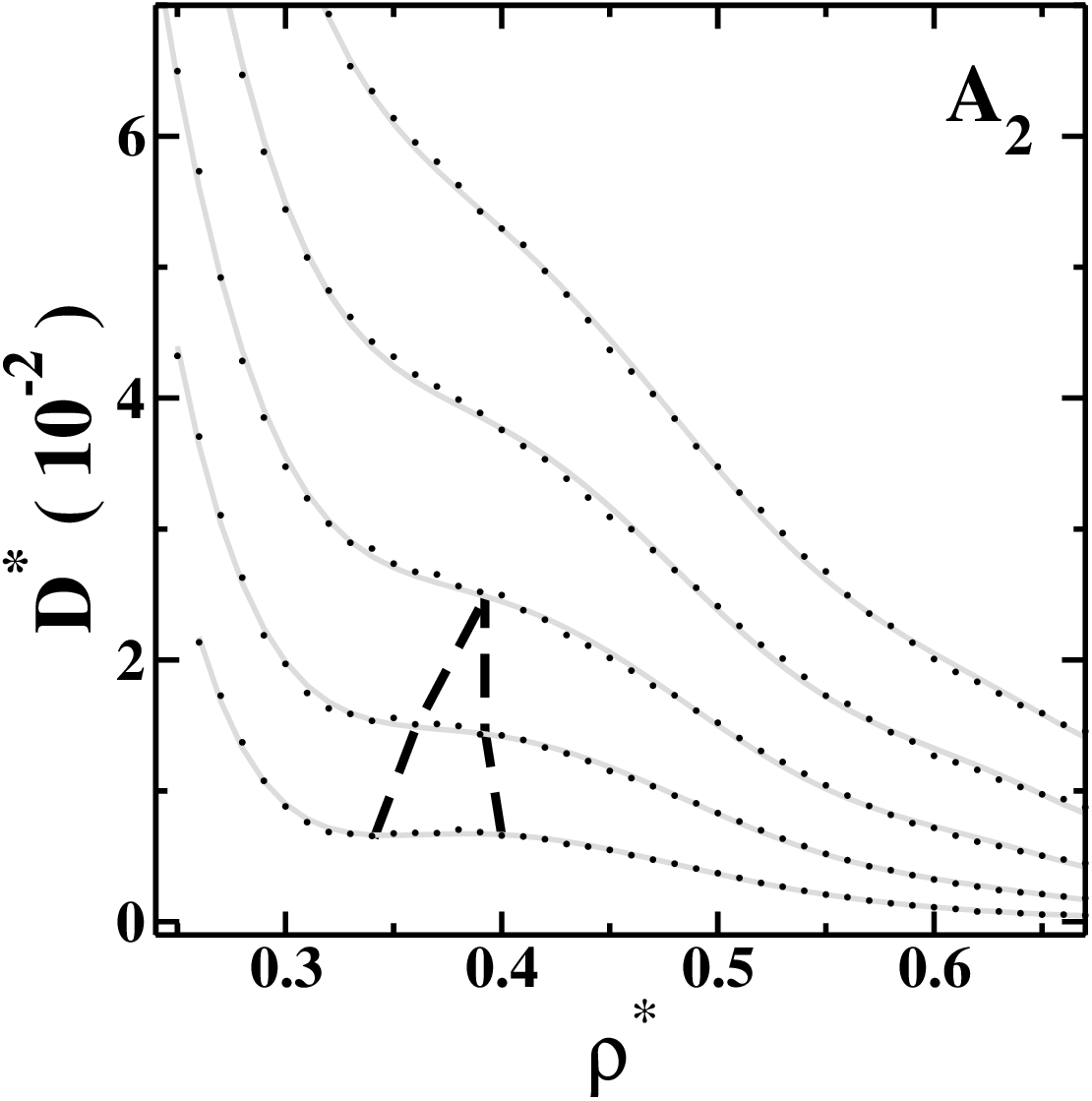}  & \includegraphics[clip,width=5cm]{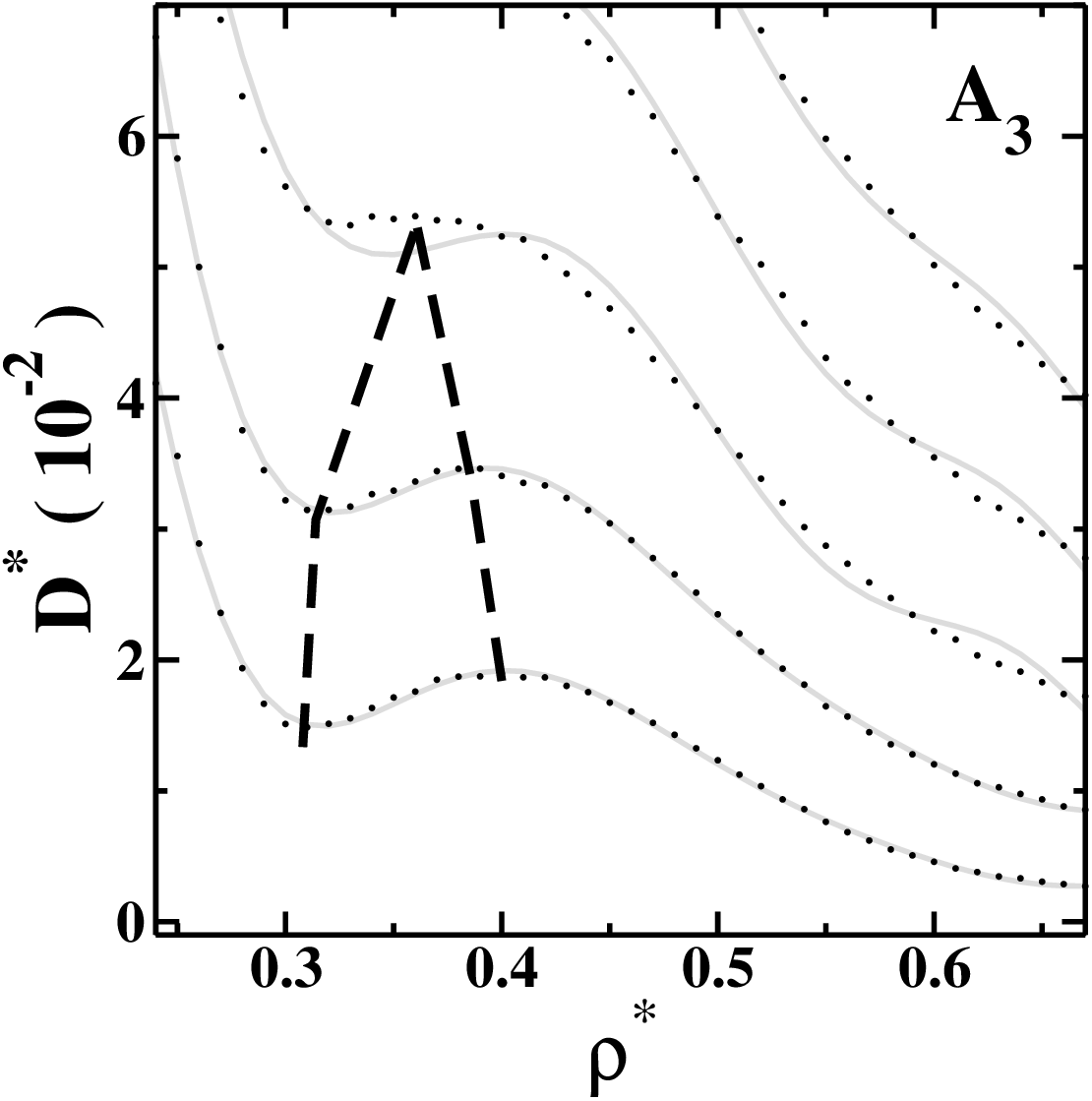} \\
    \includegraphics[clip,width=5cm]{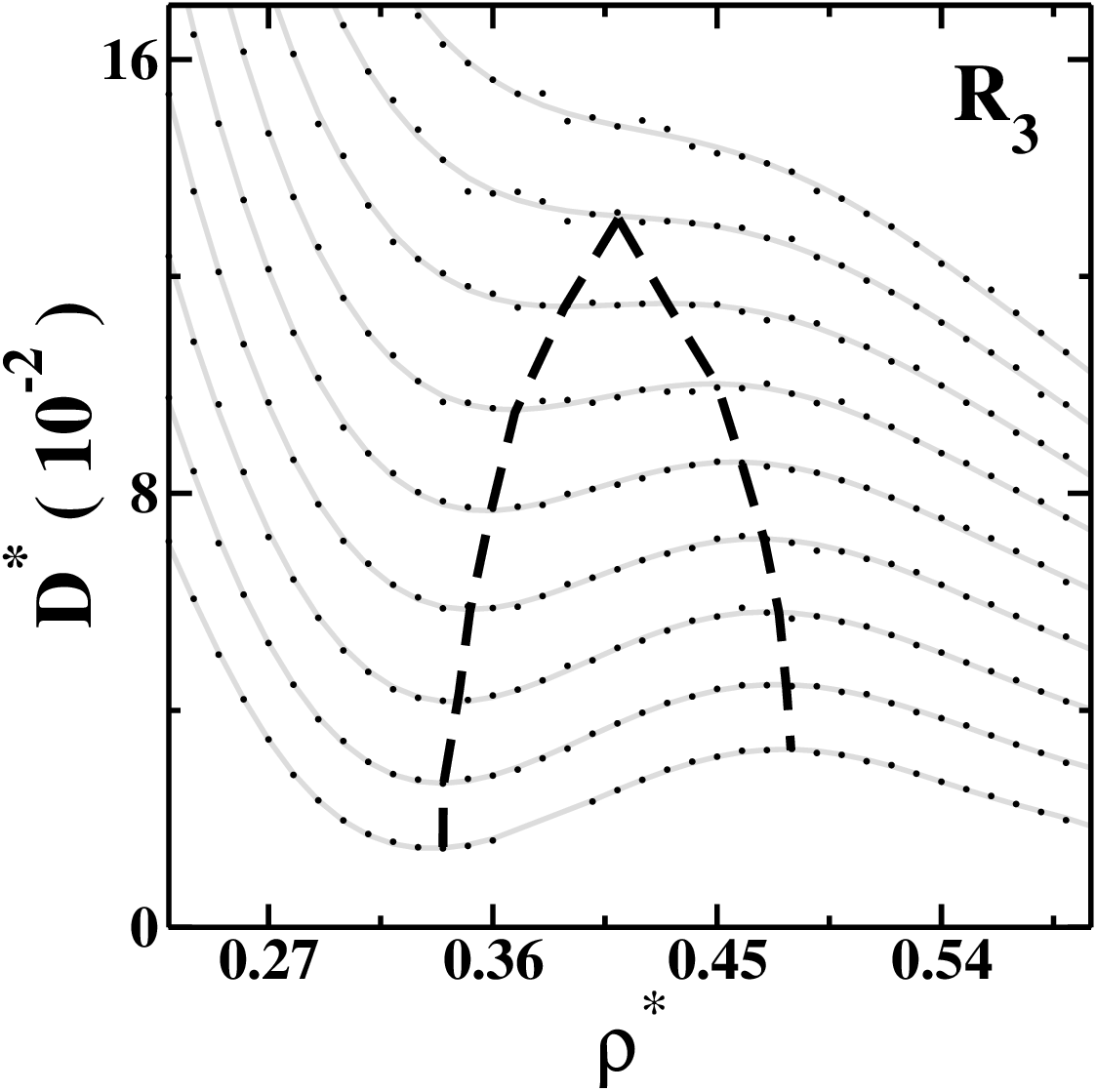}  & \includegraphics[clip,width=5cm]{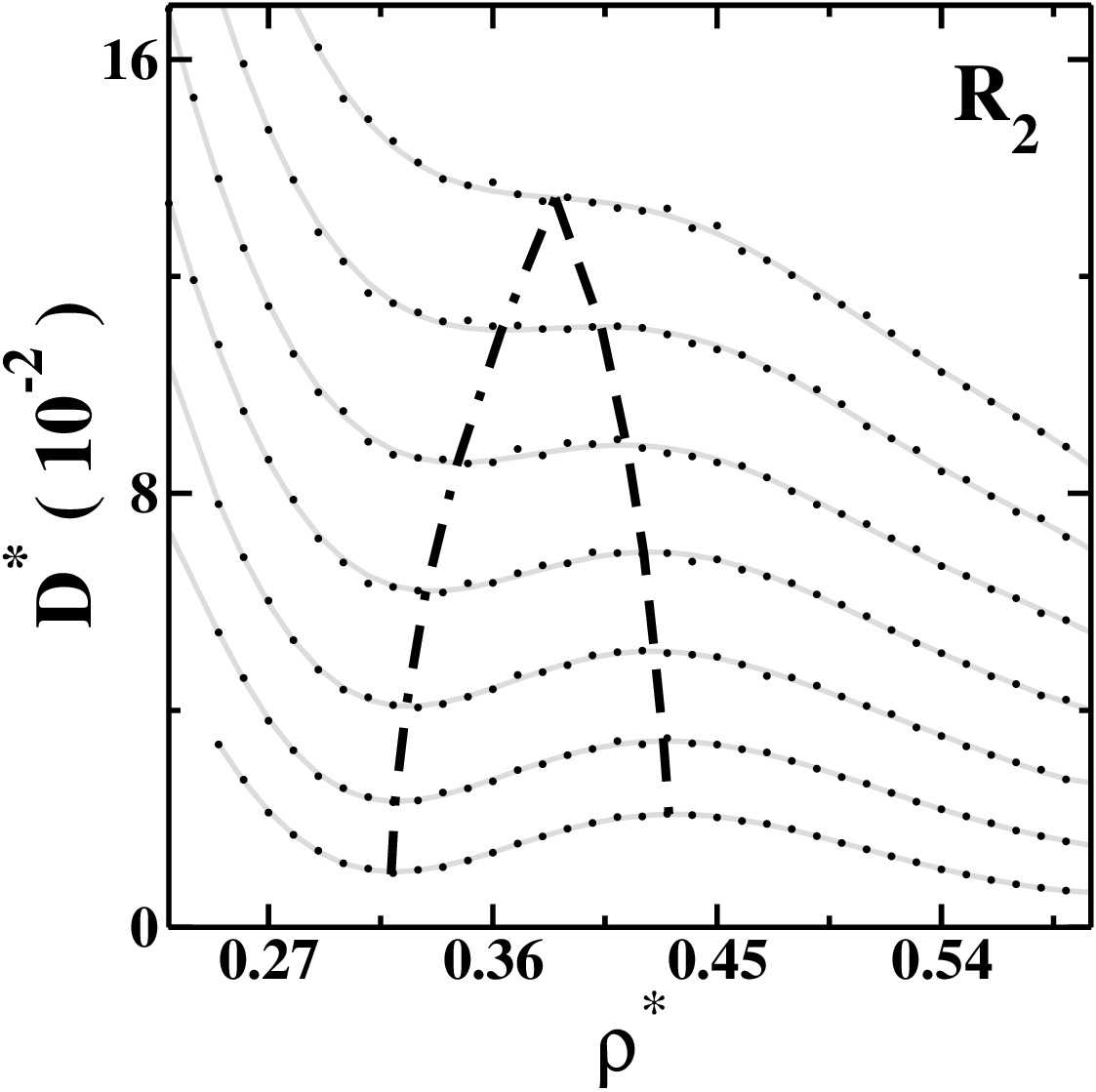}  & \includegraphics[clip,width=5cm]{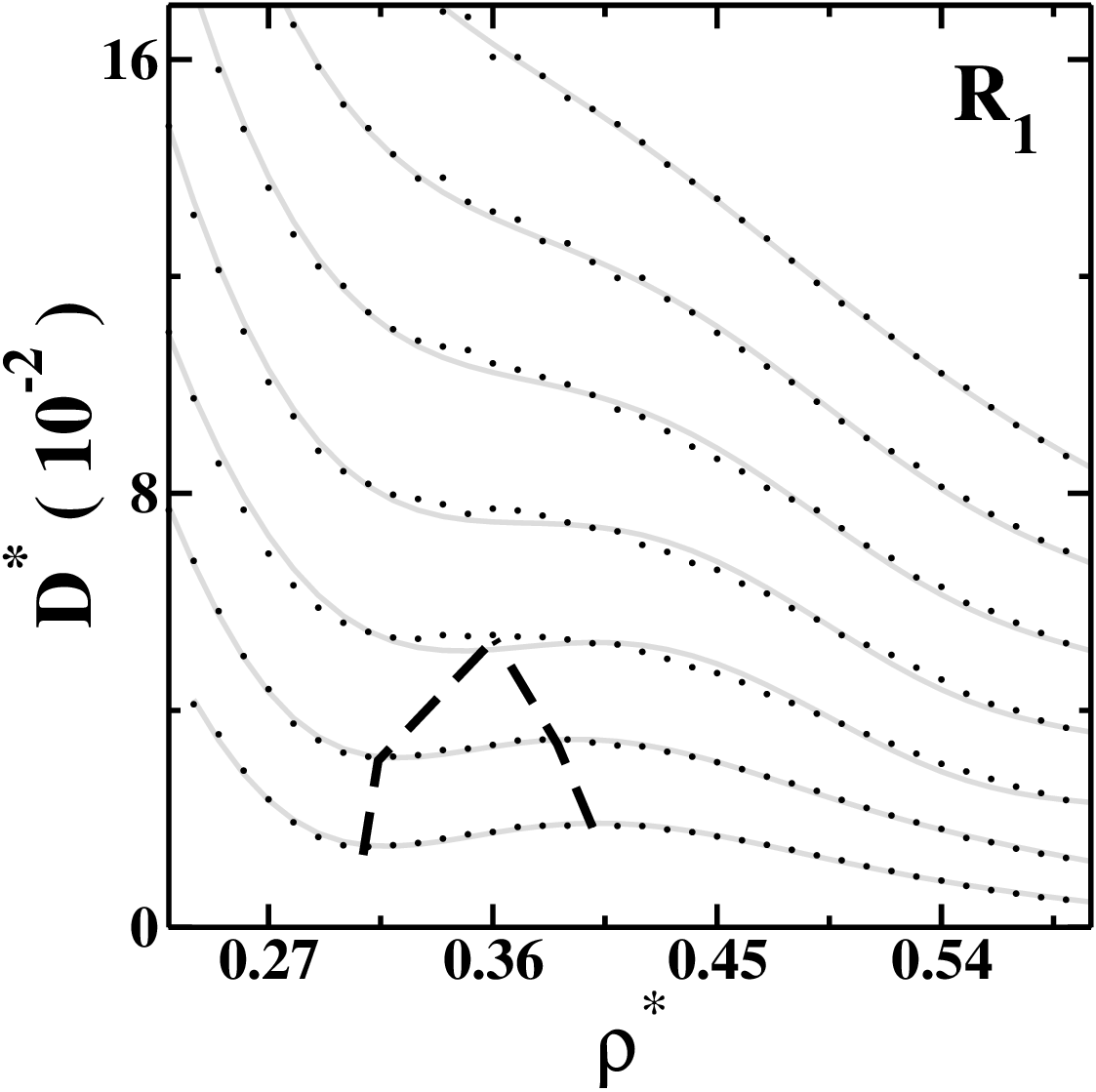} \\
  \end{tabular}
  \caption{Diffusion coefficient as a function of density. The dots
    are the simulation data and the solid lines are polynomial
    fits. The dashed lines connect the densities of minimal and
    maximal diffusivity that limit the diffusion anomalous region.}
  \label{fig:Drho} 
\end{figure}

The Fig.~\ref{fig:Drho} shows the graphs
of  the dimensionless translational diffusion
coefficient as function of density for all families, $S$, $A$ and
$R$. The solid gray lines are a polynomial fits to the data obtained by
the simulations (the dots in the Fig.~\ref{fig:Drho}).  The diffusion
coefficient follows the same trend of the TMD line. This result is not
surprising since the hierarchy of the anomalies suggests that the
mechanism for the presence of a TMD line is related to the mechanism
for the existence of a maximum and minimum diffusion coefficient.

We also test the effects that changes
in  the two length scales have in the location in the
pressure-temperature phase diagram of the structural anomalous region.

The translational order parameter is defined as~\cite{Sh02,Er01,Er03}
\begin{equation}\label{eq_trans}
  t = \int_0^{\xi_c} \left| g(\xi) - 1 \right| d\xi
\end{equation}
where $\xi = r \rho^{\frac{1}{3}}$ is the distance $r$ in units of the
mean interparticle separation $\rho^{-\frac{1}{3}}$, $\xi_{c}$ is the
cutoff distance set to half of the simulation box times~\cite{Ol06b}
$\rho^{-\frac{1}{3}}$, $g(\xi)$ is the radial distribution function
which is proportional to the probability of finding a particle at a
distance $\xi$ from a referent particle. The translational order
parameter measures how structured is the system.
For an ideal gas it is $g = 1$ and $t = 0$, while for 
the  crystal phase it is $g\neq 1$ over long distances
resulting in a large $t$. Therefore for normal fluids $t$ increases with the
increase of the density.

\begin{figure}[ht]
  \begin{centering}
    \begin{tabular}{ccc}
      \includegraphics[clip,width=5cm]{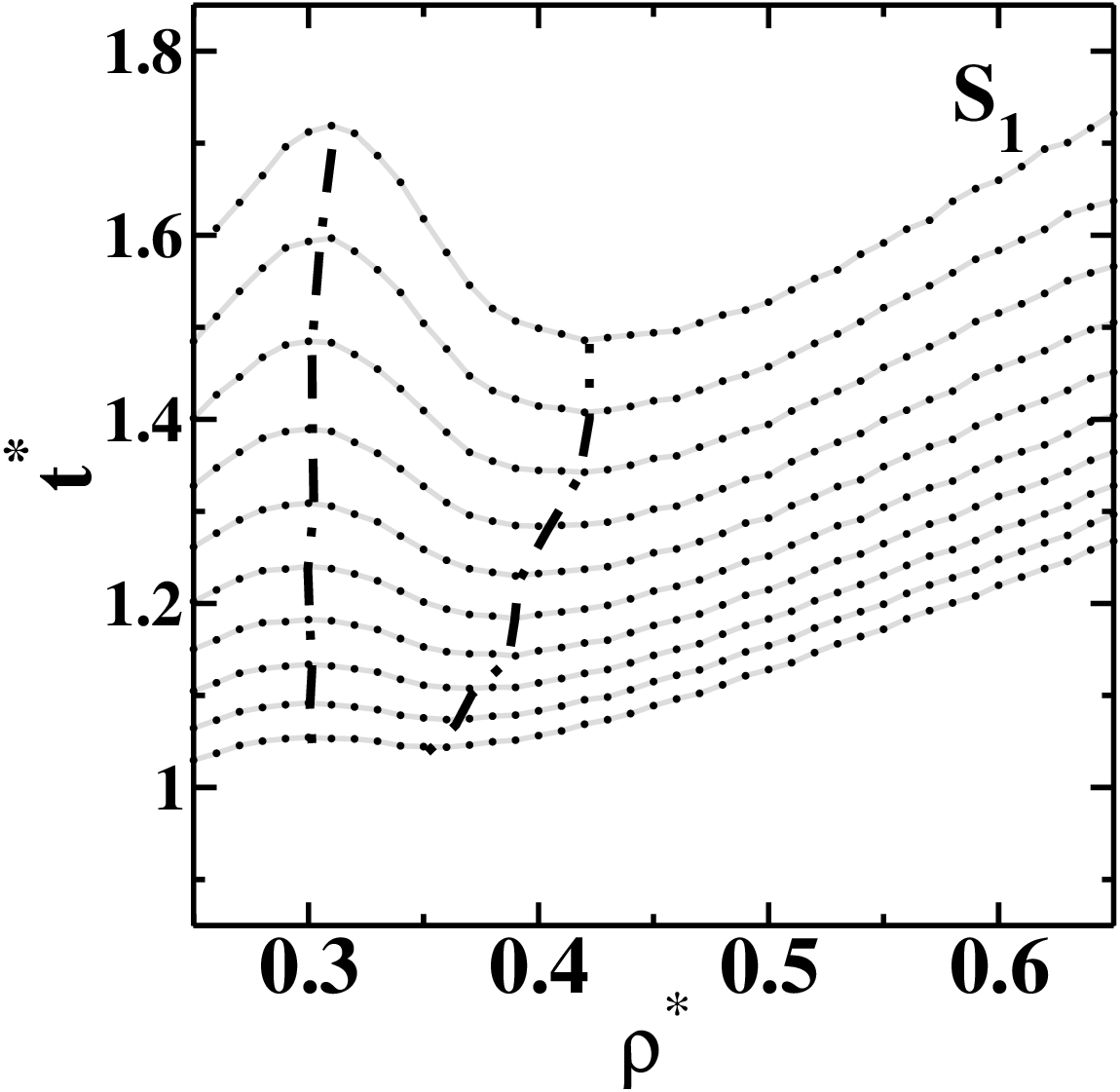}  & \includegraphics[clip,width=5cm]{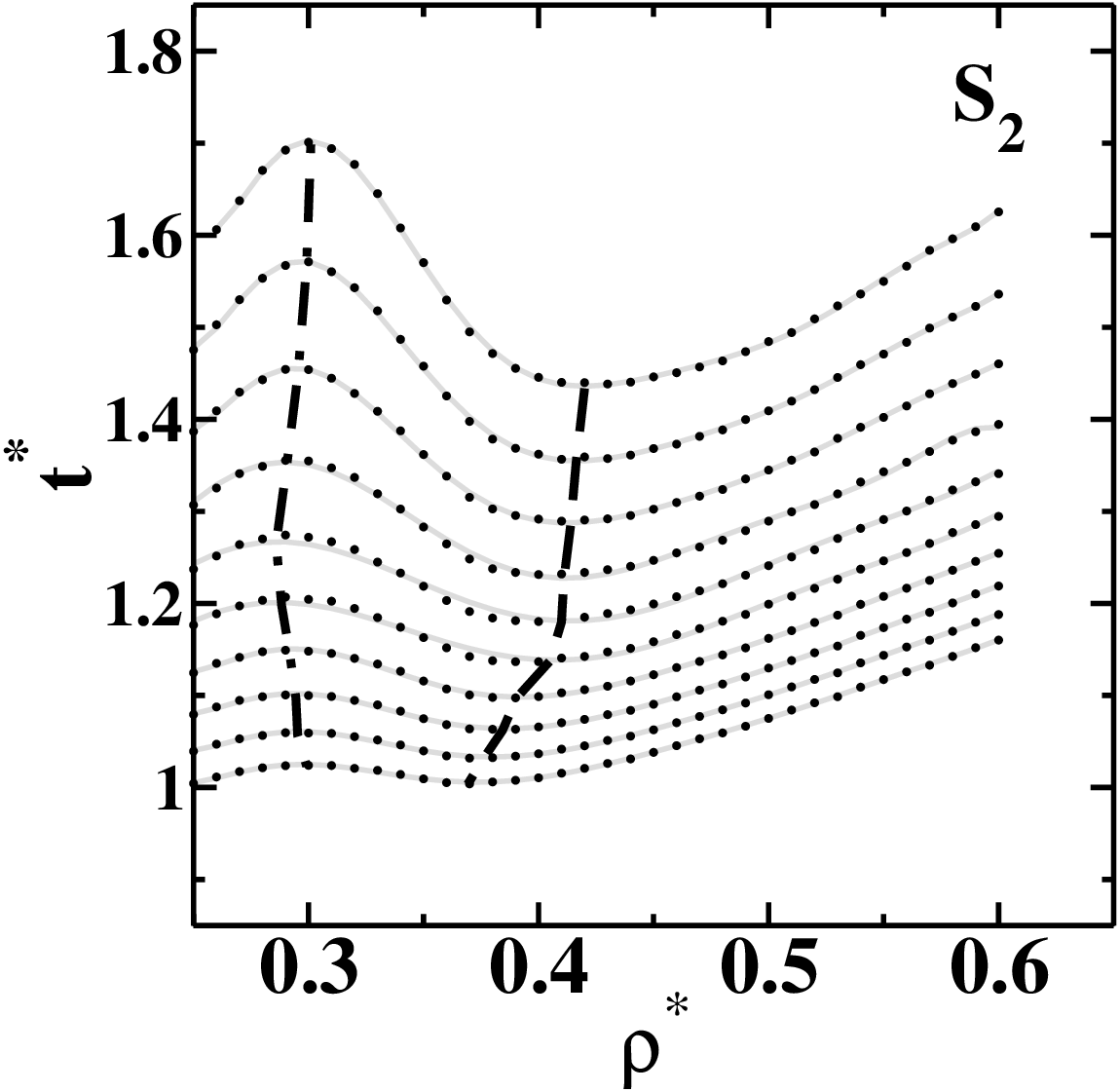}  & \includegraphics[clip,width=5cm]{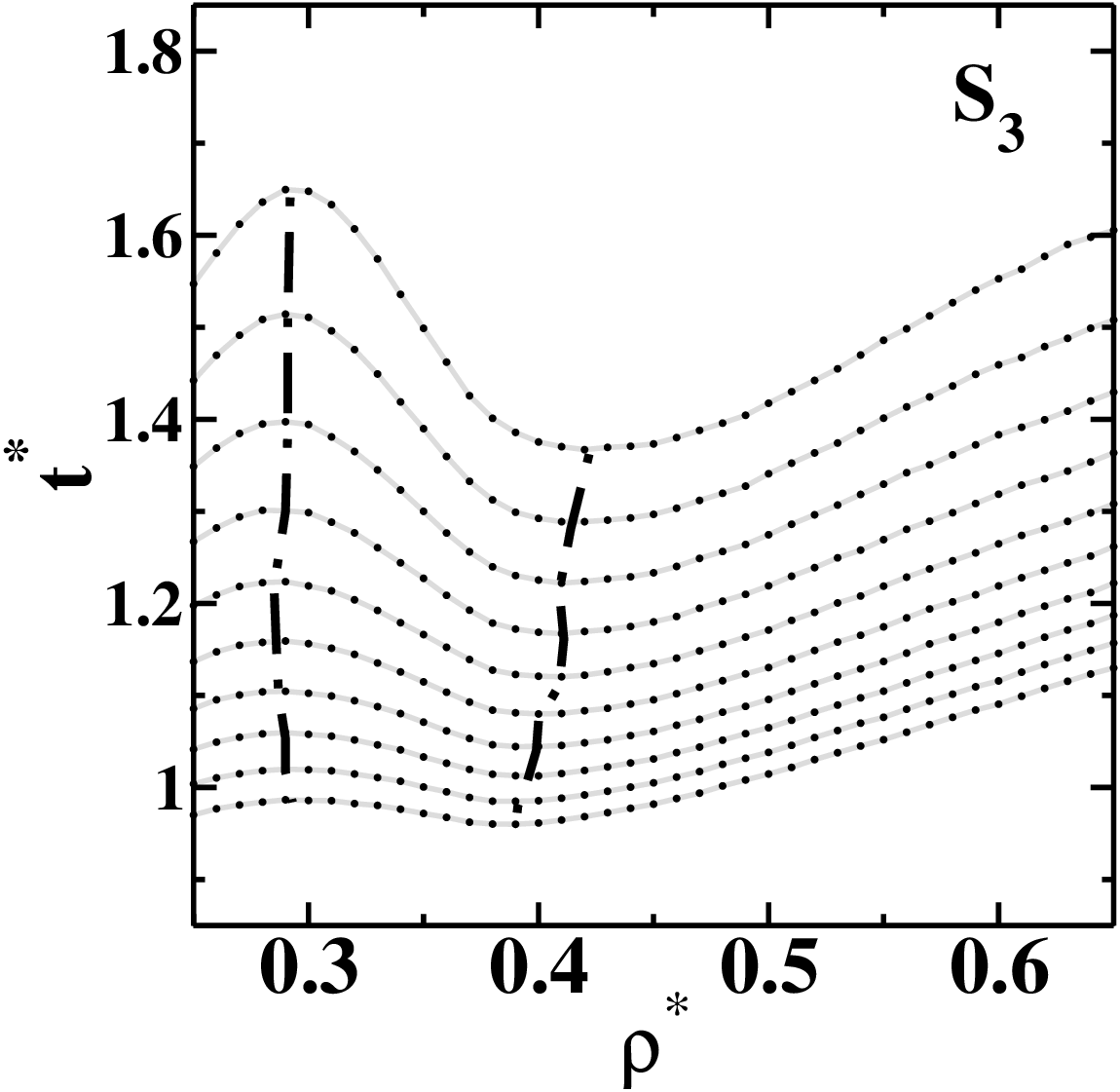} \\
      \includegraphics[clip,width=5cm]{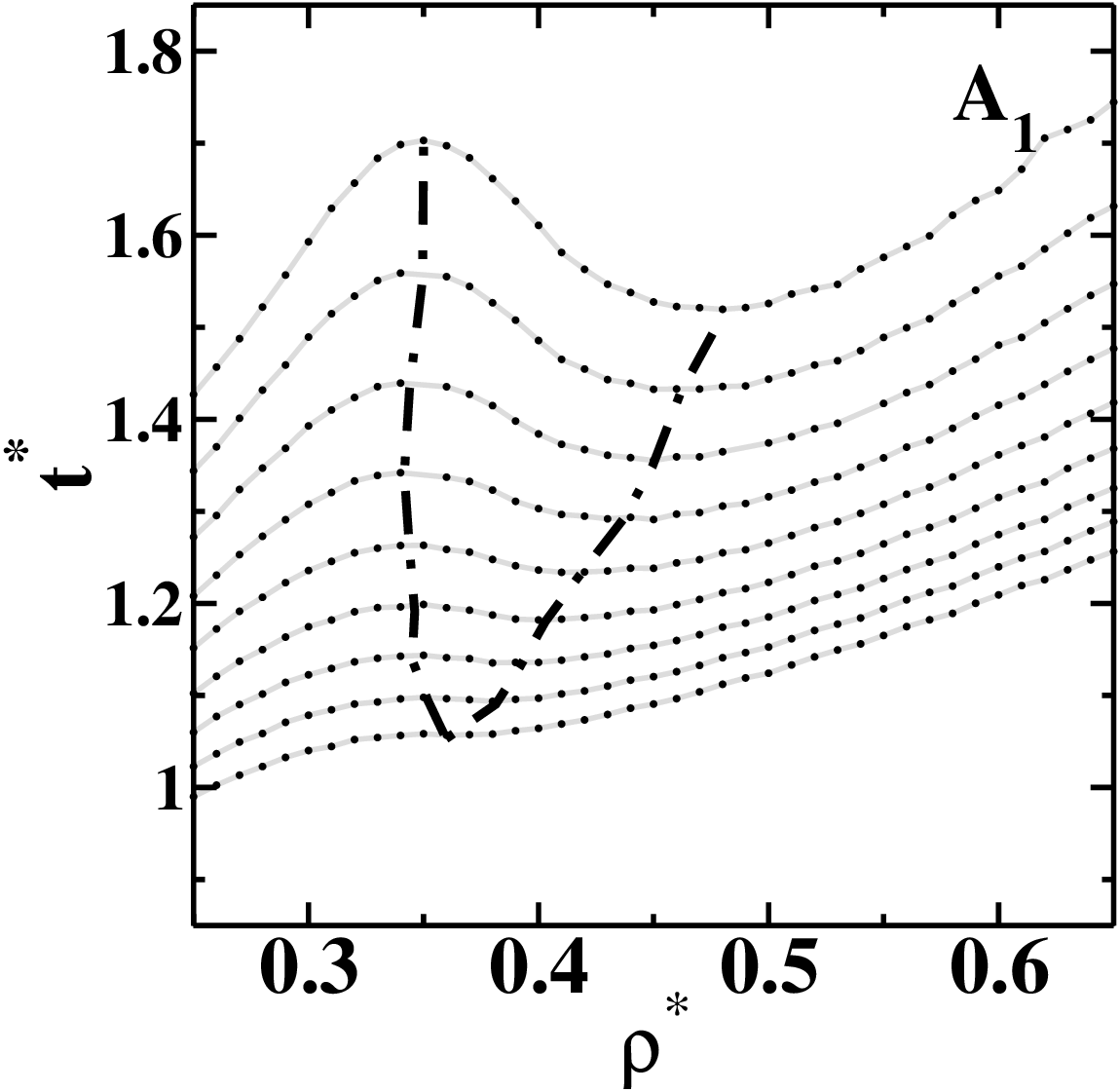}  & \includegraphics[clip,width=5cm]{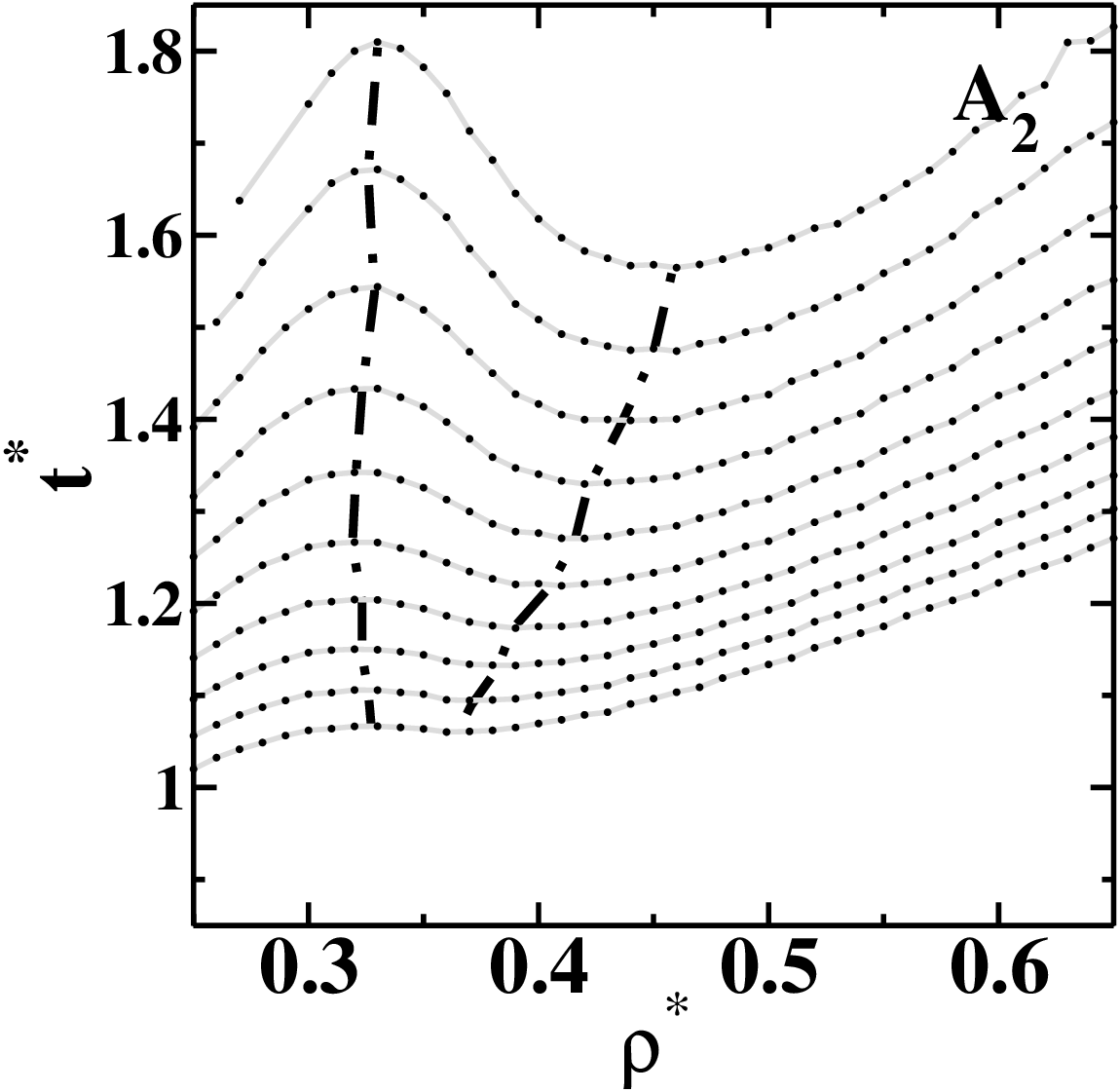}  & \includegraphics[clip,width=5cm]{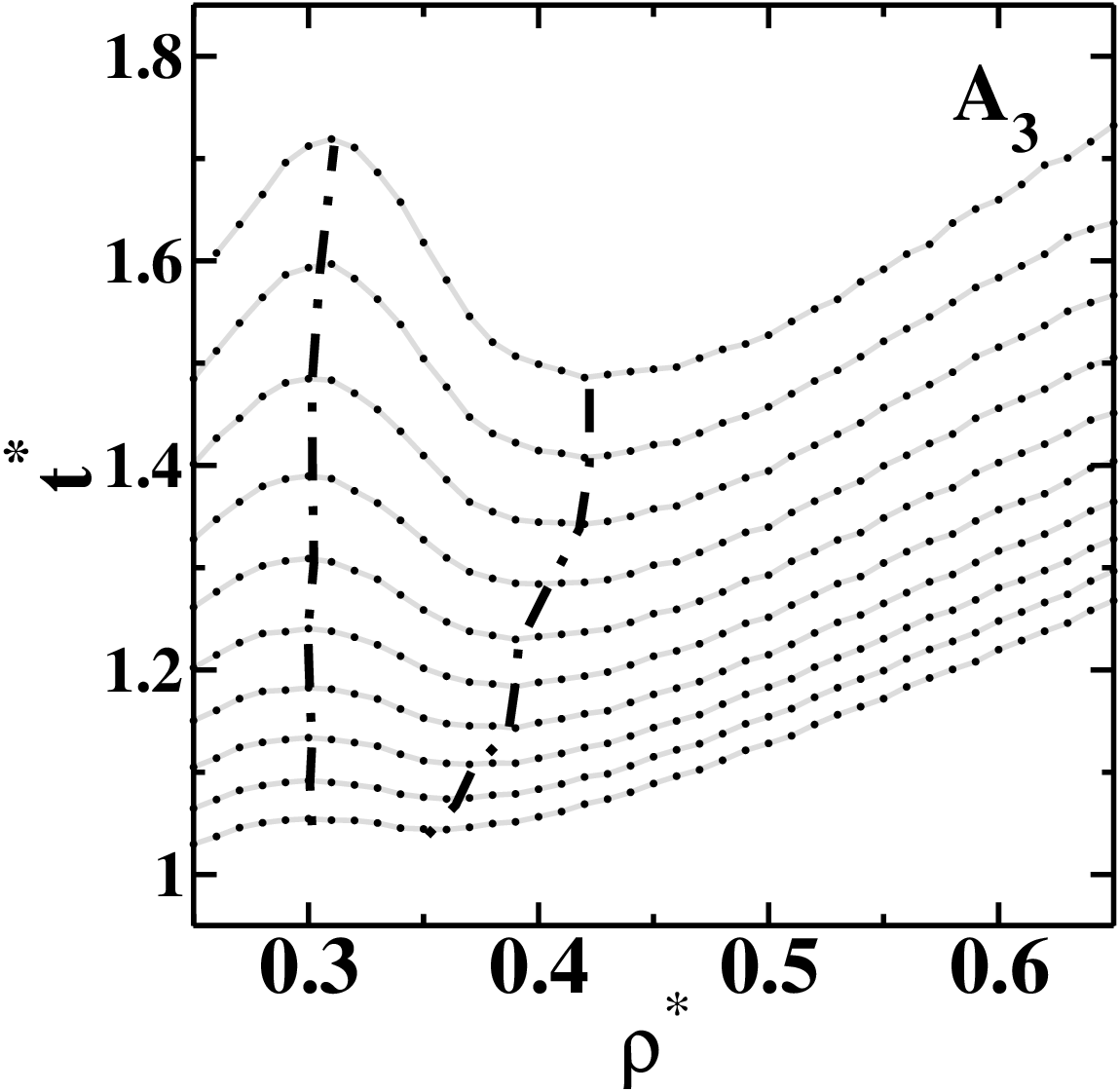} \\
      \includegraphics[clip,width=5cm]{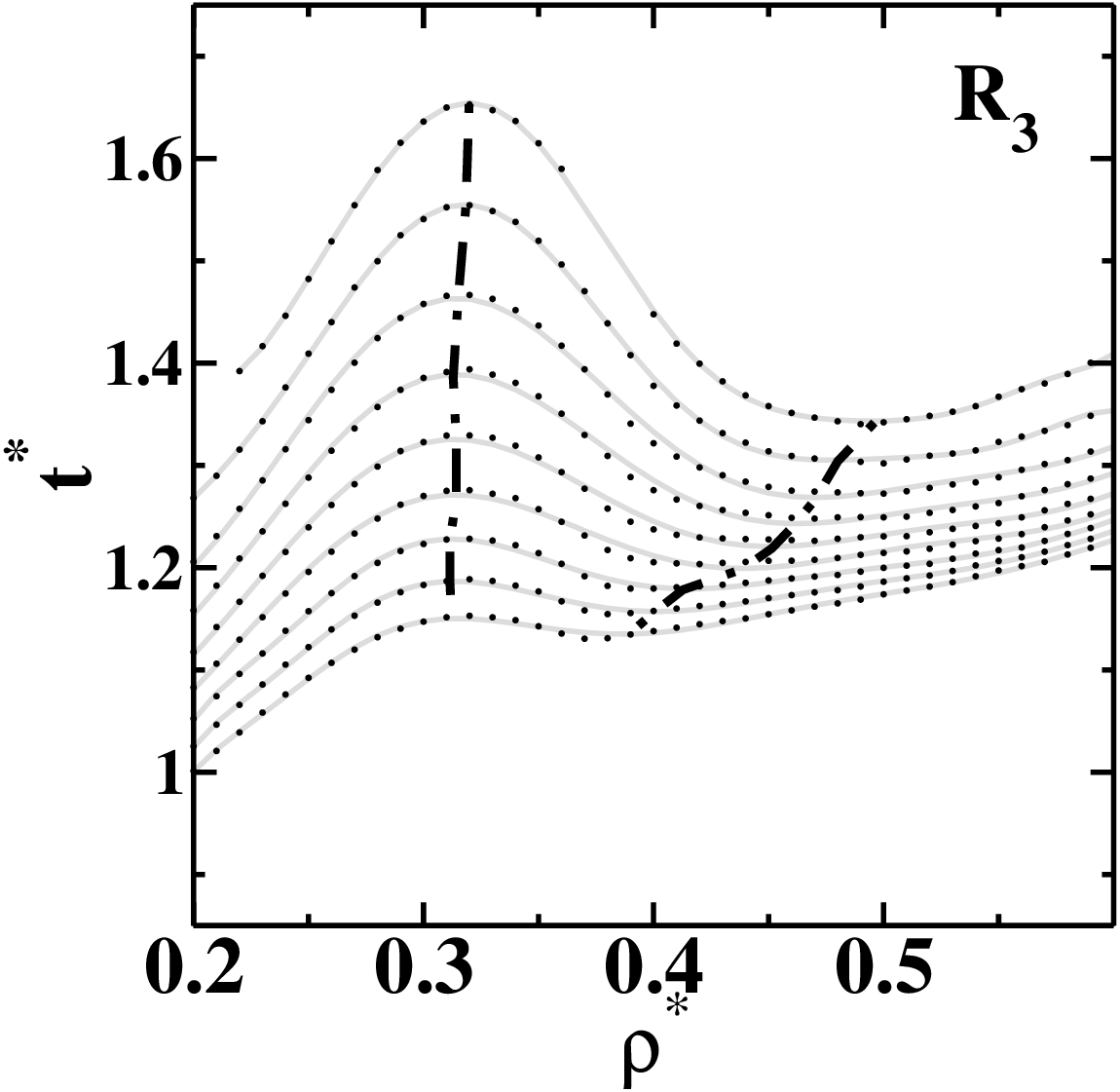}  & \includegraphics[clip,width=5cm]{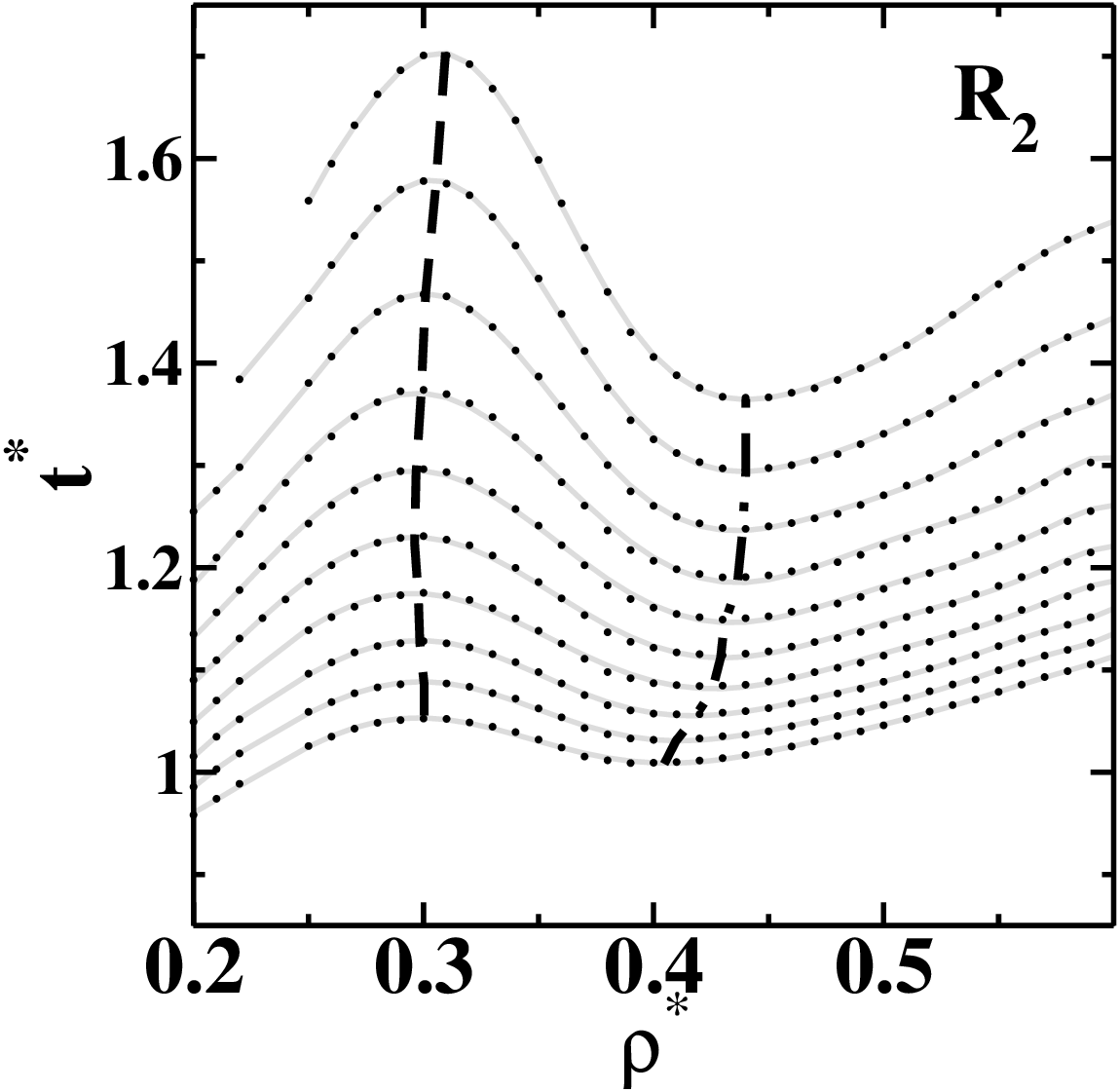}  & \includegraphics[clip,width=5cm]{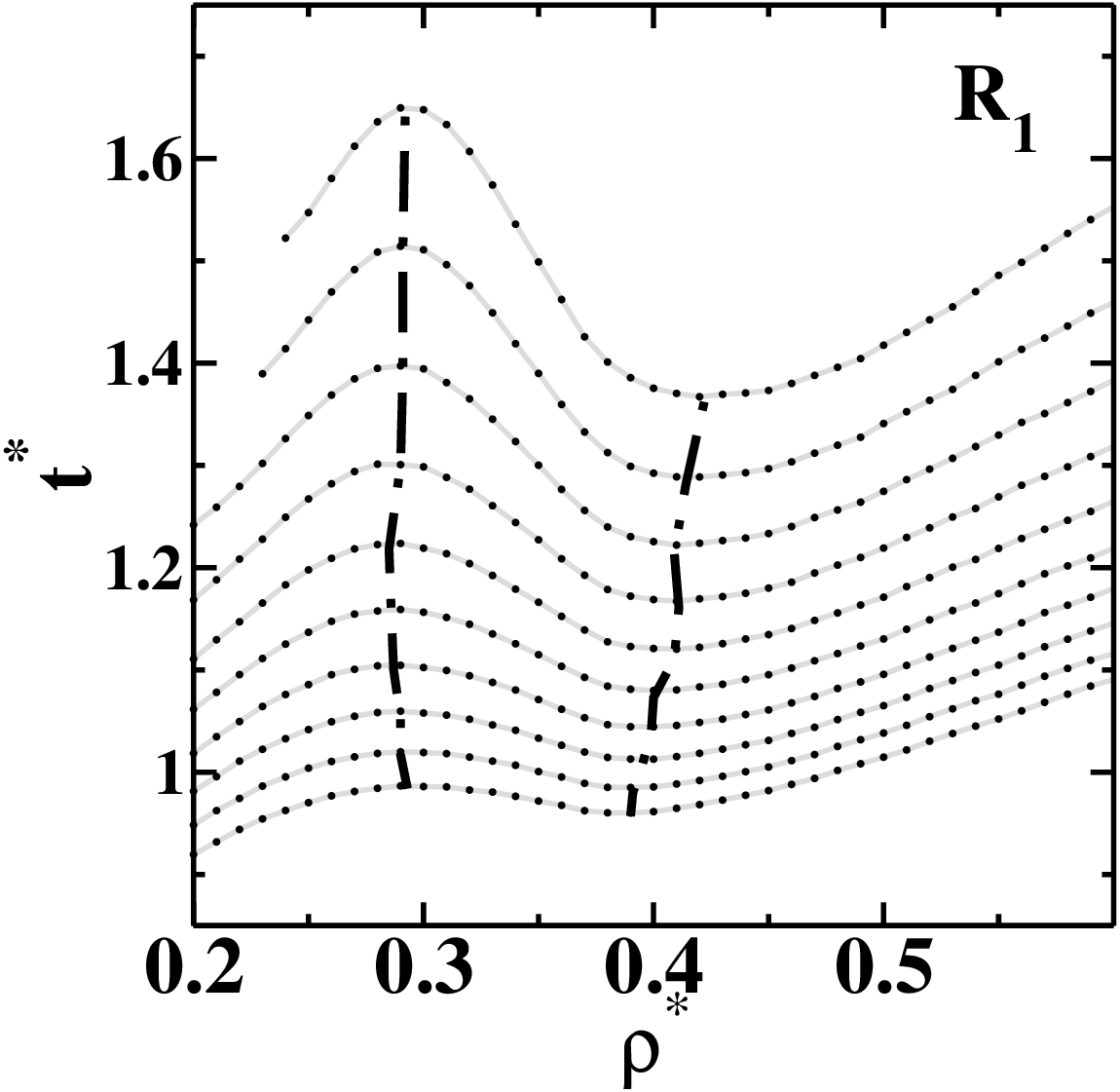} \\
    \end{tabular}
  \end{centering}
  \caption{The translational order parameter a function of density for
    fixed temperatures
    $T^*=0.6,\,0.7,\,0.8,\,0.9,\,1.0,\,1.1,\,1.2,\,1.3,\,1.4,\,1.5$
    (from top to bottom) for the families $S$ (on top), $A$ (on
    middle) and $R$ (on bottom). The dot-dashed lines locate the
    maximal and minimal in $t^*$.}
  \label{fig:tRho} 
\end{figure}

The graphs in Fig.~\ref{fig:tRho} illustrate the translational order
parameter versus density for the potentials studied. The 
dot-dashed lines show the maximum and minimum in
the values of $t$ that limit the region of anomalous behavior.
These extrema are also shown as dot-dashed lines
in Figs.~\ref{fig:PT}. The values at  the pressure-temperature
phase diagram for the different potentials follow the same
trend as the TMD and diffusion anomalous regions.

\subsection{Radial distribution function}

The density anomaly can be related to the structure by analyzing the
behavior of the radial distribution function. For a two length scales
potential the $g(r)$ has two peaks: one at the closest scale, $r_1$,
and another at the furthest scale, $r_2$~\cite{Ol06b}.

Recently it has been suggested that a signature
of the presence of  TMD line would be given
by the radial distribution
function as follows. At fixed temperature as the density
is increased the radial distribution
function of the closest scale, $g(r_1)$, would increase its value
while the radial distribution function of the furthest scale, $g(r_2)$, would
decrease. This
can be represented by the rule~\cite{Ol08a,Vi10}
\begin{eqnarray}
  \Pi_{1,2}=\left.\frac{\partial g(r)}{\partial\rho}\right|_{r_{1}}\times\left.\frac{\partial g(r)}{\partial\rho}\right|_{r_{2}} <0\;\;.
  \label{eq:condition}
\end{eqnarray}

\begin{figure}[hb]
  \begin{centering}
    \begin{tabular}{ccc}
      \includegraphics[clip,width=5cm]{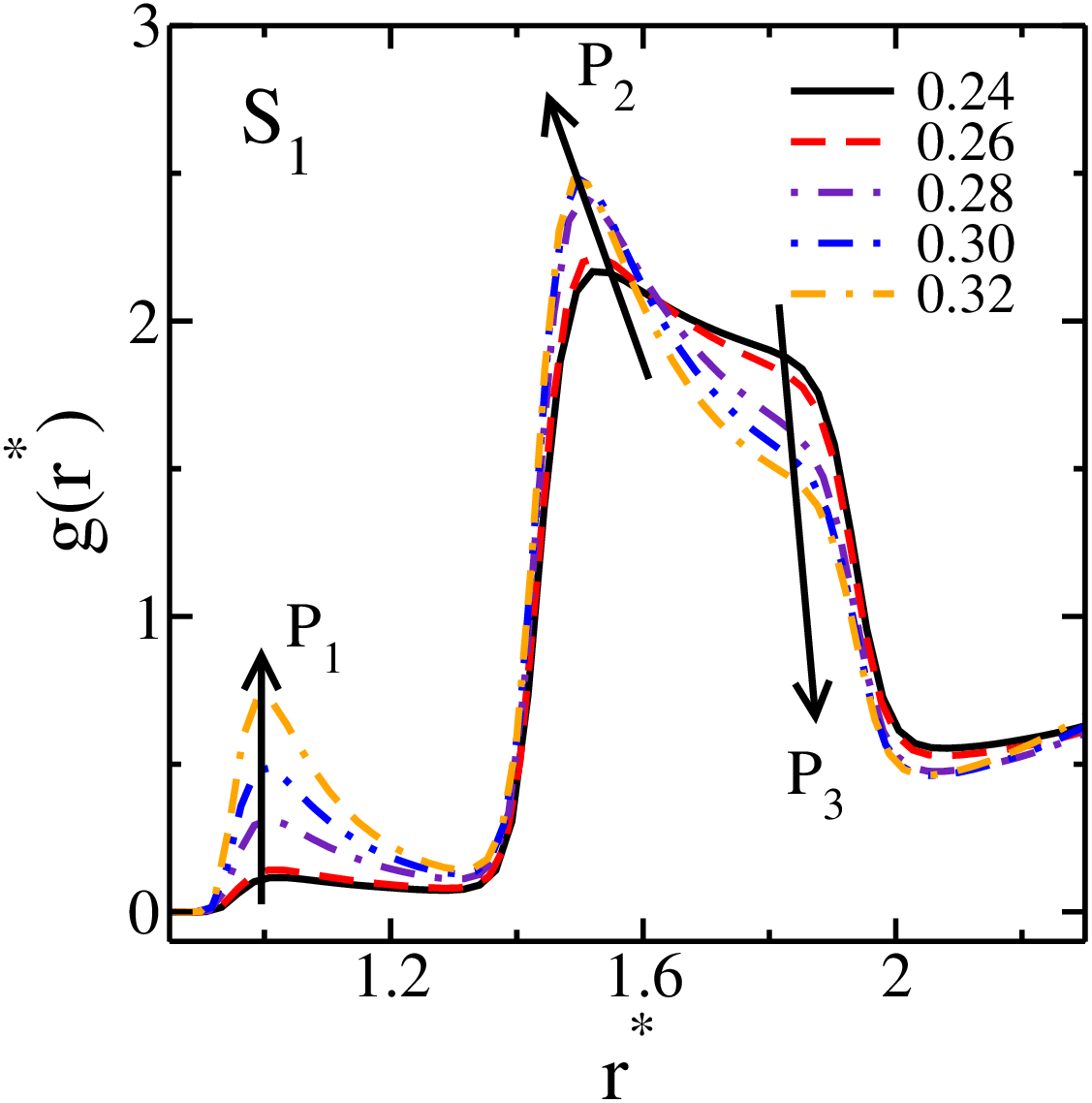}  & \includegraphics[clip,width=5cm]{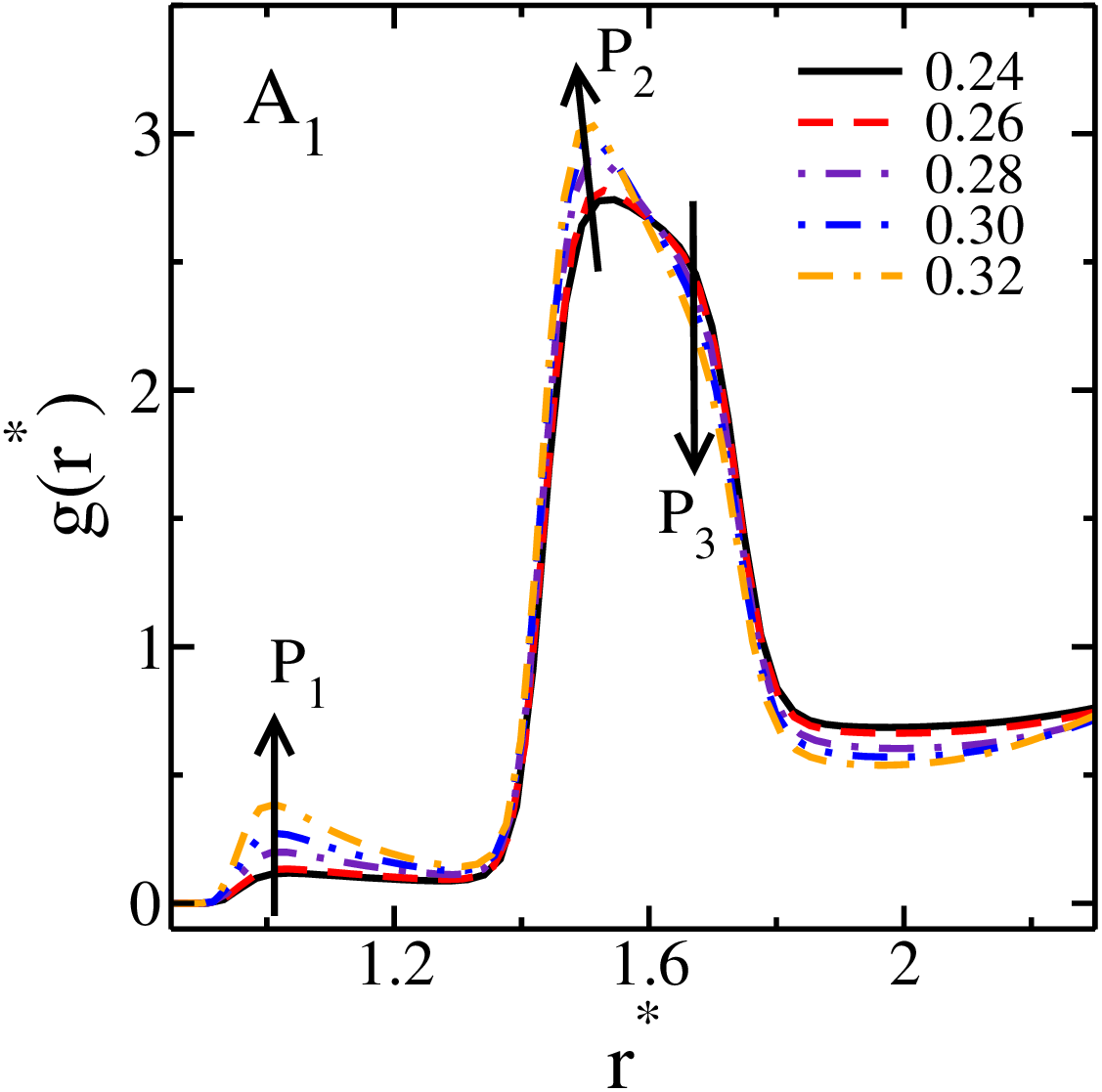}  & \includegraphics[clip,width=5cm]{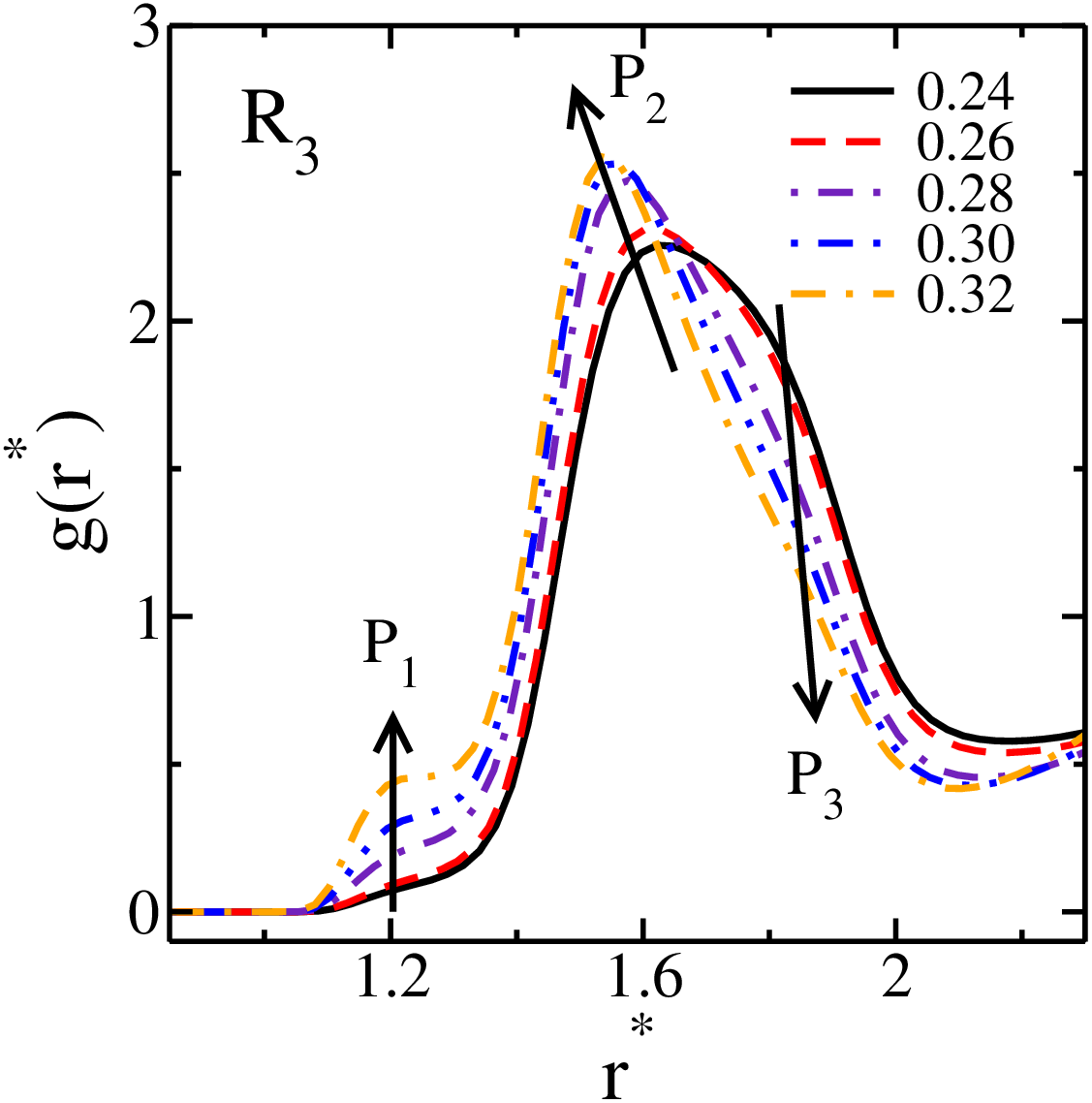}\\
      \includegraphics[clip,width=5cm]{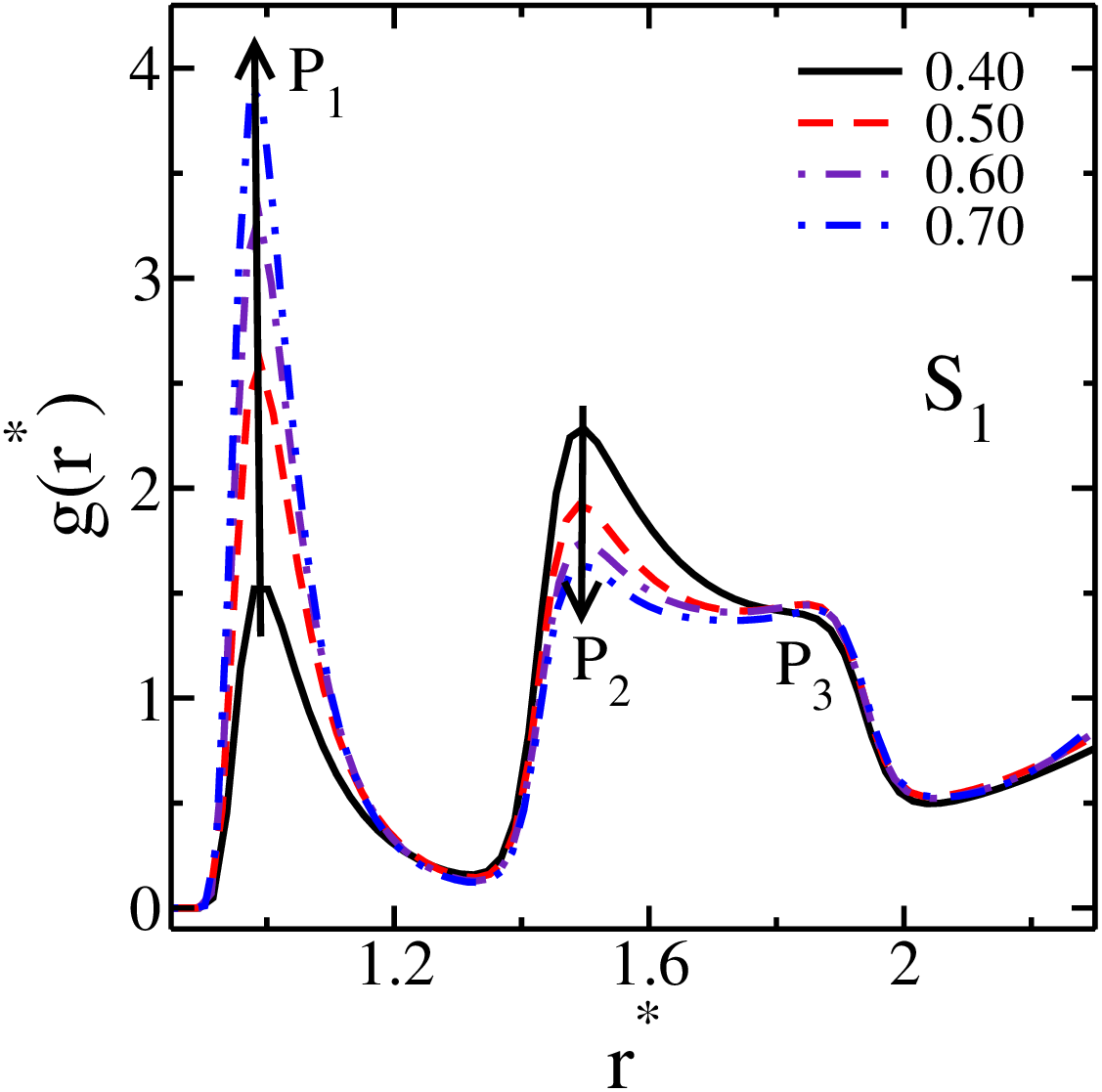}  & \includegraphics[clip,width=5cm]{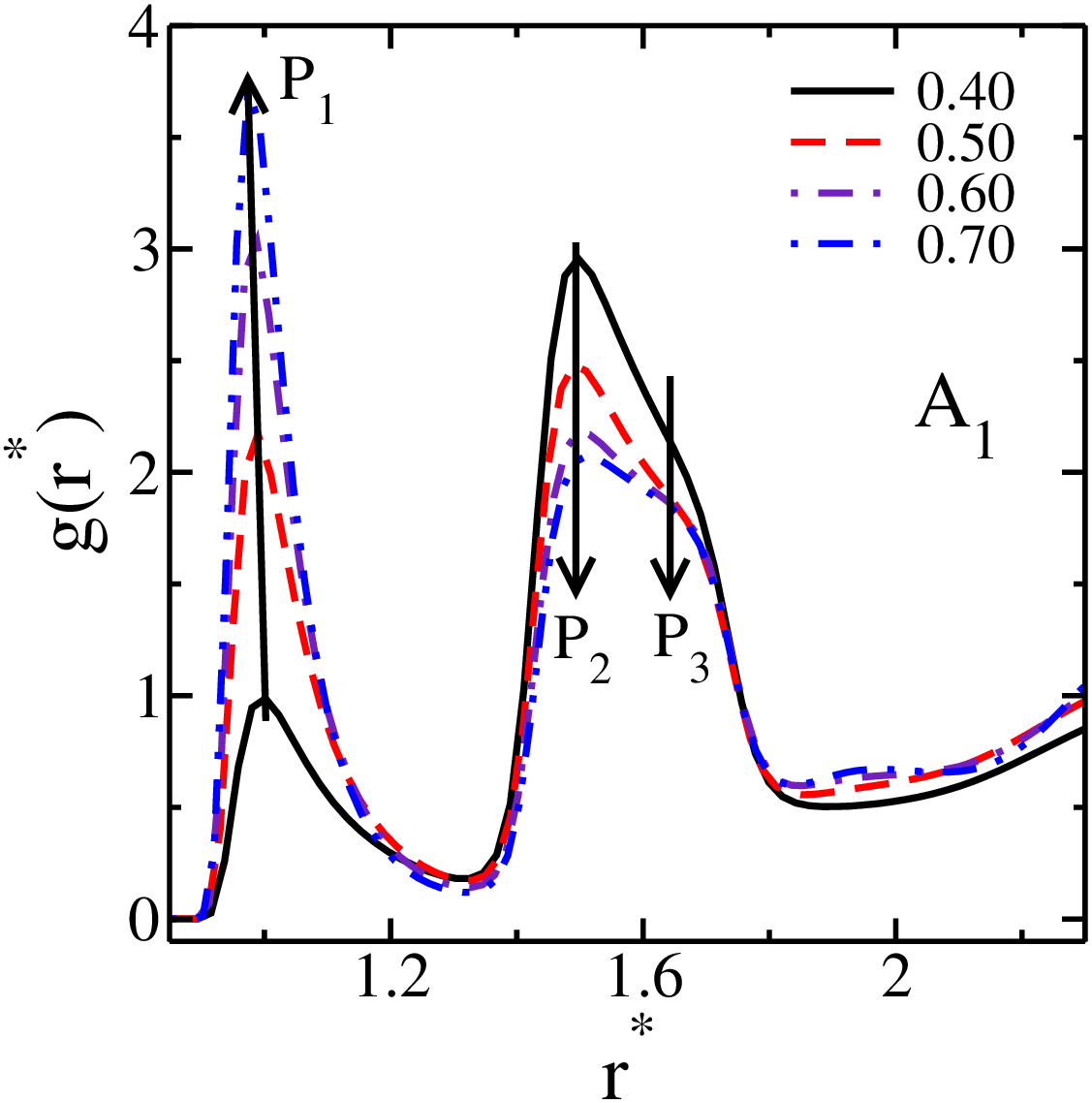}  & \includegraphics[clip,width=5cm]{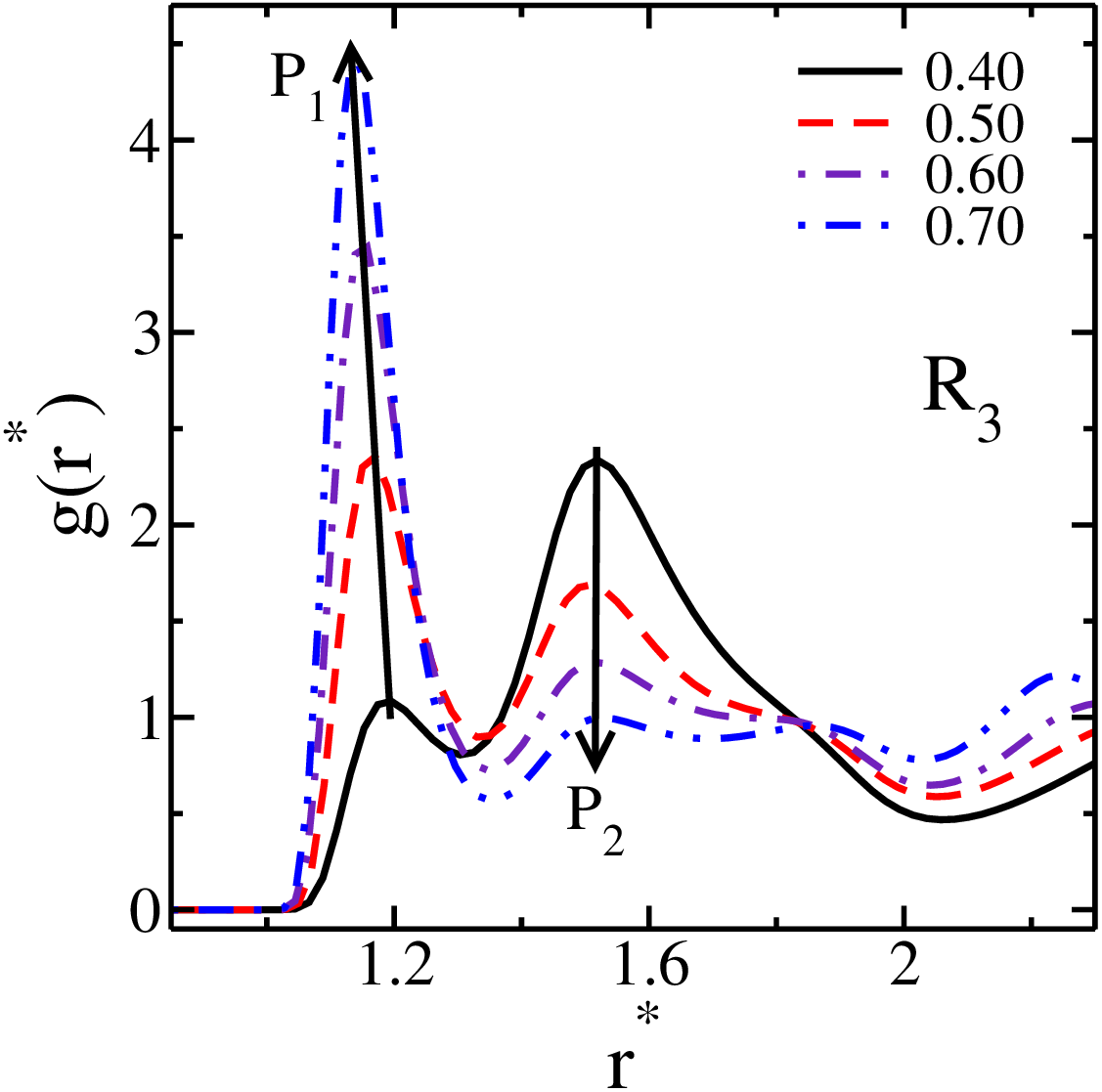}
    \end{tabular}
  \end{centering}
  \caption{Radial distribution as a function of the reduced distance for 
    selected cases in the three families of potentials for
    $T^*=0.8$. For all the families, for $\rho^* < 0.40$ the 
first and second peaks of $g(r)$
    increase with the increase of density. For $\rho^* \geq 0.40$ the first peak increases while
    second peak decreases with the increase of density.}\label{fig:gr} 
\end{figure}

The physical picture behind this condition~\cite{Ol08a} is that for a
fixed temperature as density increases particles that are located at
the attractive scale, $r_2$, move to the repulsive scale,
$r_1$. Figures~\ref{fig:gr} illustrate a typical radial distribution
functions for fixed $T^*$ as $\rho^*$ is varied. These graphs show
that the picture of particles changing
length scales due to pressure increase
is valid for densities beyond a threshold density
$\rho^*_{min}$.


\begin{figure}[ht]
  \begin{centering}
    \begin{tabular}{ccc}
      \includegraphics[clip,width=5cm]{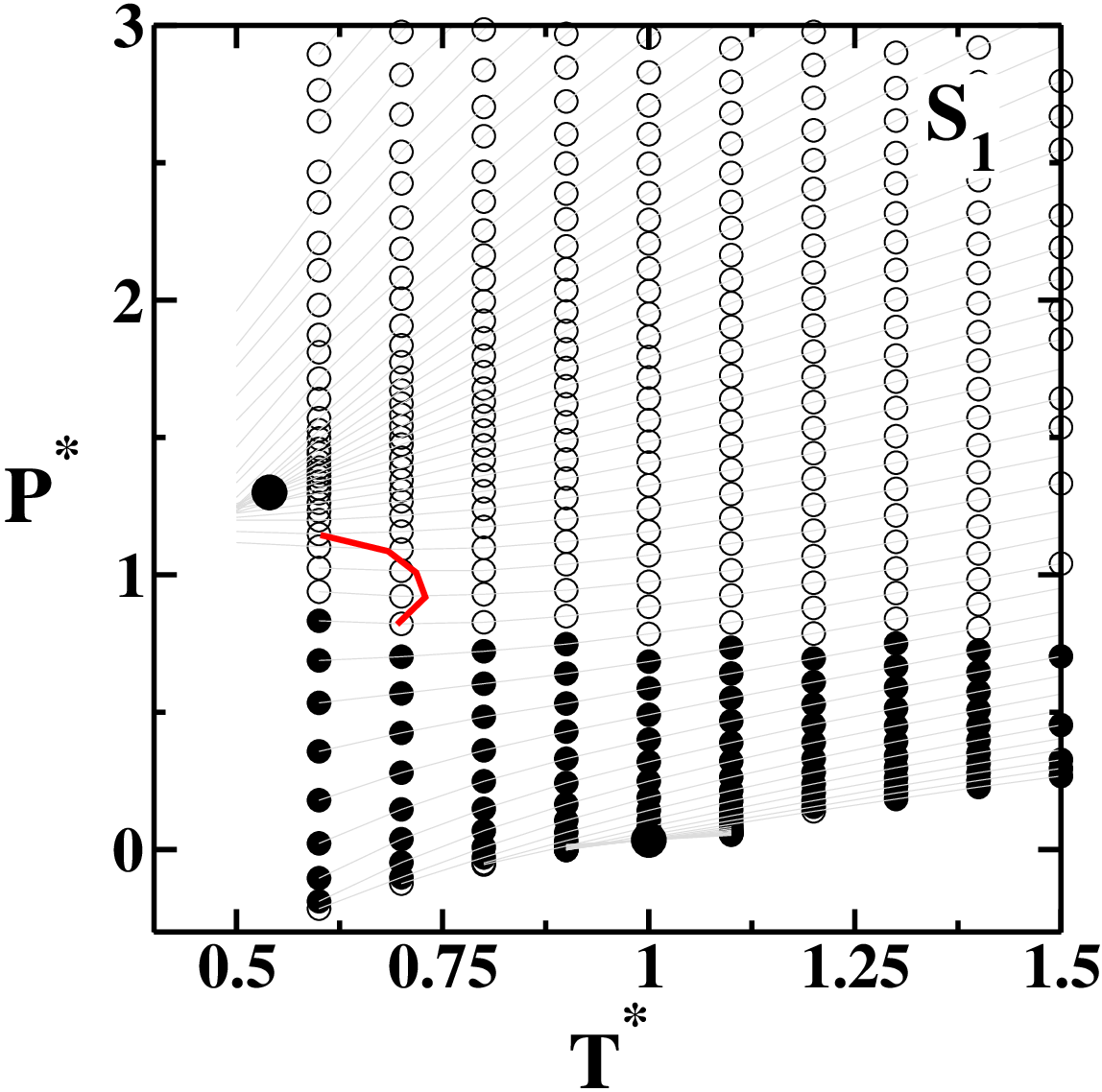}  & \includegraphics[clip,width=5cm]{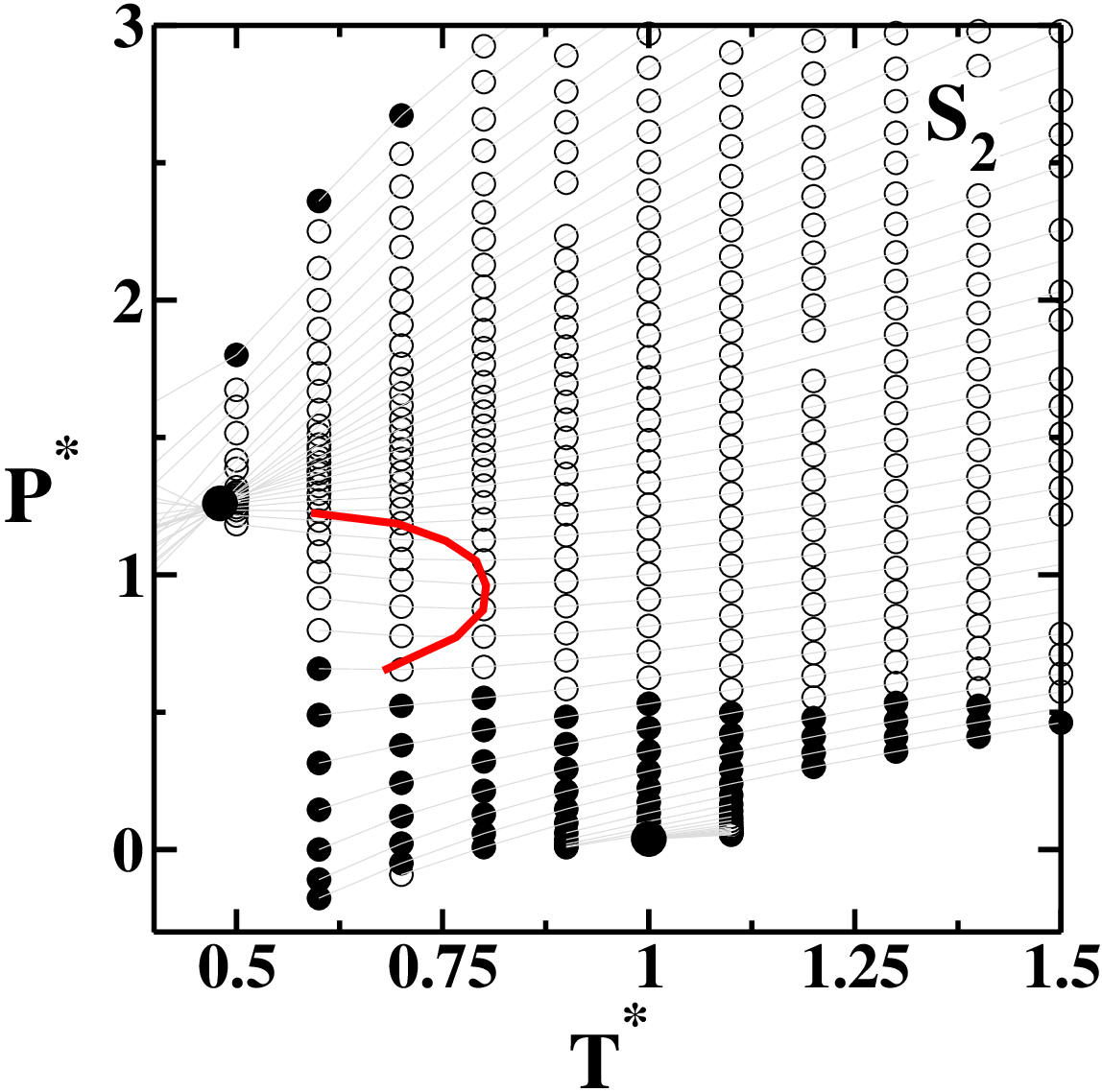}  & \includegraphics[clip,width=5cm]{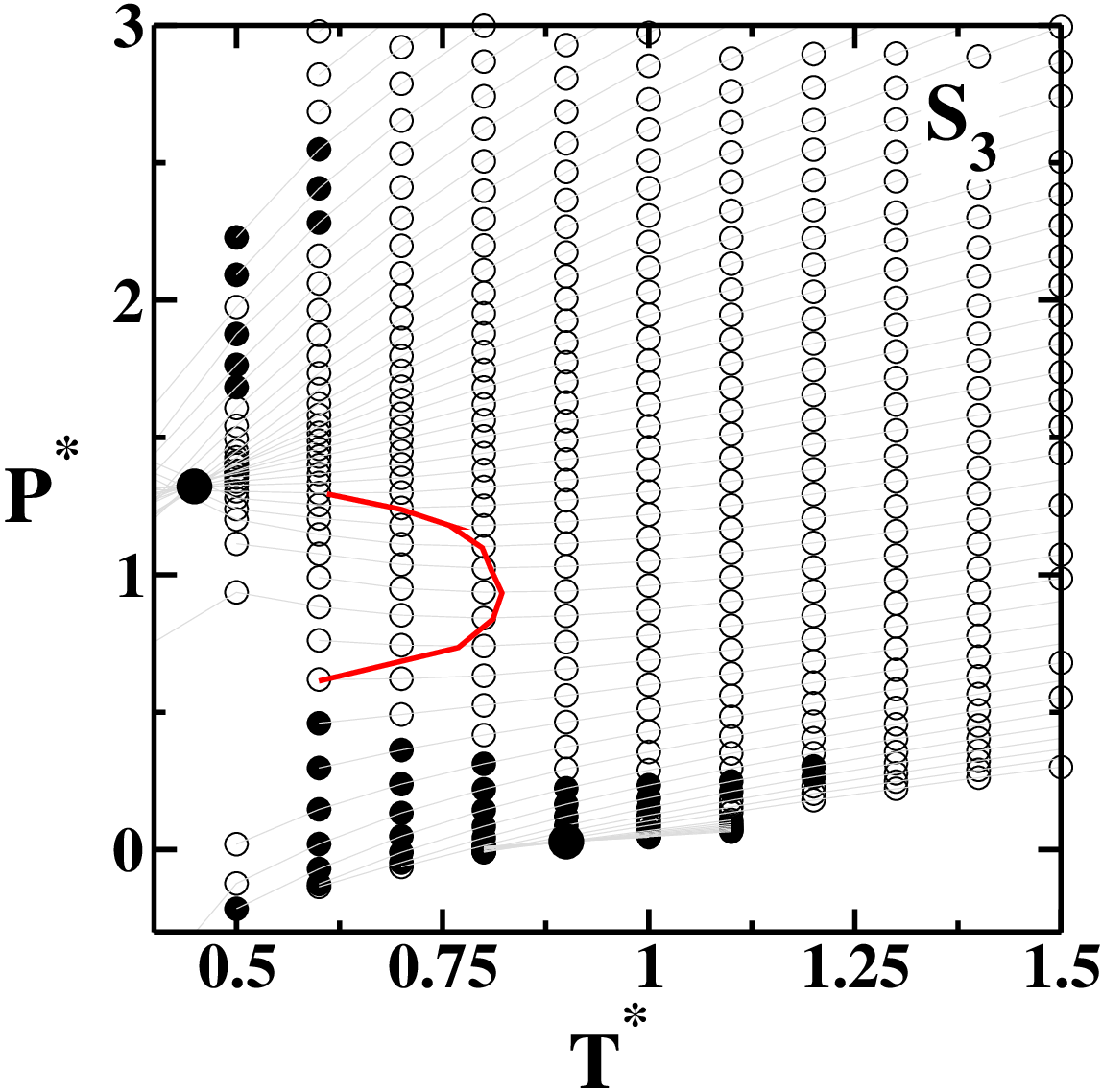} \\
      \includegraphics[clip,width=5cm]{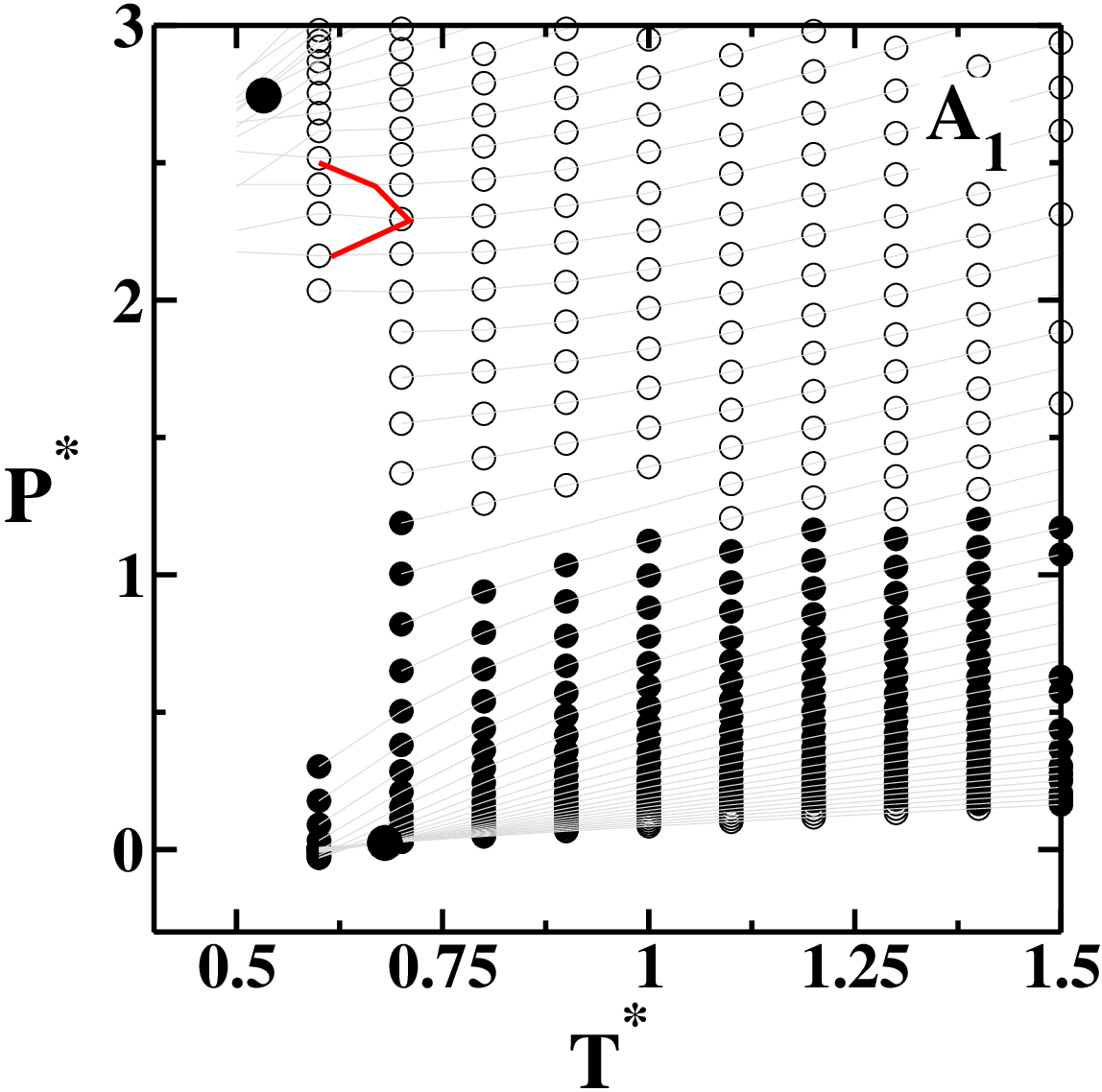}  & \includegraphics[clip,width=5cm]{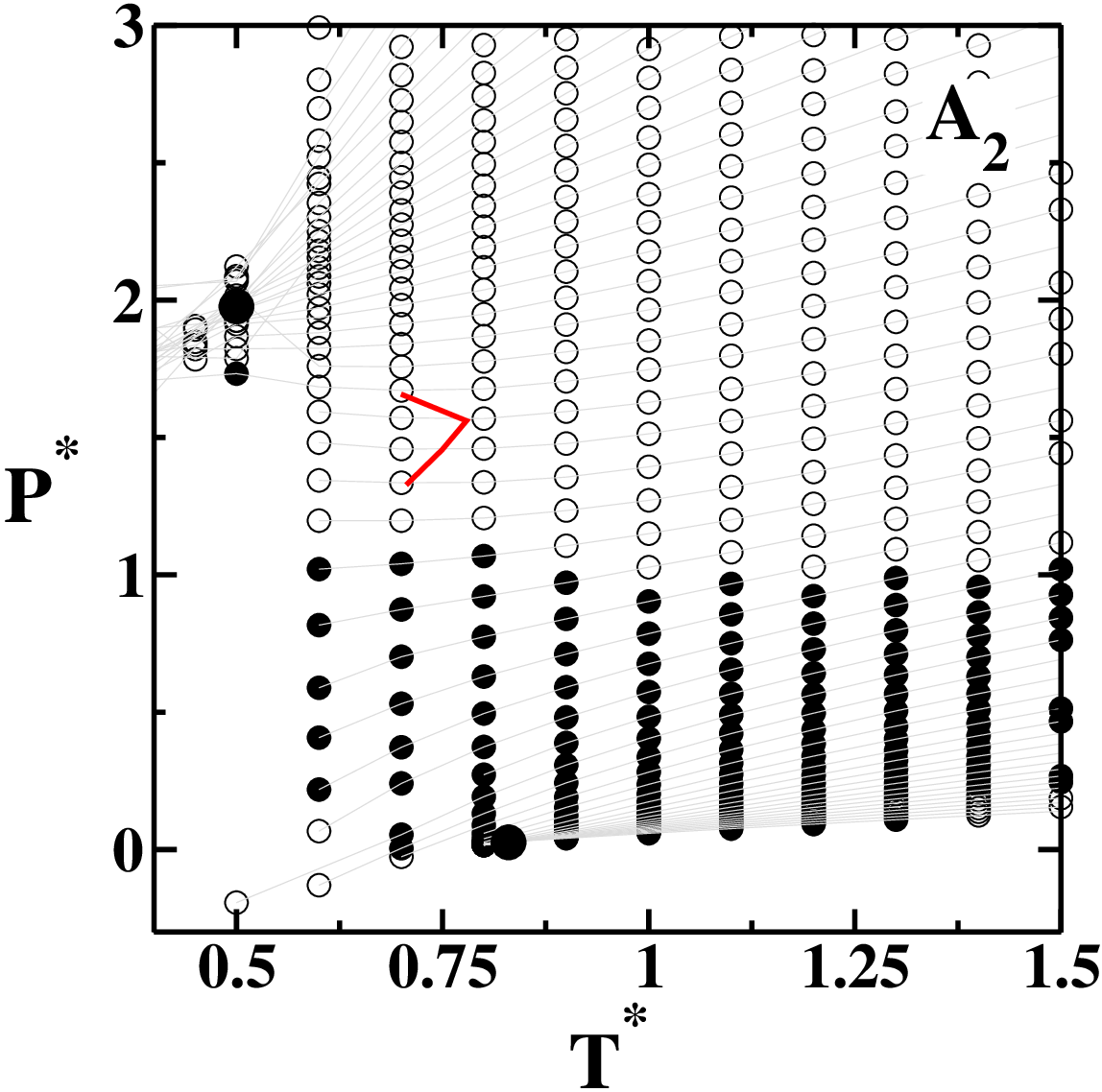}  & \includegraphics[clip,width=5cm]{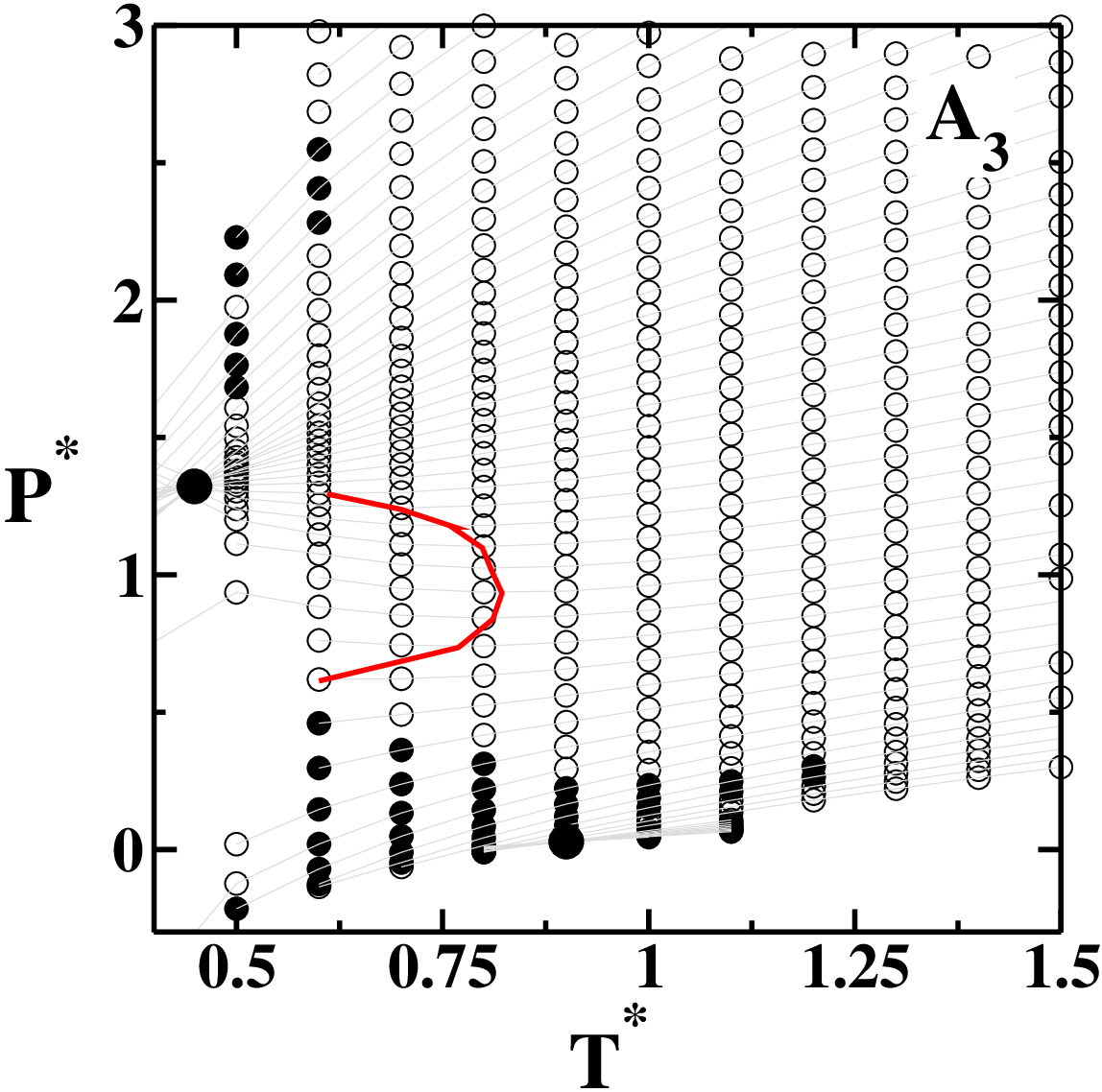} \\
      \includegraphics[clip,width=5cm]{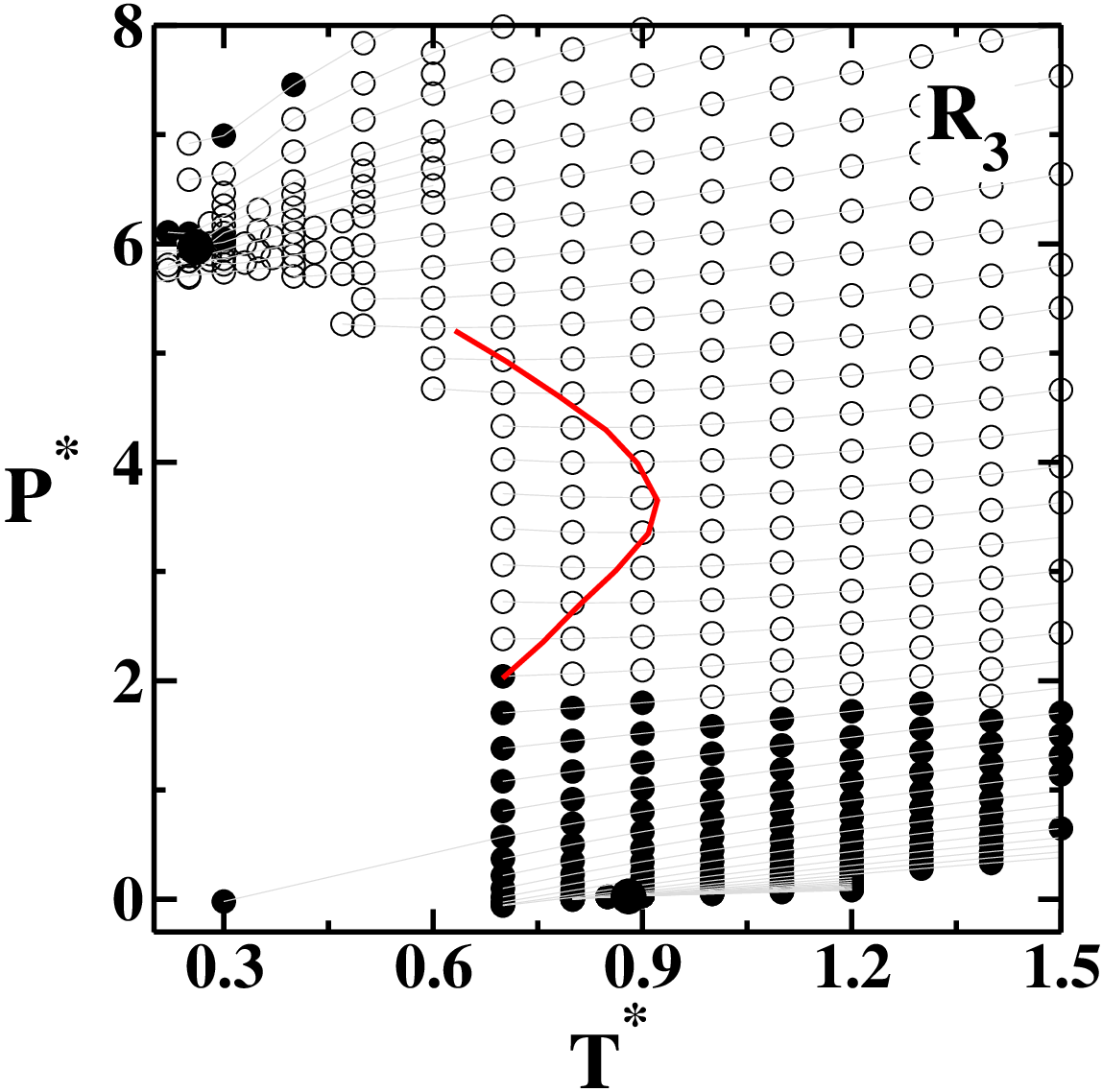}  & \includegraphics[clip,width=5cm]{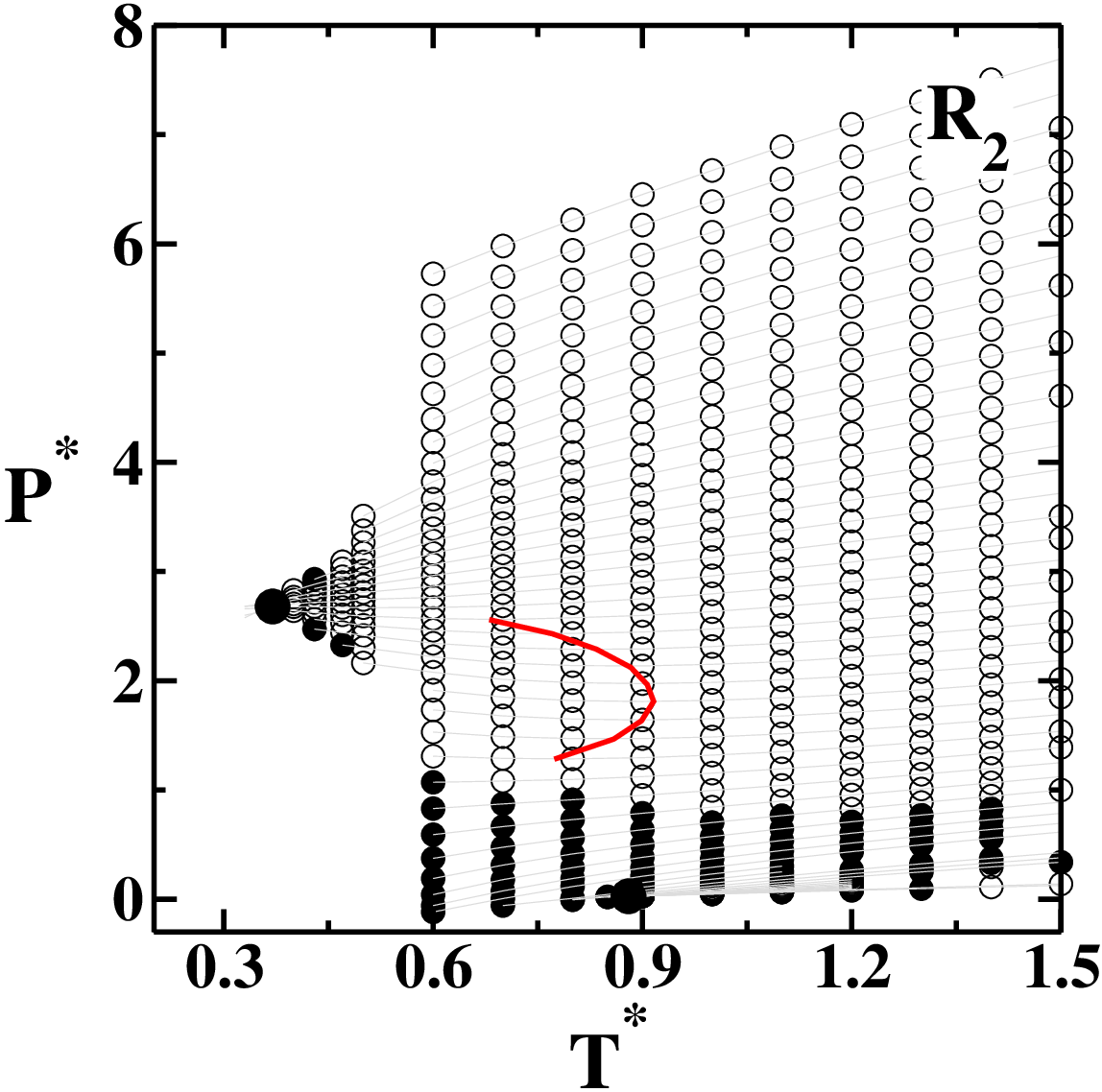}  & \includegraphics[clip,width=5cm]{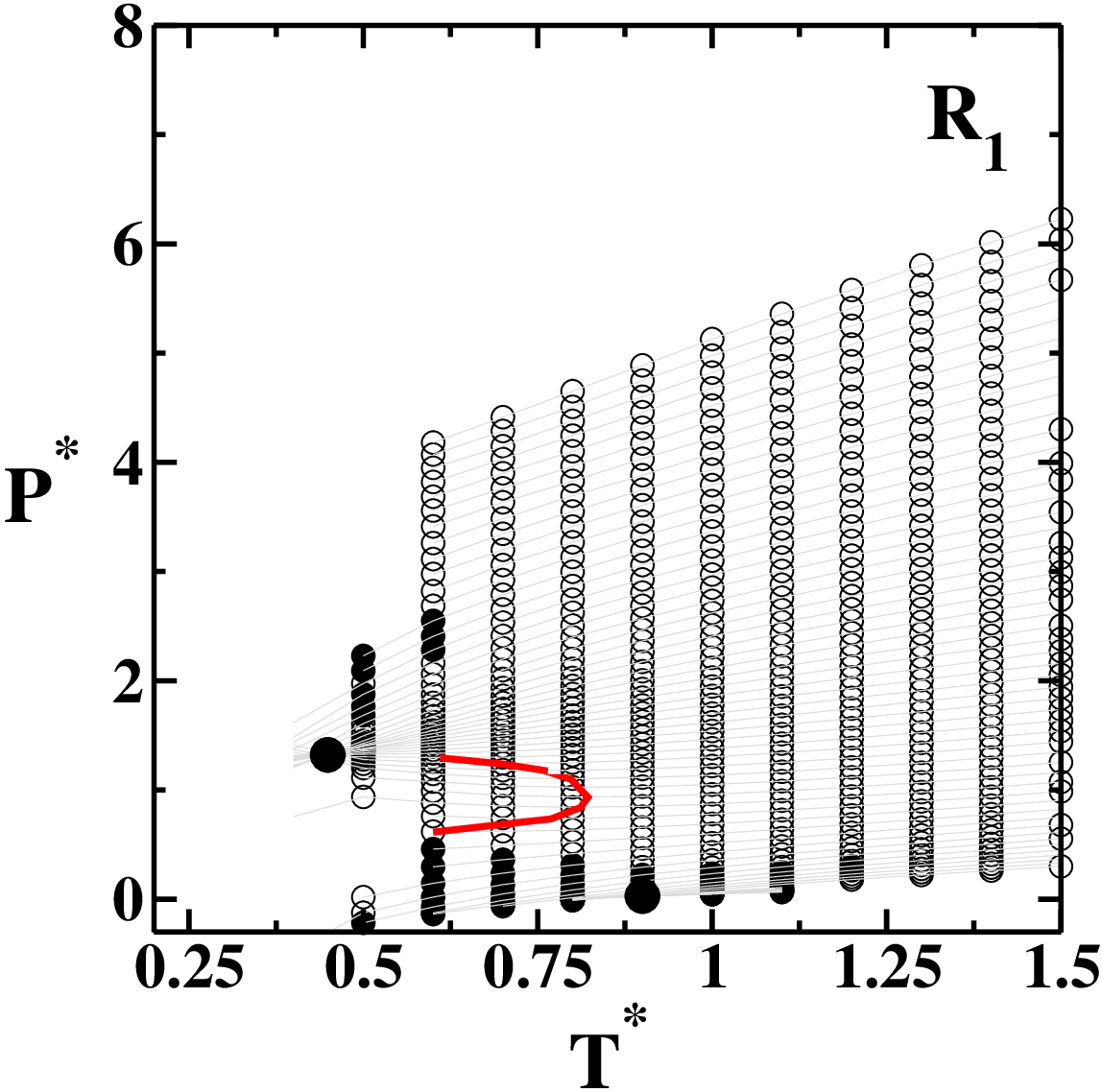} \\
    \end{tabular}
  \end{centering}
  \caption{Pressure-temperature phase diagram for the $S$, $A$ and $R$ families of potentials, illustrating as opened circles the regions where the condition
Eq.~\ref{eq:condition} is obeyed. } 
 \label{fig:p.vs.t-condition} 
\end{figure}

The regions identified by the
radial distribution function as 
fulfilling the condition Eq.~\ref{eq:condition}
are illustrated as opened circles in Fig.~\ref{fig:p.vs.t-condition}.  The
solid curve shows the TMD line. All the stable state points with
density equal or higher the minimum density at the TMD line verify the
relation $\Pi_{1,2}\left(\rho,T\right)<0$.  This result gives support 
to our assumption that the
presence of anomalies is related to particles moving from the furthest
scale, $r_2$, to closest length scale, $r_1$.


\begin{figure}[ht]
  \begin{centering}
    \begin{tabular}{ccc}
      \includegraphics[clip,width=5cm]{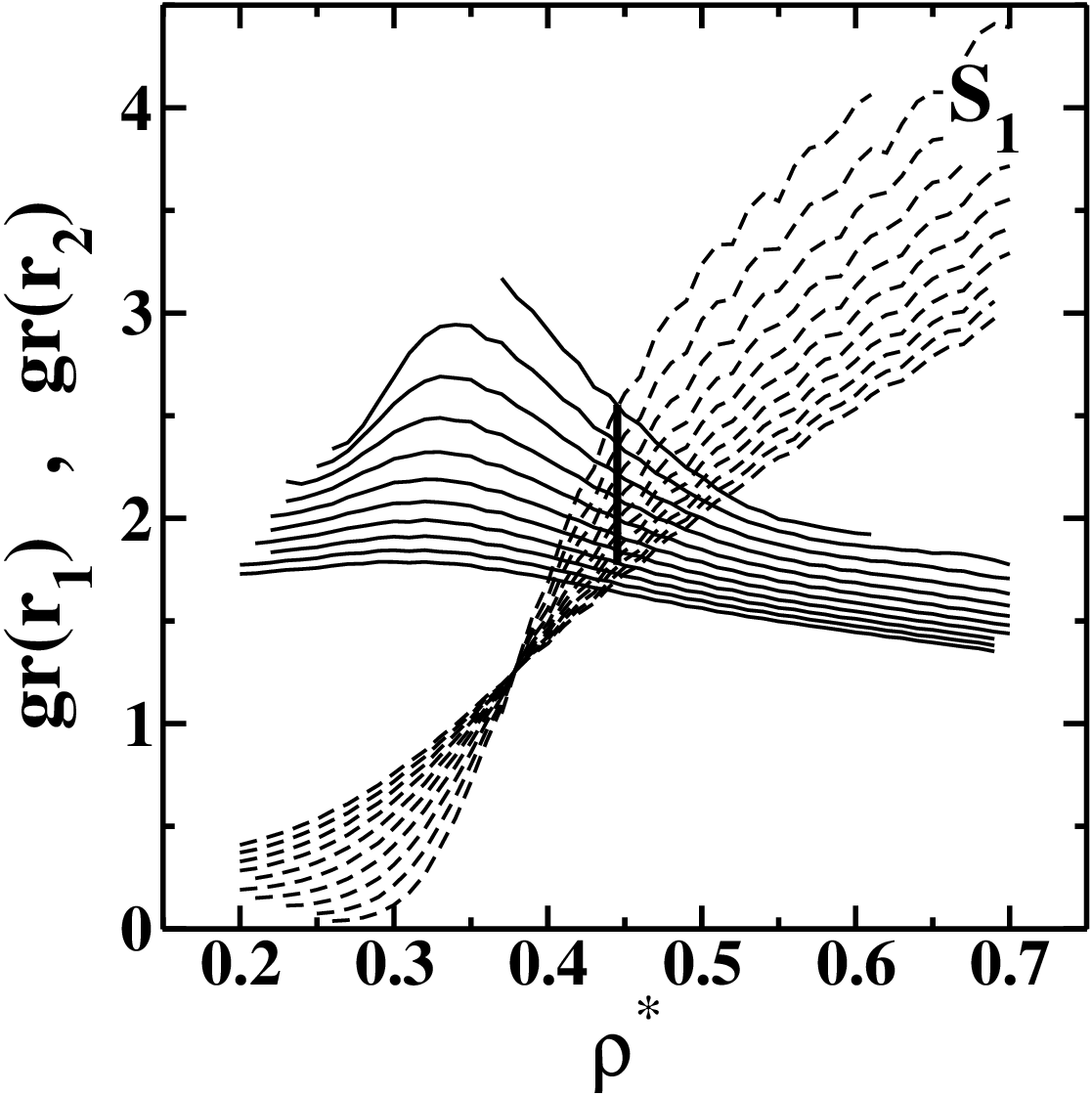}  & \includegraphics[clip,width=5cm]{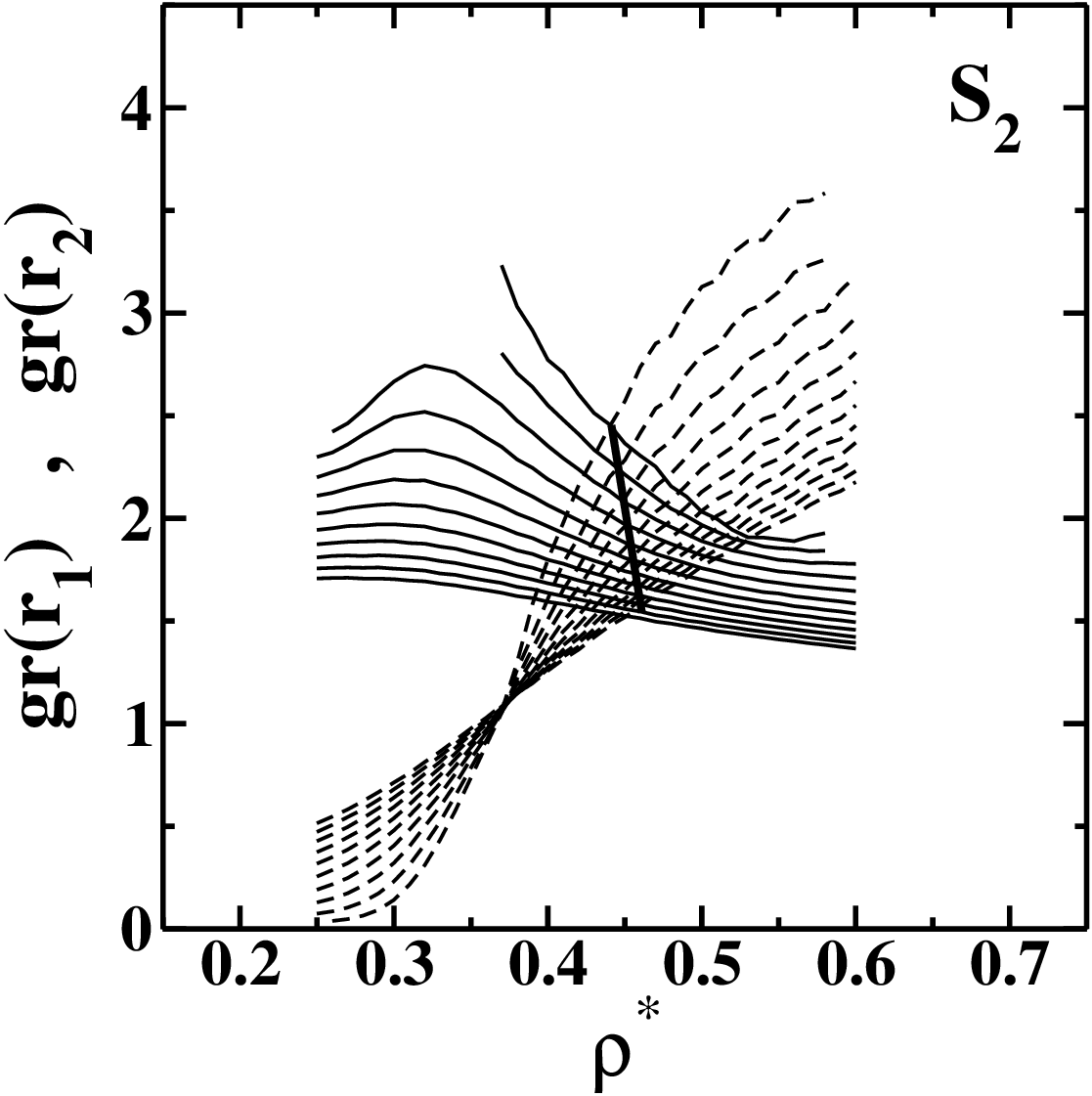}  & \includegraphics[clip,width=5cm]{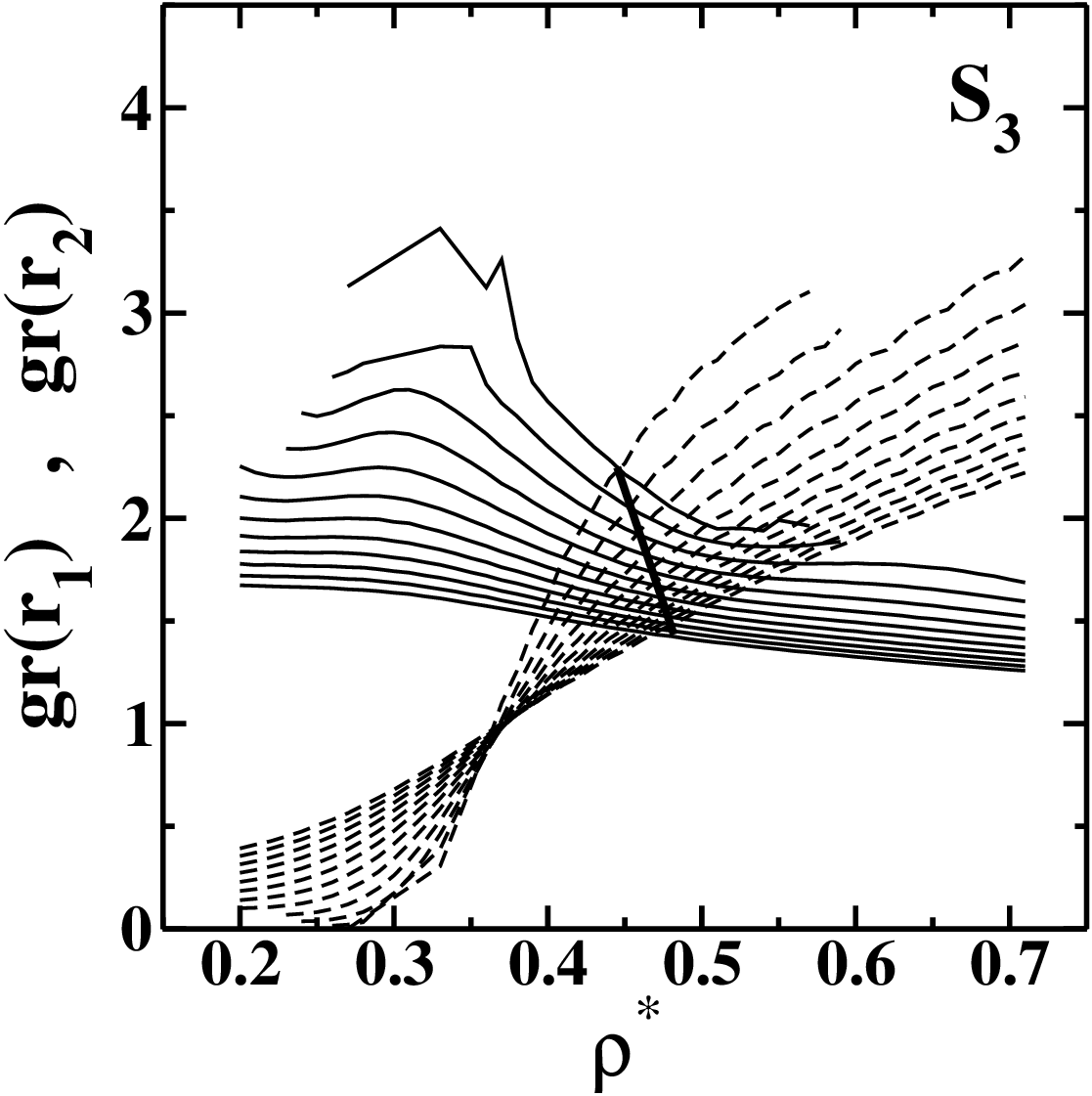} \\
      \includegraphics[clip,width=5cm]{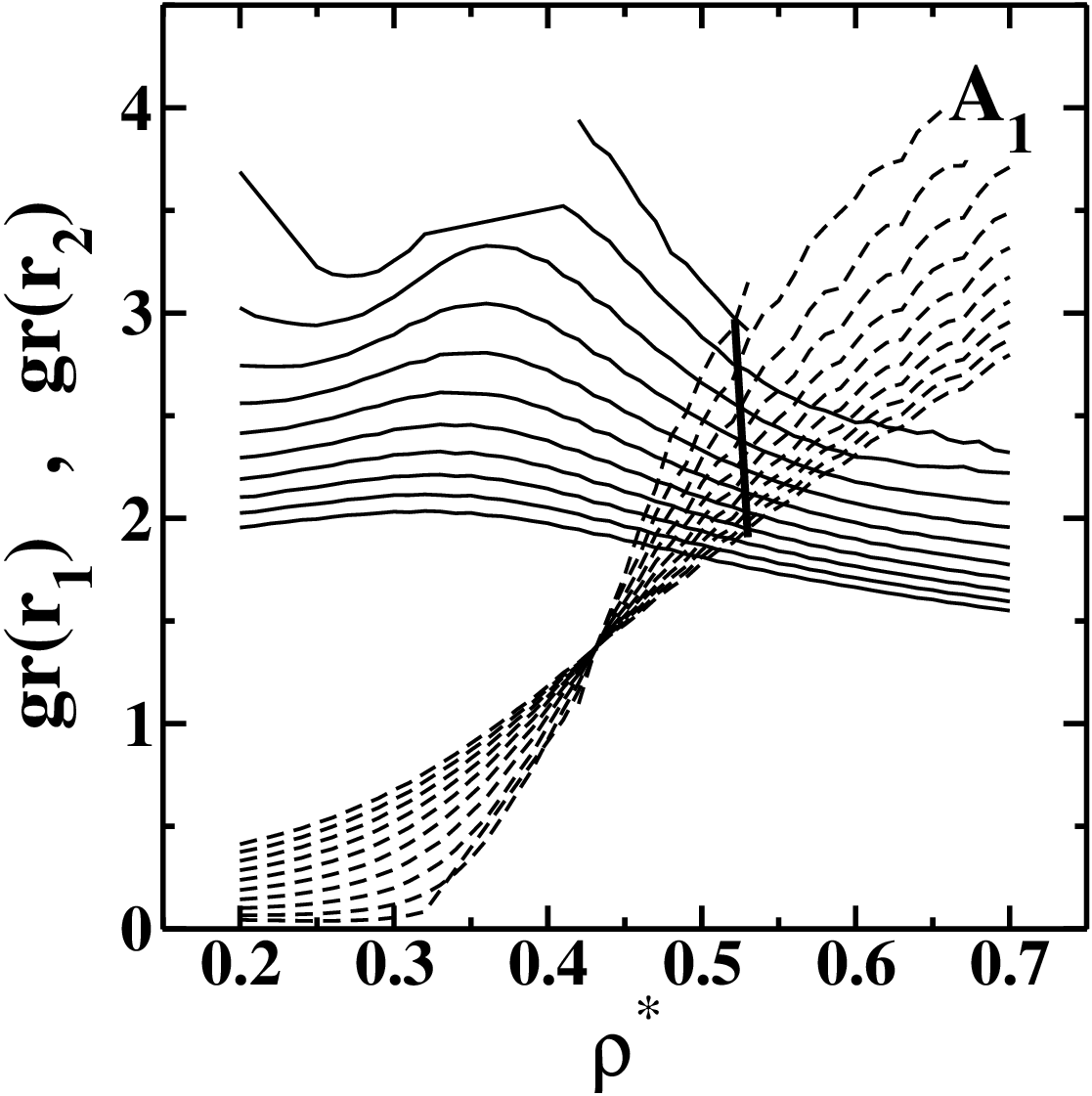}  & \includegraphics[clip,width=5cm]{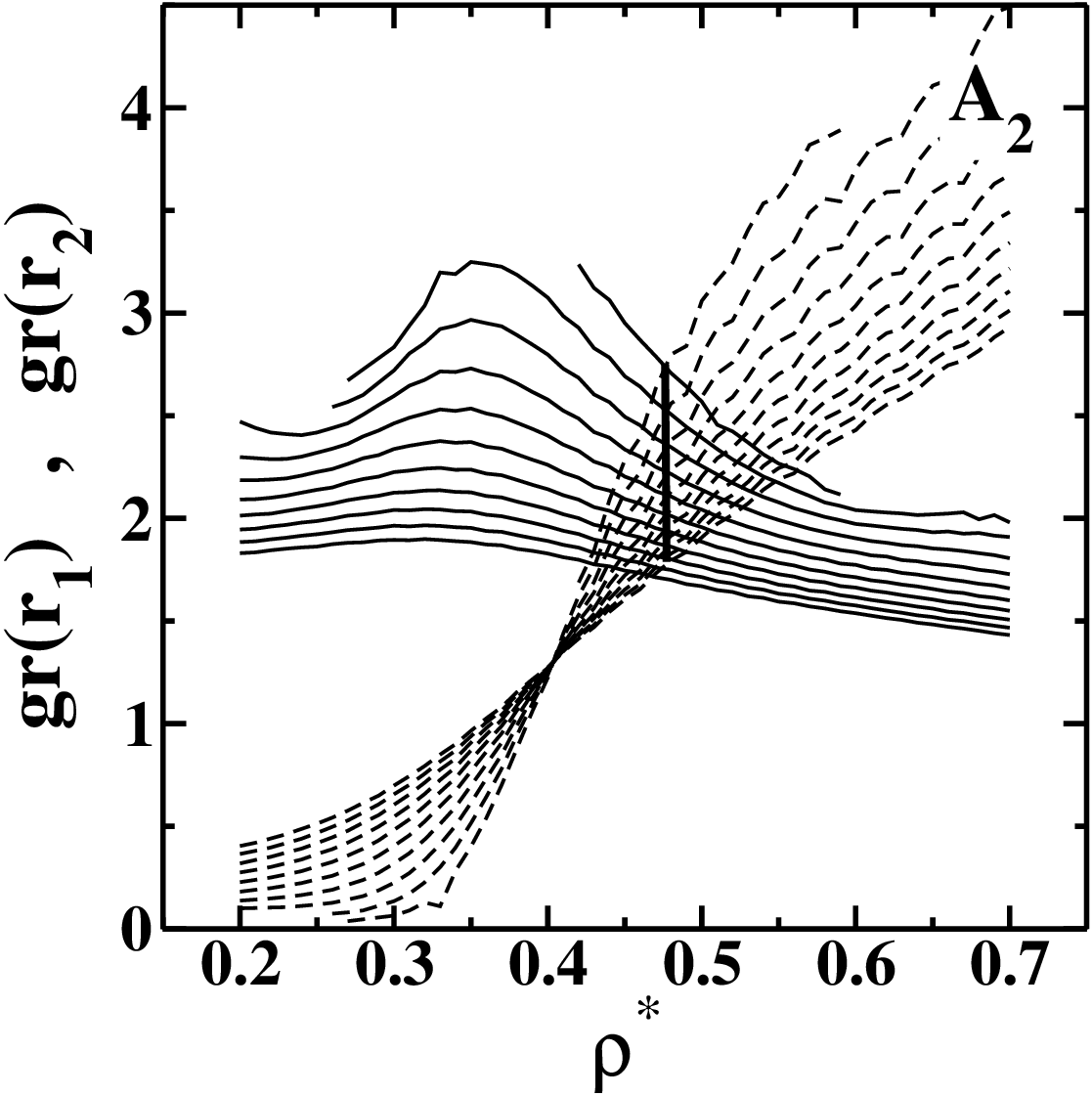}  & \includegraphics[clip,width=5cm]{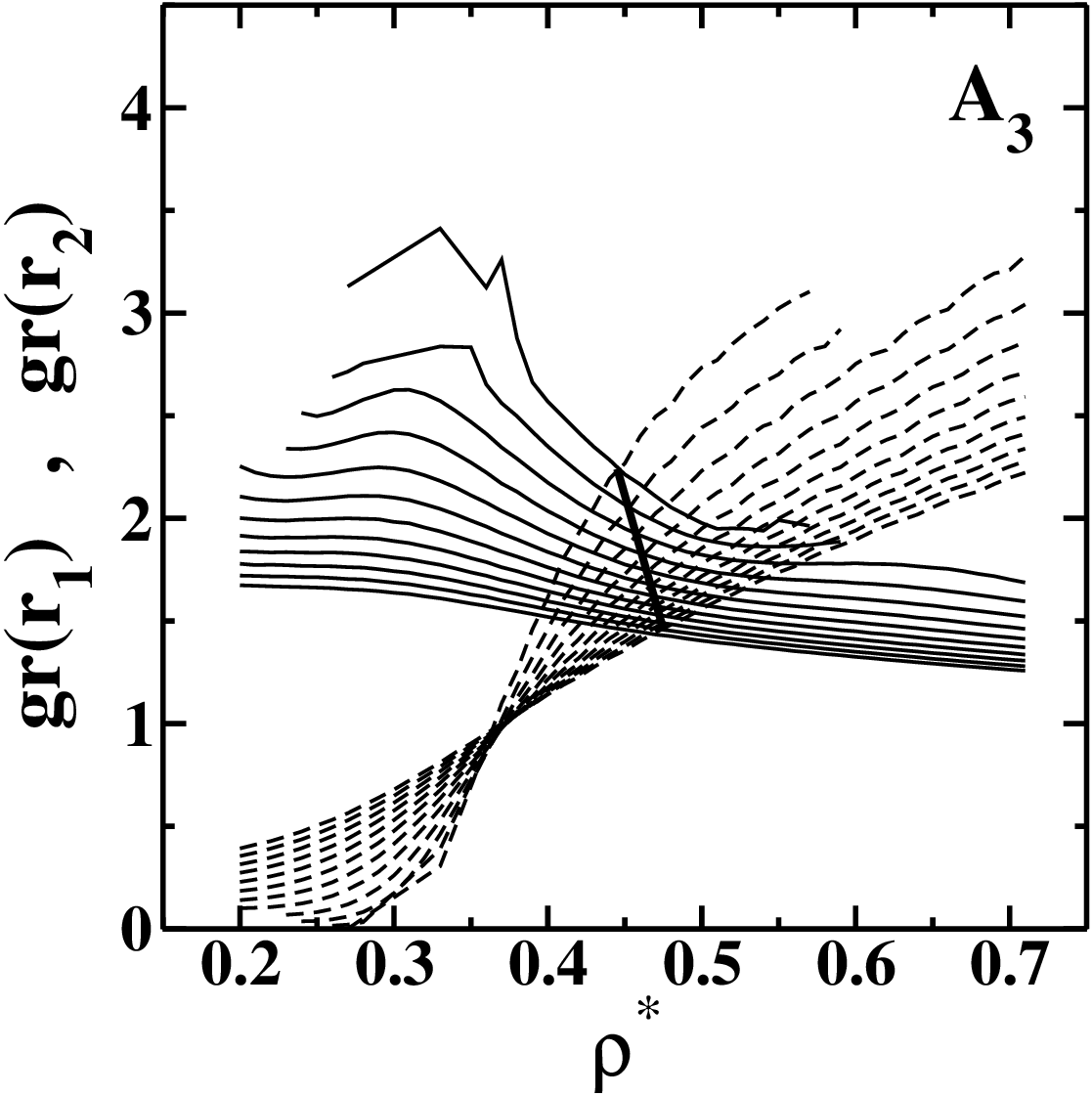} \\
      \includegraphics[clip,width=5cm]{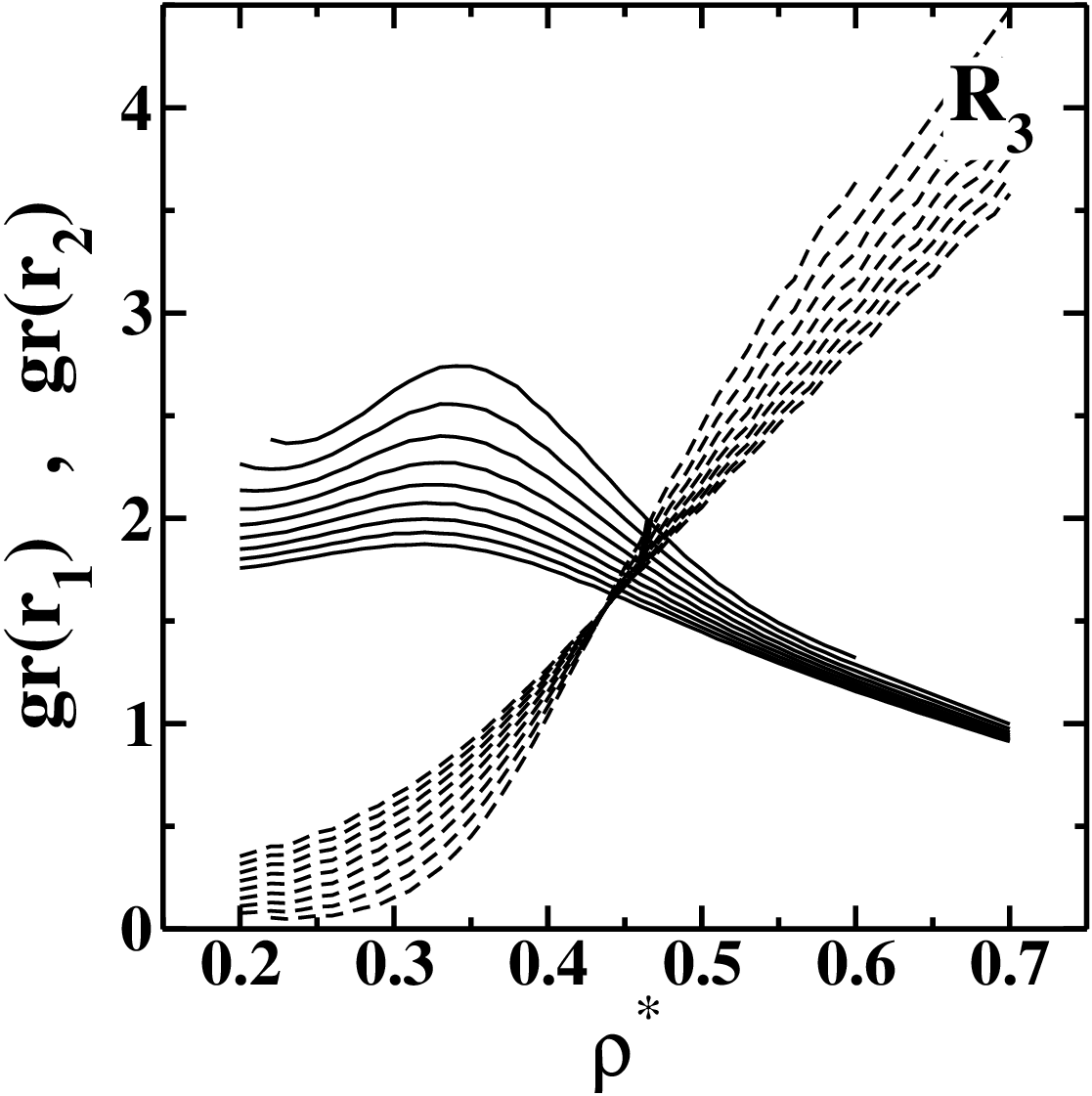}  & \includegraphics[clip,width=5cm]{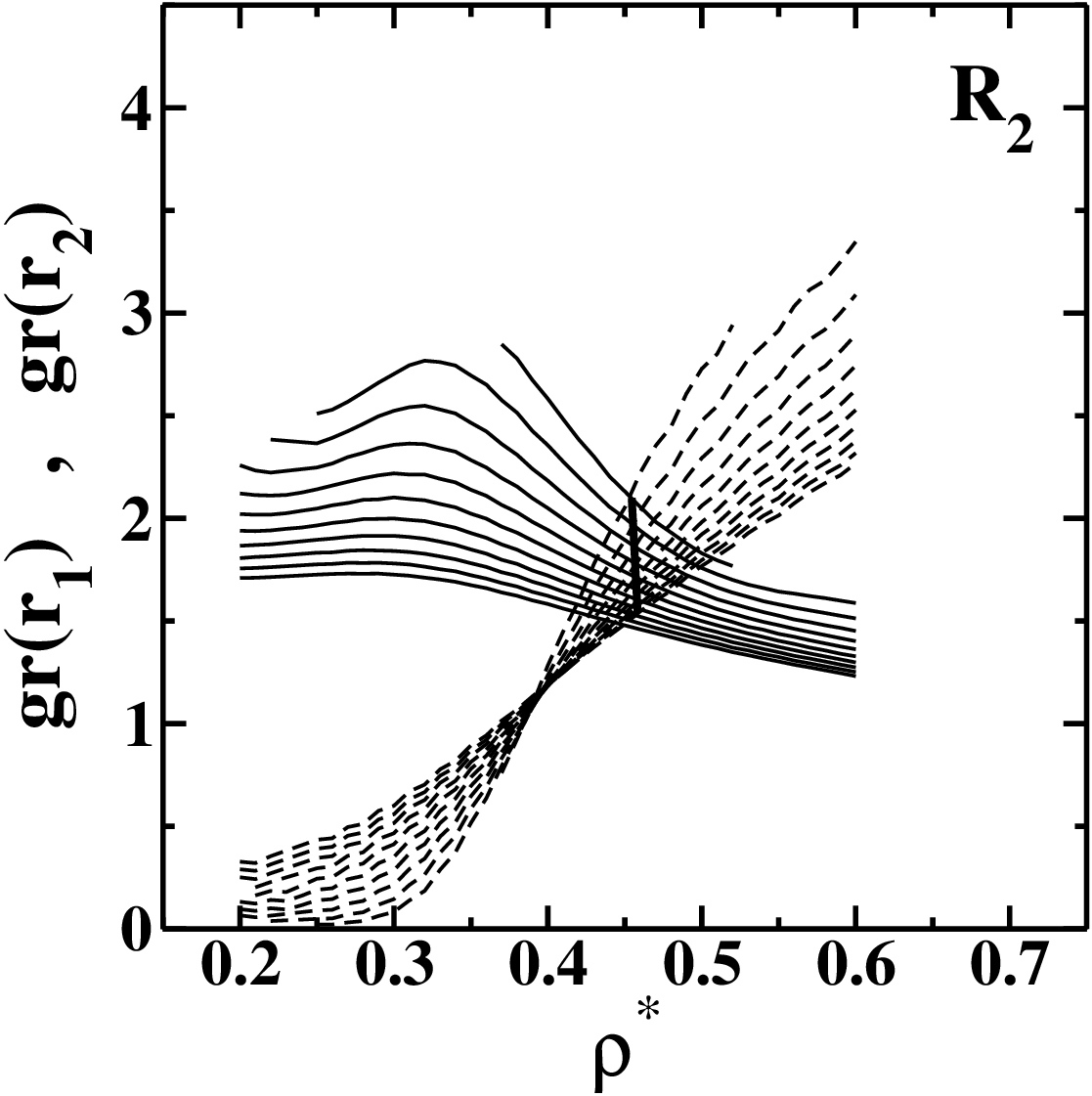}  & \includegraphics[clip,width=5cm]{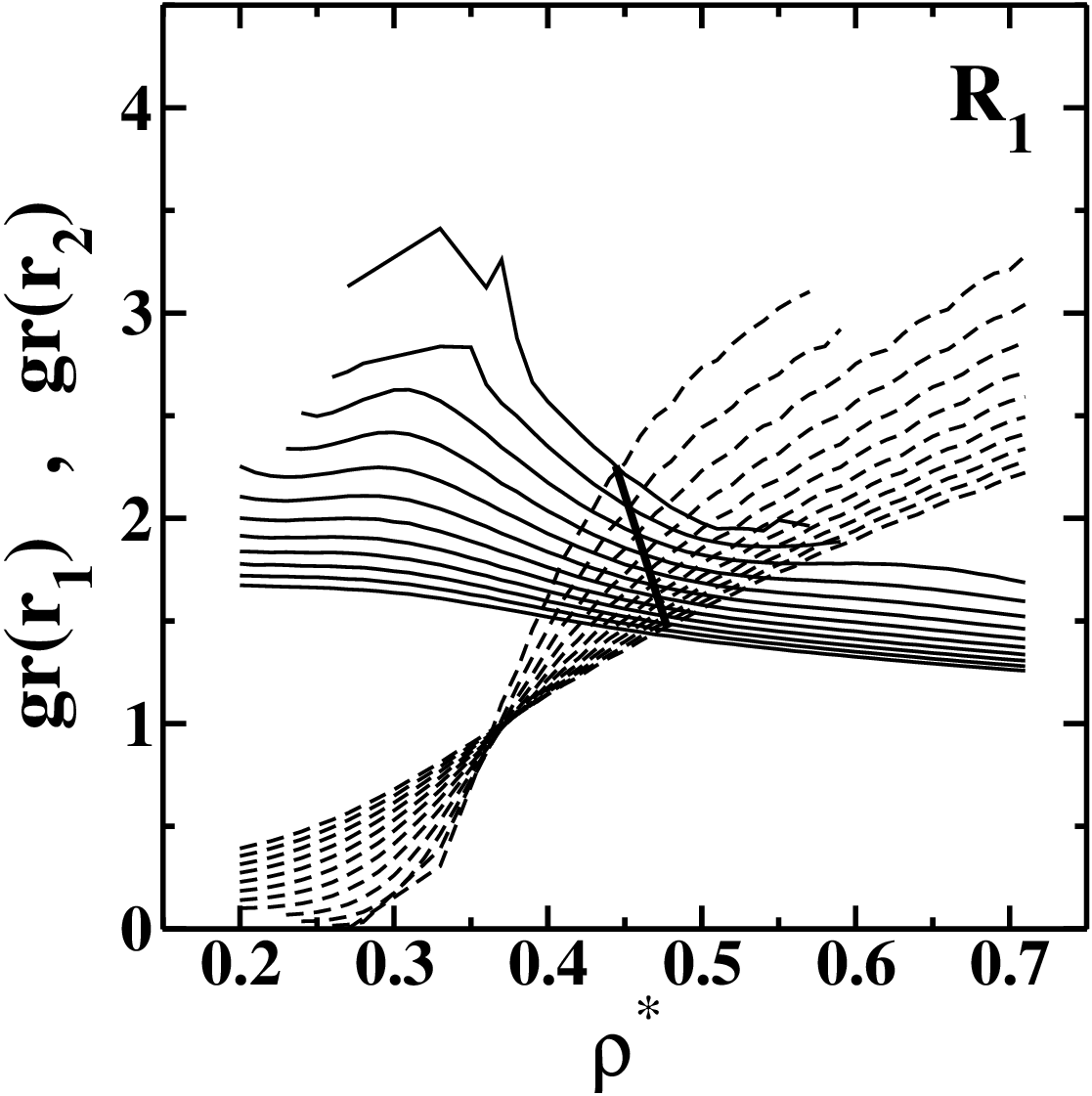} \\
    \end{tabular}
  \end{centering}
  \caption{$g(r_{1})$ and $g(r_{2}$) for the $S$, $A$ and $R$ families
of potentials as a function of the 
reduced density. The temperatures are  $T^*=0.4 \rightarrow 1.5$
 (from top to bottom $\rho^*>0.6$). The solid line connects the
    points, for different temperatures, where $g(r_{1})=g(r_{2})$.}
  \label{fig:grcros} 
\end{figure}

The Figs.~\ref{fig:grcros} show the value of the 
radial distribution function at the
closest, $g(r_1)$ (dashed lines), and at the furthest scale,
$g(r_2)$ (solid lines), as a function of the reduced density, $\rho^*$.
For the closest scale $g(r_1)$ is monotonic with density while
the value
for the $g(r_2)$ for a fixed temperature
increases with the density for densities
below the $\rho^*<\rho^*_{min}$ and decreases for densities
above this threshold. This behavior, also shown in 
the Fig.~\ref{fig:gr},
corroborates the condition
stated in the Eq.~\ref{eq:condition} and supports
the idea that particles move from one scale to
the other by compression at $\rho>\rho_{min}$~\cite{Ol08a}.

Besides to the move of particles
from the attractive scale to the repulsive
scale for $\rho^*>\rho^*_{min}$ as 
the pressure (density) is increased, 
particles also move from one scale to the
other due to the increase of temperature~\cite{Ba09} 
for $\rho^*<\rho^*_{min}$.
 At constant density, $\rho^*<\rho^*_{min}$,
the radial distribution function of  the attractive scale, $g(r_2)$,
decreases with the increase of the temperature while
$g(r_1)$ increases with
the increase of temperature, indicating
that particles move from one scale to the 
other due to thermal effects.  At the density $\rho^*_{min}$, the
value of $g(r_1)$ is
independent of the temperature.

\begin{figure}
  \begin{centering}
    \includegraphics[width=8cm]{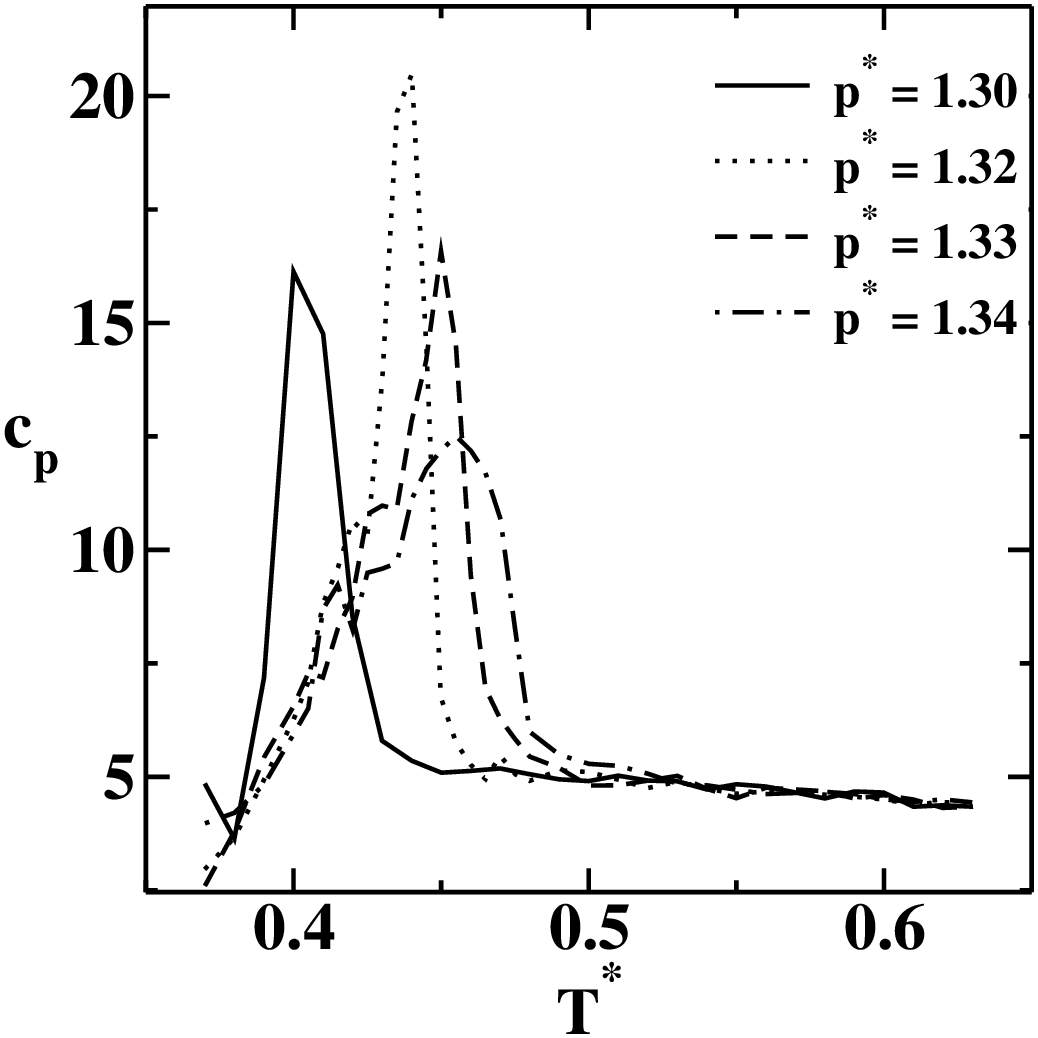}
  \end{centering}
  \caption{Isobaric specific heat for the $S_{1}$ potential
for different temperatures for different pressures.}
  \label{fig:cp}
\end{figure}

What is the meaning of the density $\rho_{min}$ in which $g(r_{1})$ is
independent of temperature? As it was pointed
in the previous paragraph, for $\rho^*<\rho^*_{min}$ particles move
from the furthest scale, $r_2$, to the closest scale, $r_1$, using
thermal energy as the temperature is increased. In this case $g(r_1)$
increases with temperature. For $\rho^*>\rho^*_{min}$ particles move
from $r_2$ to $r_1$, using the increase of pressure
pressure (or density) as illustrated
by the Eq.~\ref{eq:condition}.  From statistical point 
of view, the two mechanisms
governing the behavior for $\rho^*>\rho^*_{min}$ and
$\rho^*<\rho^*_{min}$ are quite different. While increasing
temperature affects particles individually, increasing the density or
the pressure affects the particles as clusters or networks. Then,
as the potential becomes more soft, the 
threshold density $\rho^*_{min}$ beyond which the particles
move form one scale to the other by compression should decrease as 
observed in Fig.~\ref{fig:grcros}. Therefore $\rho^*_{min}$ is 
the threshold between these two mechanisms present in systems
that have density anomaly.

In addition to these low density  limit, the density anomalous
systems also have a high density threshold, $\rho^*_{max}(T)$.
Fig.~\ref{fig:grcros}  illustrates as a solid thick line the temperatures
and densities,  $\rho^*_{max}(T)$,  in which $g(r_1)=g(r_2)$. 
Since $g(r)$ is related with the
number of particles at distance $r$, for $\rho^*<\rho^*_{max}(T)$
more particles are in the attractive scale, $r_2$, while
for $\rho^*>\rho^*_{max}(T)$ more particles
are at the repulsive scale $r_1$.  Therefore, the thick solid line is a
boundary  between the high density liquid
and the low density liquid. Analysis of the stability indicates that
no real phase transition is observed across this line.

 In
order to understand what what happens in the region of the
pressure-temperature phase diagram of the $\rho^*_{max}(T)$,
the behavior of the specific heat in this region
was
analyzed. Fig.~\ref{fig:cp} shows the curves of isobaric specific
heat, for different pressures as a function of the temperature for the
potential $S_{1}$.   The peak of $c_p$ for
each one of the potentials analyzed in this manuscript coincide with
the region $\rho^*_{max}(T)$ where $g(r_1)=g(r_2)$.  This result 
indicates that the
structure in the TMD region already build the liquid arrangements
required for the liquid-liquid phase separation.

\section{\label{sec:conclusions}  Conclusions}

In this paper we have studied three families of core-softened
potentials that exhibit two length
scales, one repulsive, $r_1$, and another
attractive, $r_2$.  

We had observed that the region in the
pressure-temperature phase diagram occupied 
by the TMD is quite sensitive to the slope between
the two length scales. As the slope increases
the region decreases. 
We  also found that the region in the
pressure-temperature phase diagram where
the density, diffusion and structural anomalous
behavior is observed  shifts to lower pressures and
shrinks  the 
attractive scales or the repulsive scales  become wider.
 Our results suggests the competition between two
length scales are the relevant mechanism for the existence of the TMD.

In an attempt to confirm this assertion we showed in this family of
potentials that the condition $\Pi_{1,2}<0$ seems to be associated
with the presence of anomalous behavior. In addition we also observed
that the peaks of the radial distribution function at each length
scales exhibit very distinct behavior with density and temperature
suggesting two complementary mechanisms for the competition between
the two scales.

At low densities,  $\rho^*<\rho^*_{min}$ particles move from $r_2$ to $r_1$
with the increase of temperature, using thermal
energy. For densities above $\rho^*_{min}$ the
increase of $g(r_1)$ is associated with the increase in the 
pressure (density). In
this interval of densities at a certain density and temperature the
radial distribution function at the first length scale equals the
value at the second scale, namely $g(r_1)=g(r_2)$. This point can be
identified with the Widom line.

The relation between the radial distribution function and the Widom
line, believed to be the onset of the liquid-liquid phase transition
give the support to the idea that the Widom line separates two
structural distinct regions that are also separated by a fragile-strong
transition~\cite{Xu05}.

We expect that his result will not only shade some light in the
definition of what is the shape an effective core-softened would have
in order to held anomalies but also would serve to reinforce the ideas
of linking dynamic transitions and thermodynamic properties.

\section*{ACKNOWLEDGMENTS}

We thank for financial support the Brazilian science agencies CNPq and
Capes. This work is partially supported by CNPq, INCT-FCx.

\end{document}